\documentclass[12pt]{article}

 \pdfoutput=1

\usepackage{epsfig}
\usepackage{color,graphicx}
\usepackage{cite}
\usepackage{amssymb,amsmath}
\usepackage{a4wide}
\usepackage{amssymb}
\usepackage{amsmath}
\usepackage{graphicx}
\usepackage{mathdots}
\usepackage{youngtab}
\usepackage{caption}
\usepackage{subcaption}
\usepackage{enumerate}

\newcommand{\be}{\begin{equation}}
\newcommand{\ee}{\end{equation}}
\newcommand{\bea}{\begin{eqnarray}}
\newcommand{\eea}{\end{eqnarray}}

\def\drawbox#1#2{\hrule height#2pt 
        \hbox{\vrule width#2pt height#1pt \kern#1pt 
              \vrule width#2pt}
              \hrule height#2pt}

\def\Fund#1#2{\vcenter{\vbox{\drawbox{#1}{#2}}}}
\def\Asym#1#2{\vcenter{\vbox{\drawbox{#1}{#2}
              \kern-#2pt       
              \drawbox{#1}{#2}}}}

\def\funda{\Fund{6.5}{0.4}}
\def\asymm{\Asym{6.5}{0.4}}

\def\symm{\funda\kern-0.4pt\funda}

\begin{document}

\begin{center}  

\vskip 2cm 

\centerline{\Large {\bf 5-Brane Webs, Symmetry Enhancement, and Duality}}

\vskip 0.2cm

\centerline{\Large {\bf in 5d Supersymmetric Gauge Theory}}

\vskip 1cm

\renewcommand{\thefootnote}{\fnsymbol{footnote}}

   \centerline{
    {\large \bf Oren Bergman${}^{a}$}\footnote{bergman@physics.technion.ac.il},
    {\large \bf Diego Rodr\'{\i}guez-G\'omez${}^{b}$}\footnote{d.rodriguez.gomez@uniovi.es}
    {\bf and}
    {\large \bf Gabi Zafrir${}^{a}$}\footnote{gabizaf@techunix.technion.ac.il}}

\vspace{1cm}
\centerline{{\it ${}^a$ Department of Physics, Technion, Israel Institute of Technology}} \centerline{{\it Haifa, 32000, Israel}}
\vspace{1cm}
\centerline{{\it ${}^b$ Department of Physics, Universidad de Oviedo}} \centerline{{\it Avda. Calvo Sotelo 18, 33007, Oviedo, Spain }}
\vspace{1cm}

\end{center}

\vskip 0.3 cm

\setcounter{footnote}{0}
\renewcommand{\thefootnote}{\arabic{footnote}}   
   
\begin{abstract}
We present a number of investigations of 5d ${\cal N}=1$ supersymmetric gauge theories that make use 
of 5-brane web constructions and the 5d superconformal index.
These include an observation of enhanced global symmetry in the 5d fixed point theory corresponding to 
$SU(N)$ gauge theory with Chern-Simons level $\pm N$,
enhanced global symmetries in quiver theories, and
dualities between quiver theories and non-quiver theories.
Instanton contributions play a crucial role throughout.
%
 \end{abstract}
 
 \newpage
 
\tableofcontents

\section{Introduction}

Interacting quantum field theories in 5d are non-renormalizable and therefore
do not generically exist as microscopic theories.
Nevertheless there is compelling evidence that there exist ${\cal N}=1$ supersymmetric interacting 
fixed point theories in 5d, some of which have relevant deformations corresponding to ordinary 
gauge theories with matter \cite{Seiberg:1996bd,Morrison:1996xf,Douglas:1996xp,Intriligator:1997pq,Aharony:1997ju}.
This follows from the special properties of gauge theories
with eight supersymmetries in five dimensions, namely that the pre-potential function 
of the low-energy effective theory on the Coulomb branch ${\cal F}(\phi)$ is one-loop exact and at most cubic.
The effective gauge coupling is then schematically given by
\be
\label{eff_coupling}
\frac{1}{g_{eff}^2(\phi)} = \frac{\partial^2{\cal F}}{\partial\phi^2} = \frac{1}{g_0^2} + c |\phi| \,,
\ee
where the number $c$ has both tree-level contributions, related to a bare 5d Chern-Simons term,
as well as one-loop contributions coming from integrating out massive gauge and matter degrees
of freedom.
If $c>0$ one can remove the UV cutoff, namely take $g_0\rightarrow\infty$,
without encountering a singularity on the moduli space.
Since $g_{eff}(0)\rightarrow\infty$, this corresponds to a strongly-interacting 
fixed point. 
All the gauge groups and matter content satisfying this condition were classified in \cite{Intriligator:1997pq}.

A number of these theories can be realized using brane configurations in string theory.
For example D4-branes in Type I' string theory realize a 5d ${\cal N}=1$ gauge theory with
an $Sp(N)$ gauge group, a hypermultiplet in the antisymmetric representation,
and $N_f$ hypermutiplets in the fundamental representation, where $N_f$ is the number of D8-branes
near the D4-branes \cite{Seiberg:1996bd}. 
For $N_f\leq 7$ the background can be arranged such that the effective Yang-Mills coupling diverges
at the origin of the Coulomb branch,
and this corresponds to the superconformal fixed point theory.
The Type I' construction suggests that 
these theories exhibit a non-perturbative enhancement of the global symmetry from $SO(2N_f)\times U(1)_T$ to $E_{N_f +1}$,
where $U(1)_T$ is the topological symmetry associated to the instanton number current $j_T = *\mbox{Tr}(F\wedge F)$.

This was recently confirmed from the field theory viewpoint by computing the 5d superconformal index  
via localization \cite{KKL}.
This is a remarkable computation, which required taking into account non-perturbative
contributions from instanton operators in the gauge theory.
In the final result, the index can be expressed in terms of characters of $E_{N_f+1}$,
beautifully showing the enhanced global symmetry.
The results of \cite{KKL} can also be extended to other 5d superconformal gauge theories, 
such as the so-called $\tilde{E}_1$ theory, which does not exhibit enhancement \cite{Bergman:2013ala,Iqbal:2012xm}.
In this paper we will use the superconformal index to study several other 5d superconformal gauge theories,
some of which exhibit non-perturbatively enhanced global symmetries.

Another useful tool in the study of ${\cal N}=1$ 5d theories has been $(p,q)$5-brane web configurations in 
Type IIB string theory \cite{Aharony:1997ju}.
These configurations can describe both ordinary gauge theories with 5d UV fixed points, as well as superconformal
theories that do not have a gauge theory description.
In this construction the parameters and moduli of the gauge theory are described by the relative positions 
of the 5-branes. This allows one to consider a continuation ``past infinite coupling" in the gauge theory
by appropriately shifting the positions of the 5-branes.
In some cases this leads to another gauge theory, which we will refer to as the dual gauge theory.
Thus a single strongly-interacting superconformal theory may be deformed to two different 
weakly-interacting IR gauge theories.
This is different from the usual sense of duality in lower dimensions, which relates different UV theories that 
flow to the same IR theory. 

Quiver gauge theories, namely theories with product gauge groups and bifundamental matter fields, 
are examples where the idea of continuation past infinite coupling is relevant.
Such theories were originally ruled out by the argument based on (\ref{eff_coupling}),
since they always become strongly coupled somewhere out on the Coulomb branch \cite{Intriligator:1997pq}.
On the other hand 5-brane web constructions of quiver theories indicate that in some cases
a continuation past infinite coupling is possible, and leads to a different gauge theory 
which is finitely-coupled on its Coulomb branch \cite{Aharony:1997ju,Aharony:1997bh}.\footnote{The existence
of certain quiver theory fixed points is also supported by the existence of 
large $N$ supergravity duals \cite{Bergman:2012kr}.}

While these dualities are well-motivated by 5-brane webs, they have not been systematically studied, and
the full extent of their meaning has not been explored.\footnote{A certain family of dual theories was considered in \cite{Bao:2011rc}.
The lowest rank example coincides with one of our examples.}
For example the mapping of the symmetries and charges between dual theories has not been carried out.
Here too, the superconformal index should provide a useful diagnostic.
The index essentially counts BPS operators in the superconformal theory, and is therefore protected
from continuous deformations \cite{KMMS}. Therefore the dual theories, corresponding to opposite deformations
of the fixed point theory, should have the same index.
Furthermore, by comparing the contributions to the index with given charges in the two cases,
we can derive the precise map between them.
Our present interest is in the quiver theories corresponding to D4-branes in orbifolds of Type I' 
string theory \cite{Bergman:2012kr}, specifically the even orbifolds with vector structure,
for which there exist 5-brane web constructions.
We will study a number of examples with low rank gauge groups and a small number of matter fields.
In each case, the 5-brane web description will suggest the identity of the dual gauge theory beyond infinite coupling.
We will then proceed to compute the index for the two theories as a test of the conjectured duality.

\medskip

The outline for the rest of the paper is as follows.
In section 2 we will review some basic properties of 5-brane webs,
and their interpretation in terms of 5d ${\cal N}=1$ gauge theories,
paying close attention to the case with parallel external 5-branes.
In section 3 we will review the general strategy for computing the superconformal index
for 5d gauge theories via localization, as developed in \cite{KKL}.
We will also point out some problems with this approach related to the
contribution of instantons as obtained from the Nekrasov partition function.
In section 4 we will study a set of examples of webs with parallel external 5-branes corresponding to $SU(N)$
gauge theory with a level $\pm N$ Chern-Simons term, and show that these theories
have an enhanced $SU(2)$ global symmetry.
In section 5 we study the simplest quiver theory, $SU(2)\times SU(2) + ({\bf 2},{\bf 2})$,
and confirm that its dual is $SU(3) + 2\cdot {\bf 3}$.
We also show that variants of the quiver theory exhibit global symmetry enhancement to $SU(3)$ and $SU(4)$.
In section 6 we study a number of generalizations to higher rank, extra flavor and an extra quiver node.
In each case we will compare the superconformal indices, including instanton corrections, and show that they are equal.
Section 7 contains our conclusions, and the Appendices contain the explicit formulas for the
instanton partition functions that we use and the explicit expressions for the superconformal 
indices to the highest order that we computed.

\section{5-brane web basics}

5-brane webs provide a very general approach to realizing 5d quantum field theories in the context of string theory 
\cite{Aharony:1997ju,Aharony:1997bh}.
These are planar configurations of connected $(p,q)$5-branes in Type IIB string theory, where some of the 5-branes
are internal, having a finite extent on the plane, and others are external, being semi-infinite on the plane.
The charges must sum to zero at the vertices, and
the 5-branes are oriented such that 1/4 of the background supersymmetry is preserved.
This gives, at low energy, a 5d ${\cal N}=1$ quantum field theory living on the internal 5-branes.
In many cases this is a gauge field theory with a known action.
The relative positions of the external 5-branes correspond to parameters of the field theory,
or equivalently to VEVs of scalars in background vector multiplets associated to the global symmetries of the theory.
The number of real parameters is given by the number of external 5-branes minus 3.
Overall translations of the web in the plane give the same theory, and the position of 
one external 5-brane is fixed by the positions of all the others.
Deformations of the web that keep the planar positions of the external 5-branes fixed correspond to the moduli
of the theory. Those that move the external 5-branes in directions transverse to the plane are Higgs branch
moduli, and those that do not are associated with the Coulomb branch.
The latter are identified with the number of faces in the web.
In cases where the web becomes singular, with all external 5-brane lines meeting at a point, it describes a 5d superconformal theory.
However, if any external 5-branes intersect away from the origin of the Coulomb branch the theory is not well-defined.

\begin{figure}[h]
\center
\begin{subfigure}[]{0.4\textwidth}
\center
\includegraphics[width=\textwidth]{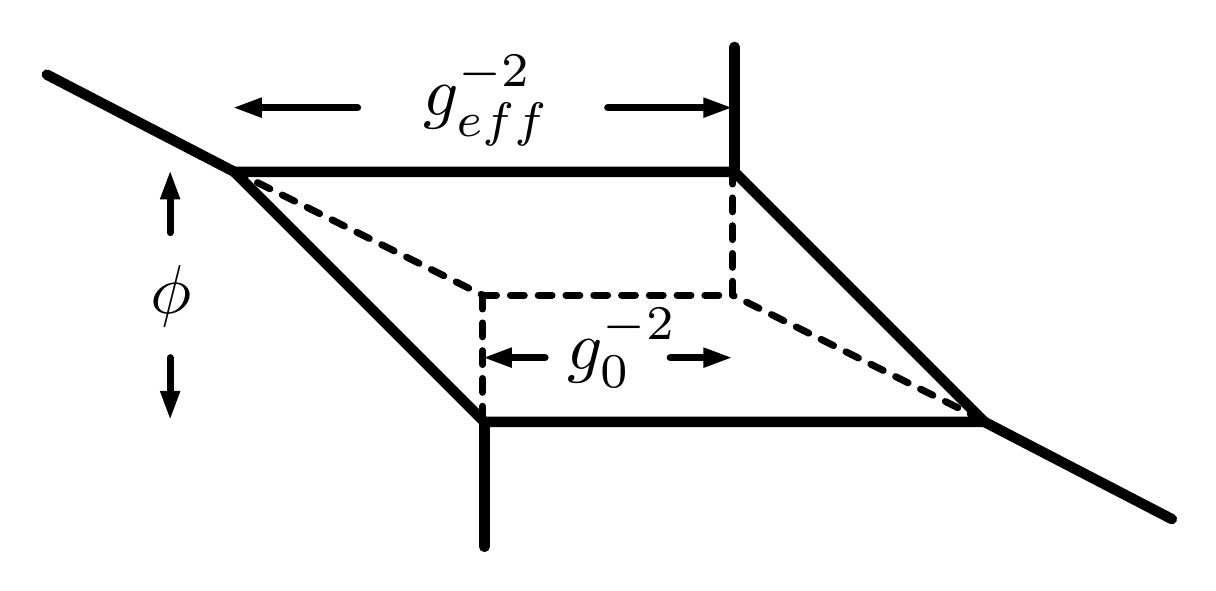} 
\caption{}
\label{}
\end{subfigure}
\hspace{40pt}
\begin{subfigure}[]{0.3\textwidth}
\center
\includegraphics[width=\textwidth]{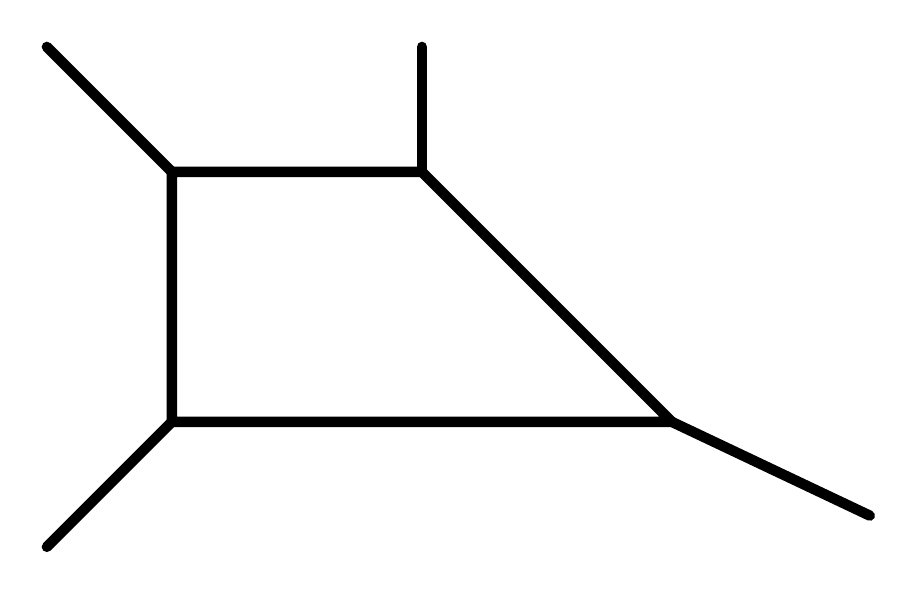} 
\caption{}
\label{}
\end{subfigure}
\caption{$SU(2)$ webs. (a) $SU(2)_0$ ``$E_1$ theory", (b) $SU(2)_\pi$ ``$\tilde{E}_1$ theory"}
\label{web_SU(2)}
\end{figure}

The simplest example of a 5d superconformal theory is the pure supersymmetric $SU(2)$ theory.
The 5-brane web of this theory, together with the bare and effective coupling and Coulomb modulus, 
is shown in Fig.~\ref{web_SU(2)}a.\footnote{This web is related to the Type I' configuration by T-duality.
The O8-plane becomes two O7-planes at antipodal points on the circle, and the D4-brane becomes 
a pair of D5-branes wrapping it. Each O7-plane is then resolved into a pair of 7-branes, resulting in the 
5-brane web, with the four external 5-branes ending on the four 7-branes.
The charges of the 7-branes depend on the $SL(2,\mathbb{Z})$ frame.}
This web corresponds to a particular $SL(2,\mathbb{Z})$ frame, 
in which the charges of the external 5-branes are $(0,1)$ and $(2,-1)$.
(We have also fixed $g_s=1$ and the 10d Type IIB axion to zero.)
In other frames, related by $SL(2,\mathbb{Z})$ to this one, the web looks different but the theory is identical.
An $SL(2,\mathbb{Z})$-inequivalent $SU(2)$ web is shown in Fig.~\ref{web_SU(2)}b.
This web describes a different fixed point theory known as the
$\tilde{E}_1$ theory \cite{Morrison:1996xf}, in which the global symmetry is not enhanced.
The difference in the gauge theories is the value of a discrete $\theta$ parameter associated with 
$\pi_4(SU(2))=\mathbb{Z}_2$ \cite{Douglas:1996xp}. The $E_1$ theory corresponds to $\theta = 0$, and the $\tilde{E}_1$ theory 
to $\theta = \pi$.
There is actually one more $SL(2,\mathbb{Z})$-inequivalent web, which has parallel external 5-branes.
We will discuss it below.

\begin{figure}[h]
\center
\begin{subfigure}[]{0.35\textwidth}
\center
\includegraphics[width=\textwidth]{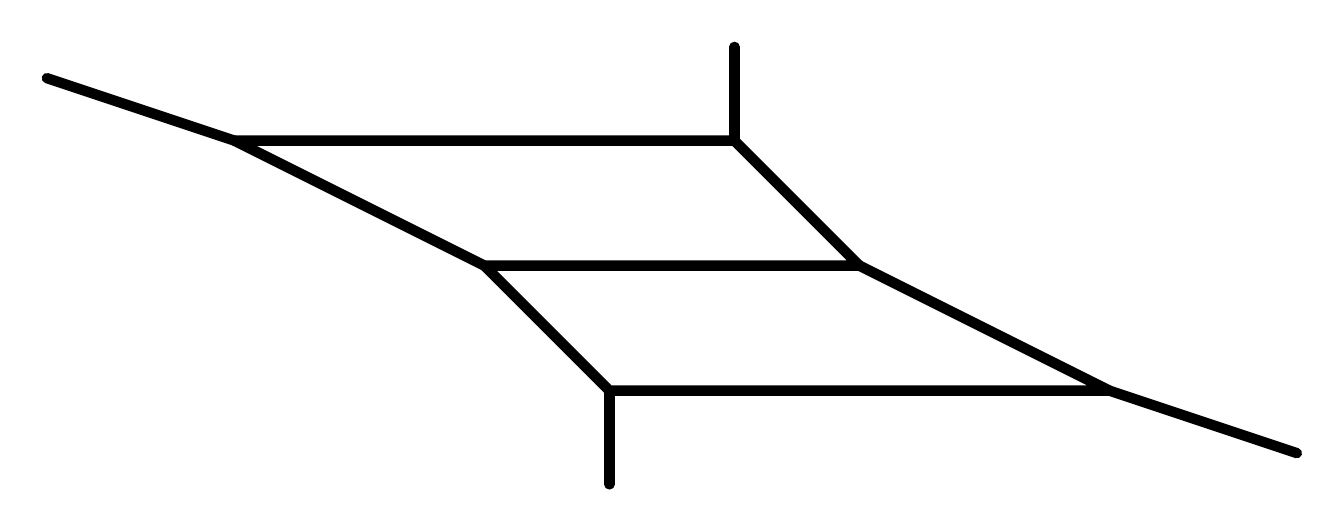} 
\caption{}
\label{}
\end{subfigure}
\begin{subfigure}[]{0.3\textwidth}
\center
\includegraphics[width=\textwidth]{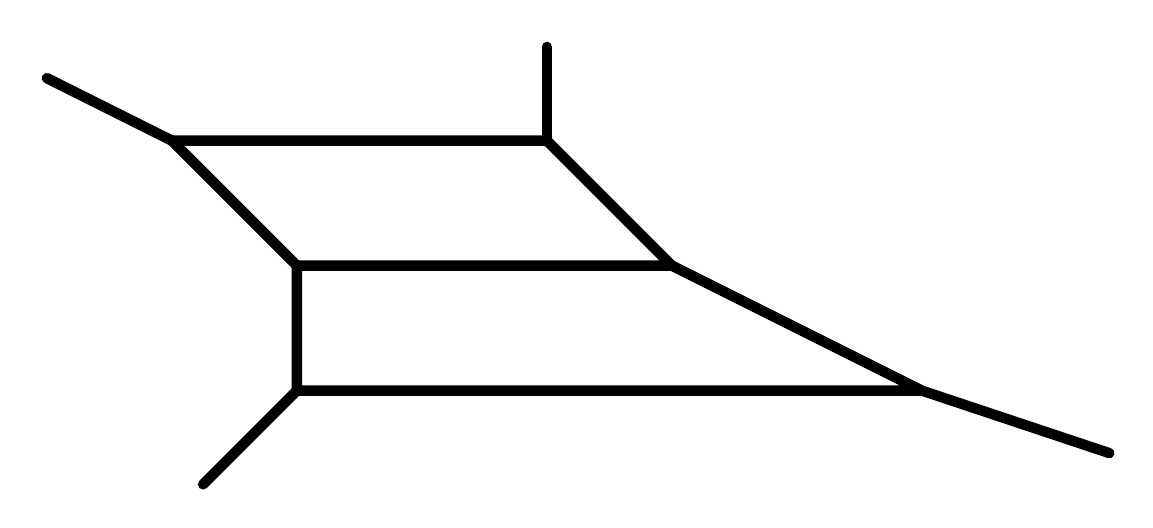} 
\caption{}
\label{}
\end{subfigure}
\begin{subfigure}[]{0.3\textwidth}
\center
\includegraphics[width=\textwidth]{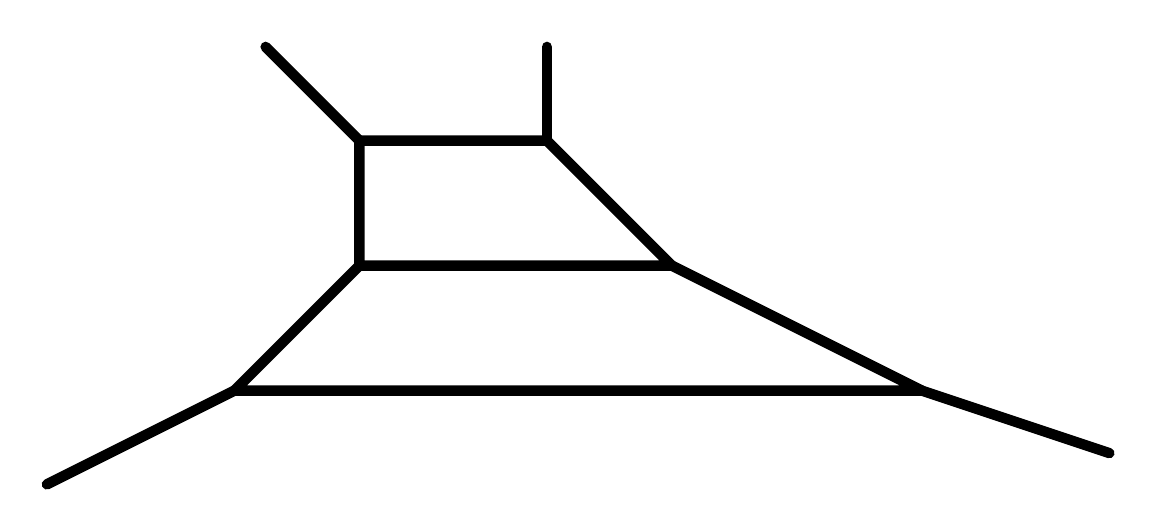} 
\caption{}
\label{}
\end{subfigure}
\caption{$SU(3)$ webs with CS level $\kappa = 0,1,2$.}
\label{web_SU(3)}
\end{figure}

This construction can be easily generalized to $SU(N)$ by including more internal D5-branes.
In this case the different $SL(2,\mathbb{Z})$ equivalence classes correspond to different CS levels $\kappa$
for the $SU(N)$ gauge field. 
In Fig.~\ref{web_SU(3)} we show the webs for pure $SU(3)$ with $\kappa = 0,1,2$.\footnote{The identification 
of the CS level associated to the web is analogous to the 3d case, where the gauge theory
on D3-branes suspended between an NS5-brane and a $(1,\kappa)$5-brane has a level $\kappa$ CS term
\cite{Kitao:1998mf,Bergman:1999na}.}
The webs for $\kappa < 0$ are related to those with $\kappa > 0$ by a $\pi$ rotation.
This corresponds to charge conjugation in the gauge theory, under which $\kappa \rightarrow -\kappa$.
Note that the web identified with $\kappa = 0$, Fig.~\ref{web_SU(3)}a, is invariant under this 
operation.
The $\kappa = N$ case is analogous to the additional $SU(2)$ web, which we will discuss below.

Adding matter is also straightforward.
Matter hypermultiplets (usually) correspond to external D5-branes. 
We will see several examples below, including one where the matter comes from other $(p,q)$5-branes.

\subsection{Continuation past infinite coupling}

A web describing $SU(N)$ with $N_f$ matter multiplets has $N_f+1$ real parameters.
These can be identified with the gauge coupling parameter $m_0 = 1/g_0^2$ and the $N_f$ masses of the hypermultiplets.
We denote the coupling parameter $m_0$ to stress that it too is a real mass parameter, corresponding to
the mass of the instanton particle at the origin of the Coulomb branch.
These mass parameters correspond to relevant deformations of the 5d superconformal theory.
Giving any of them nonzero values, by moving external 5-branes, 
generates an RG flow to a different theory in the IR,
which in many cases is a free 5d supersymmetric gauge theory.
In some cases the IR theory is another interacting SCFT.

\begin{figure}[h]
\center
\begin{subfigure}[]{0.45\textwidth}
\center
\includegraphics[width=\textwidth]{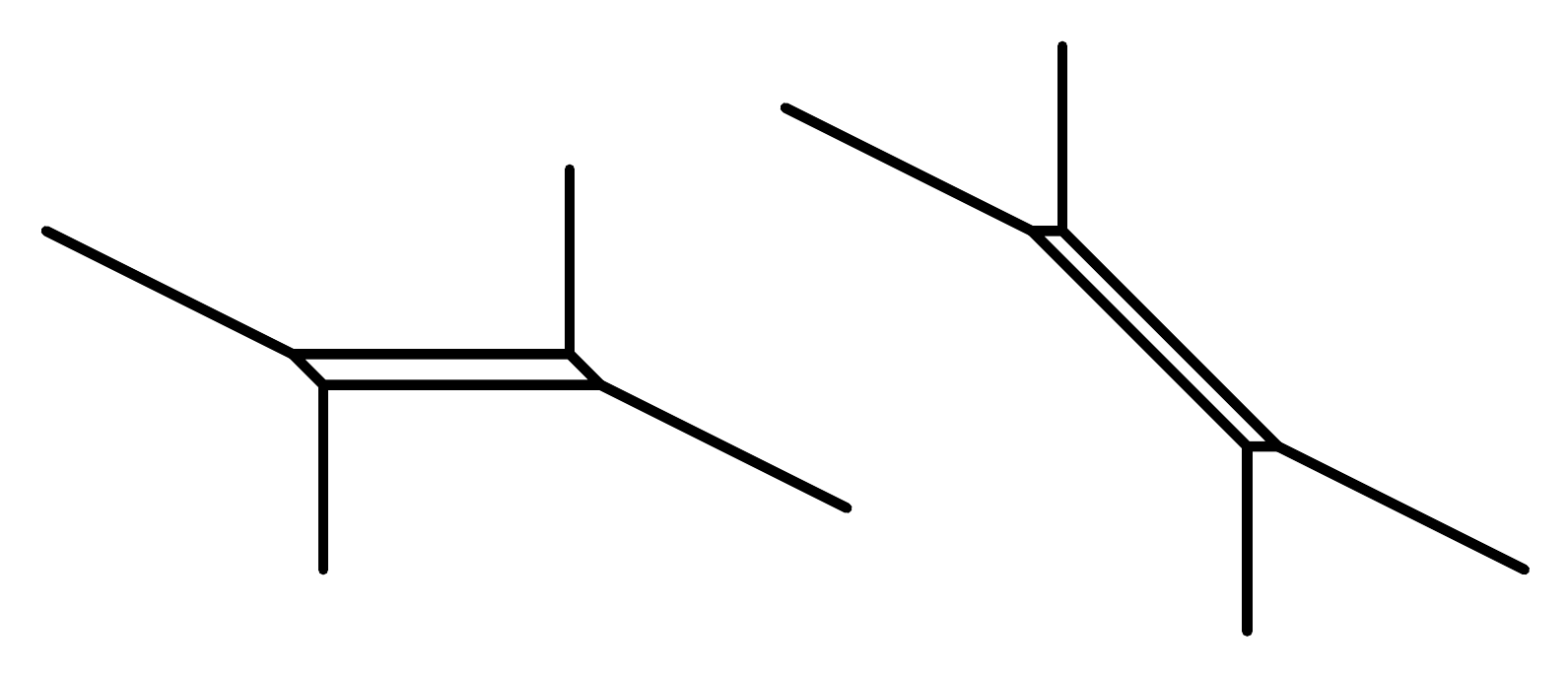} 
\caption{}
\label{}
\end{subfigure}
\hspace{1cm}
\begin{subfigure}[]{0.4\textwidth}
\center
\includegraphics[width=\textwidth]{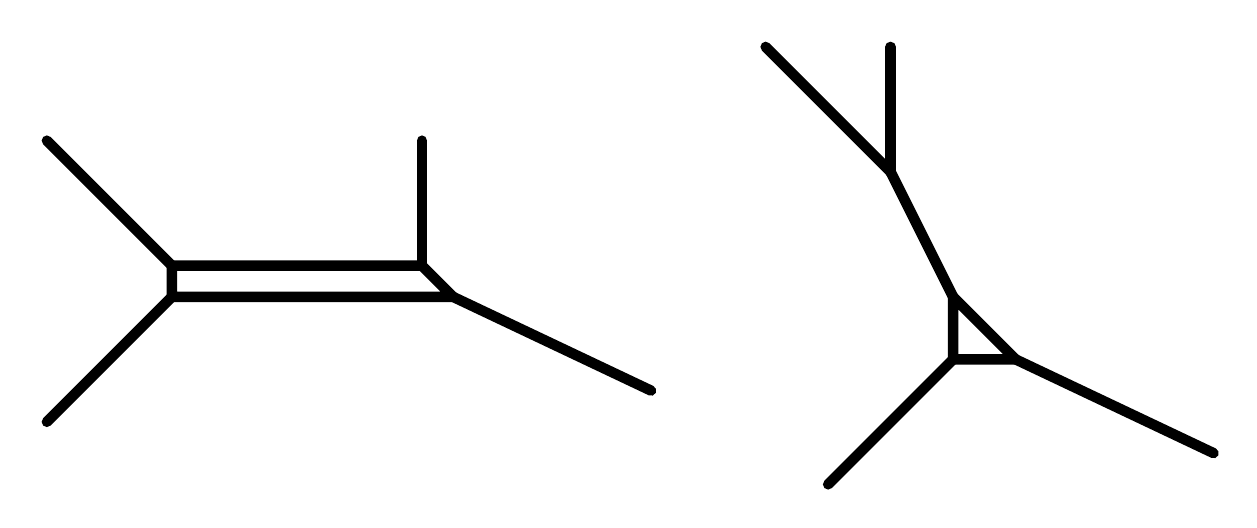}
\caption{}
\label{}
\end{subfigure}
\caption{Positive and negative deformations of the $SU(2)$ theories.}
\label{web_SU(2)_deformed}
\end{figure}

In 5d all mass parameters are real, and can take both positive and negative values.
In particular one can deform the SCFT with a negative value for the mass parameter
corresponding to the Yang-Mills coupling. This can be thought of as a continuation past 
infinite coupling of the gauge theory coming from the positive mass deformation.
In some cases this leads to a different gauge theory whose coupling is identified with minus the mass.
This is seen in the example of the $E_1$ theory.
The two deformations of the $E_1$ web, Fig.~\ref{web_SU(2)_deformed}a, 
are related by $SL(2,\mathbb{Z})$, and therefore describe the same IR-free $SU(2)$ SYM theory.
In other words, the 5d ${\cal N}=1$ $SU(2)$ YM theory with $\theta=0$ is self dual.
On the other hand for $\theta = \pi$, the continuation past infinite coupling leads to the 
interacting $E_0$ fixed point, Fig.~\ref{web_SU(2)_deformed}b.

\subsection{Parallel external legs}
\label{parallel_legs}

\begin{figure}
\centering
\begin{subfigure}[]{0.3\textwidth}
\centering
\includegraphics[width=\textwidth]{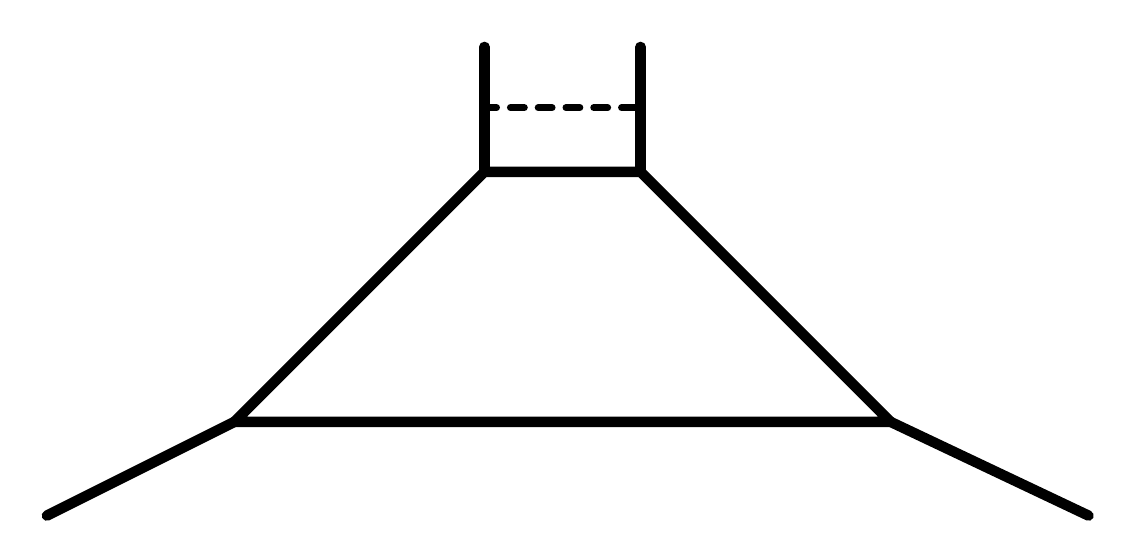} 
\caption{}
\label{}
\end{subfigure}
\hspace{2cm}
\begin{subfigure}[]{0.15\textwidth}
\centering
\includegraphics[width=\textwidth]{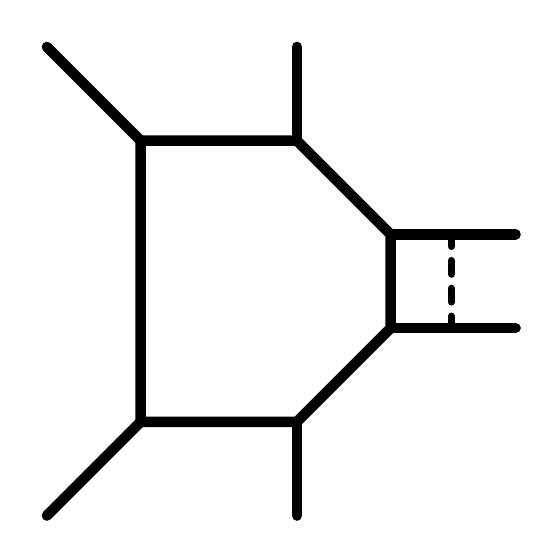} 
\caption{}
\label{}
\end{subfigure}
\caption{Webs with parallel external 5-branes. The dashed line corresponds to the decoupled state.}
\label{web_parallel_legs}
\end{figure}

The case of webs with parallel external 5-branes, {\em e.g.} Fig.~\ref{web_parallel_legs}, is special.
Originally, the question was raised about whether these actually describe 
well-defined fixed point theories, since there are light states associated to the parallel external 5-branes 
that do not obviously decouple from those on the internal 5-branes \cite{Aharony:1997bh}.
We would like to argue that these states do indeed decouple from the theory described by the rest of the web.
The reason is basically that they are uncharged under the gauge symmetry of the web.
This is seen explicitly in the webs in Fig.~\ref{web_parallel_legs}, in that 
the distance between the parallel external 5-branes,
which controls the mass of the states in question, does not depend on the size of the face, 
namely on the Coulomb modulus.
We will exhibit this more explicitly for a class of webs with parallel external 5-branes below.
This decoupling has also been argued recently from the point of view of M theory and the toric geometry dual
to the 5-brane web, where the extra states
correspond to M2-branes wrapping 2-cycles that are orthogonal to all the 2-cycles dual
to the internal 5-branes \cite{BMPTY,HKT}.

The fixed point theories corresponding to webs with parallel external 5-branes are expected to
exhibit enhanced global symmetry.
For parallel D5-branes, as in Fig.~\ref{web_parallel_legs}b, this is seen at the classical level,
and is simply the flavor symmetry associated with multiple massless matter hypermutiplets in identical 
representations.
However for other $(p,q)$5-branes, as in Fig.~\ref{web_parallel_legs}a, the enhancement of the global symmetry 
will only be apparent once one includes non-perturbative (instanton particle) corrections.
We will encounter several examples of this phenomenon in the rest of the paper.

\section{5d superconformal index basics}

The bosonic part of the 5d superconformal group is $SO(5,2)\times SU(2)_R$.
The representations are labeled by the highest weights of the $SO(5) \times SU(2)_R$ subgroup. 
We label the two weights of $SO(5)$ as $j_1, j_2$, and that of $SU(2)_R$ as $R$. 
The 5d superconformal index is defined as \cite{KKL}
\be
\mathcal{I}={\rm Tr}\,(-1)^F\,x^{2\,(j_1+R)}\,y^{2\,j_2}\,\mathfrak{q}^{\mathfrak{Q}}\,,
\label{eq:ind}
\ee
where $x,\,y$ are the fugacities associated with the superconformal group, 
and $\mathfrak{q}$ represents fugacities associated with other commuting conserved charges $\mathfrak{Q}$.
The latter are usually associated with flavor or topological symmetries.

Given a Lagrangian description of the field theory, the index can be computed from the path integral on $S^4\times S^1$
using localization. 
For 5d theories this was carried out for several cases in \cite{KKL}. 
The general result is expressed in terms of two parts.
The first is perturbative, and comes from the one loop determinant. 
This depends on the gauge group and the matter content of the theory. 
Each gauge vector multiplet contributes a one-particle index
\be
f_{vector}(x,y,\alpha) = - \frac{x (y + \frac{1}{y})}{(1 - x y)(1 - \frac{x}{y})}\sum_{\bold{R}}e^{-i\bold{R}\cdot\alpha} \,,
\label{eq:vec}
\ee
and each matter hypermultiplet contributes
\be
f_{matter}(x,y,\alpha) = \frac{x}{(1 - x y)(1 - \frac{x}{y})}\sum_{\bold{w}\in \bold{W}}
\sum^{N_f}_{i=1} (e^{i\bold{w}\cdot\alpha+im_i}+e^{-i\bold{w}\cdot\alpha-i m_i})\,. 
\label{eq:mat}
\ee
Here $m_i$ are the flavor chemical potentials, and $\alpha$ is the gauge holonomy matrix. 
The sum in (\ref{eq:vec}) is over the roots of the gauge group, and the first sum in (\ref{eq:mat}) is over the 
weights of the flavor representations.
The full perturbative contribution is given by the plethystic exponential of the one particle contributions:
\be
PE[f(\cdot)] = exp\left[\sum^{\infty}_{n=1} \frac{1}{n} f(\cdot^n)\right]\,,
\label{eq:plesh}
\ee 
where the dot represents all the variables in $f$, namely the fugacities.

The second part of the index comes from instantons localized at either the north pole or south pole (for anti-instantons) of the $S^4$.
The localization conditions allow for such configurations, and therefore they must be included in the index computation.
This is done by integrating over the Nekrasov partition function \cite{Nekrasov:2002qd,Nekrasov:2004vw}. 
The full index is given by 
\be
I(x,y,m_i,q) = \int [{\cal D}\alpha]\, PE[f_{vector} + f_{matter}] \, |\mathcal{Z}_{inst}|^2 \,,  
\label{eq:index}
\ee
where ${\mathcal Z}_{inst}$ is expressed as a power series in the instanton number $k$,
\be
\mathcal{Z}^{inst} = 1 + q Z_1 + q^2 Z_2 + \cdots \,,
\label{insum}
\ee 
where $Z_k$ is the 5d $k$-instanton partition function.
Computing these is generally the trickiest part of the calculation.
The instanton partition function $Z_k$ is generally expressed as an integral over the ``dual gauge group" of
the instanton moduli space.\footnote{In Type IIA string theory, 
the 5d instanton is a D0-brane inside a stack of D4-branes.  
The ``dual gauge group" for $k$ instantons is the gauge group living on $k$ D0-branes.}
The integrand has contributions both from the gauge multiplet and from the charged flavor hypermutiplets.
 The integral is evaluated using the residue theorem, which in general requires a pole prescription. 
 The explicit formulas for the instanton partition functions that we will use, including the pole prescriptions, are given in Appendix A.

 \subsection{Issues for instanton partition functions}
 \label{sec:issues}
 
There are a number of subtle issues in computing the instanton partition functions which
we would like to discuss here. Some of these have known analogs in 4d, and others are specific to 5d.

\subsubsection{$SU(N)$ vs. $U(N)$}
\label{sec:SU(N)vsU(N)}

Strictly speaking, the Nekrasov partition function assumes a $U(N)$ gauge symmetry.
Naively, one can then reduce to $SU(N)$ by restricting $\mbox{Tr}(\alpha)=0$, namely by setting
the fugacity of the overall $U(1)$ to 1.
However the result retains some remnants of this $U(1)$, and we must remove them in order to obtain
the $SU(N)$ instanton partition function.

One way to expose these remnants is by studying cases in which $SU(N)$ has an alternative description,
{\em e.g.} $SU(2) = Sp(1)$ or $SU(4) = SO(6)$,
which allows for an independent computation of the instanton partition functions that can be compared with the
$U(N)$ formalism.\footnote{The instanton moduli space is realized differently in the two descriptions, and this leads to
different dual gauge groups and different integrands in the instanton partition functions.
Nevertheless, the resulting partition functions must agree.}

As our test case, let us consider 
an $SU(2)$ theory with $N_f$ flavors, and compare the $k$-instanton partition function in the
 $U(2)$ description to that in the $Sp(1)$ description.
First, we find an overall sign difference which is consistent with a factor $(-1)^{k(\kappa + N_f/2)}$,
where $\kappa$ is the $U(2)$ CS level. The combination $\kappa + \frac{N_f}{2}$ is the full quantum CS level.
As usual, for an odd number of flavors $\kappa$ must be half-odd-integer.
This factor remains after we enforce the zero-trace condition, and is one remnant of the difference
between $U(2)$ and $SU(2)$. The latter, of course, does not admit a CS term.
We have not fully understood the source of this factor, but we believe, due to the dependence on the instanton number $k$, 
that it comes from the mixed CS term in $U(2)$, and more generally in $U(N)$:
\be
S_{mixed \, CS} \propto \kappa \int \hat{A} \wedge \mbox{Tr}(F\wedge F) \,.
\ee
We will include this factor in all $SU(N)$ partition functions, and in particular in $SU(2)$ partition functions computed via $U(2)$.

For $N_f > 2$ there is another discrepancy.
The instanton partition function computed using $Sp(1)$ exhibits the $SO(2N_f)$ flavor symmetry,
but in the $U(2)$ formalism it exhibits only a $U(N_f)$ subgroup. 
A similar problem was encountered in 4d 
\cite{AGT}, where it was claimed to be due to the non-decoupling of the overall $U(1)$ gauge sector.

Consider the $SU(2)$ theory with $N_f=3$.
By expanding the two partition functions computed using the $U(N)$ and $Sp(N)$ formalisms (given in Appendix A) 
in powers of $x$, a pattern emerges that suggests the relation
\be
\mathcal{Z}_{inst}^{Sp(1)+3} = PE\left[\frac{q\, x^2 \, {e^{\frac{i}{2}(m_1 + m_2 + m_3)}}}
{(1-x y)(1-\frac{x}{y})}\right] \mathcal{Z}_{inst}^{U(2)+3} \,,
\label{eq:suspftf}
\ee
where $q$ is the instanton number fugacity.
For the $U(2)$ computation we took $\kappa = \frac{1}{2}$.
The sign factor in this case is $(-1)^{k(\kappa + N_f/2)} = +1$.
Alternatively we could have taken $\kappa = -\frac{1}{2}$ and replaced $m_i \rightarrow -m_i$.
In this case the sign factor would be $(-1)^k$.
We checked this relation up to instanton number $3$.
A similar relation is found for $SU(2)$ with $N_f=4$:
\be
\mathcal{Z}_{inst}^{Sp(1)+4} = PE\left[\frac{q\, x^2 \left(e^{\frac{i}{2}\sum_{i=1}^4 m_i} + 
e^{-\frac{i}{2}\sum_{i=1}^4 m_i}\right)}{(1-x y)(1-\frac{x}{y})}\right] \mathcal{Z}_{inst}^{U(2)+4} \,.
\label{eq:SU(2)+4}
\ee
In using the $U(N)$ formalism for $SU(2)$ we need to include such a correction factor whenever
the {\em effective} number of flavors is greater than two,
for example for the $(1,1)$ instanton of the $SU(2)\times SU(2)$ linear quiver with an extra fundamental flavor
(see section \ref{extra_flavor}).

This discrepancy is related more generally to the issue of parallel external 5-branes discussed in section \ref{parallel_legs}. 
For $N_f>2$ the $SU(2)$ web necessarily has parallel external NS5-branes.
The correction factor in fact corresponds precisely to the contribution of the decoupled $U(1)$ instanton state.
This state contributes to the instanton partition function for $U(N)$, which includes the overall $U(1)$ factor,
and its contribution remains after setting $\mbox{Tr}(\alpha)=0$.
We must therefore remove it by hand, by dividing by its partition function.
For example for any $SU(N)$ web with a pair of parallel external NS5-branes, we have\footnote{This was recently also observed
from the topological vertex perspective in \cite{BMPTY,HKT}.}
\be
\mathcal{Z}_{inst}^{SU(N)} = \frac{{\mathcal Z}_{inst}^{U(N)}}{{\mathcal Z}_{inst}^{U(1)}} \,.
\label{eq:U(N)_vs_SU(N)}
\ee
It is straighforward to verify, referring to Appendix A for the $U(1)$ instanton partition function, 
that this reproduces the factors in (\ref{eq:suspftf}) and (\ref{eq:SU(2)+4}).

The $U(1)$ factor is not invariant under the flavor symmetry, or under $x\rightarrow 1/x$, 
which is part of the conformal symmetry. 
This is true more generally for the $U(N)$ instanton partition function.
However the ratio in (\ref{eq:U(N)_vs_SU(N)}) is invariant under both.
This provides us with useful consistency checks for the final results, which must respect both symmetries.

\subsubsection{Antisymmetric and bifundamental matter}
\label{AS_BF_issues}

The inclusion of matter in the antisymmetric representation presents a problem.
This is apparent already in the results of \cite{KKL} for $Sp(N)$ with an antisymmetric hypermultiplet (see eq.~(\ref{ASSPplus}-\ref{ASSPminuse}) in Appendix A).
The problem is that the contribution of the antisymmetric matter to the instanton partition function includes 
an infinite tower of states with growing representations under the $SU(2)_M$ matter global symmetry.
This is possible only if there exists an $SU(2)_M$ charged bosonic zero mode in the instanton moduli space.
However this is in conflict with the ADHM construction, in which matter multiplets contribute only fermionic zero modes,
leading to a finite number of representations under the matter global symmetry.
For example, this is what is seen for matter in the fundamental representation.

The authors of \cite{KKL} argued, for a different reason, that one must subtract a particular term
from the instanton partition function for $Sp(N)$ with an antisymmetric hypermultiplet.
Recall that this is the theory described by D4-branes in Type I' string theory.
In this picture the instantons correspond to D0-branes.
The point made in \cite{KKL} is that in addition to the Higgs branch corresponding to the instanton moduli space,
the D0-brane quantum mechanics also has a Coulomb branch, namely an extra bosonic zero mode, whose contribution should be removed.
Removing it leads to a finite number of representations.

The same problem appears for $Sp$-type quiver theories with bifundamental matter, in evaluating partition functions
for di-group instantons.
This is not surprising, since these theories correspond to orbifolds of the $Sp(N)$ theory with antisymmetric matter.
However it is not clear how to deal with the problem in these cases.
For $Sp(1)$ quivers we can use the $U(2)$ formalism to deal with such instanons, but for higher $N$
we do not yet have a solution.

A similar problem arises for $SU(N)$ with antisymmetric matter (see eq.~(\ref{ASSU}) in Appendix A).
In this case the matter contribution to the instanton index contains an infinite tower of increasing representations
of both the global and gauge symmetry.
The resolution in this case is not known either.
Some of the examples we discuss below are of this form.
In these examples we are not able yet to incorporate instanton contributions.

\section{Enhanced global symmetry in $SU(N)$ theory}

\begin{figure}[h]
\centering
\begin{subfigure}[]{0.2\textwidth}
\centering
\includegraphics[width=\textwidth]{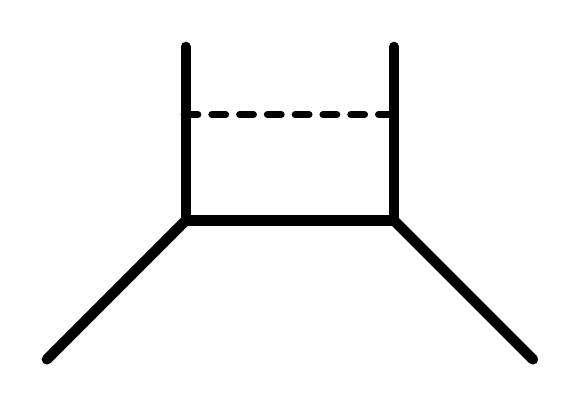} 
\caption{}
\label{}
\end{subfigure}
\hspace{0.5cm}
\begin{subfigure}[]{0.3\textwidth}
\centering
\includegraphics[width=\textwidth]{webSU22.pdf} 
\caption{}
\label{}
\end{subfigure}
\hspace{0.5cm}
\begin{subfigure}[]{0.4\textwidth}
\centering
\includegraphics[width=\textwidth]{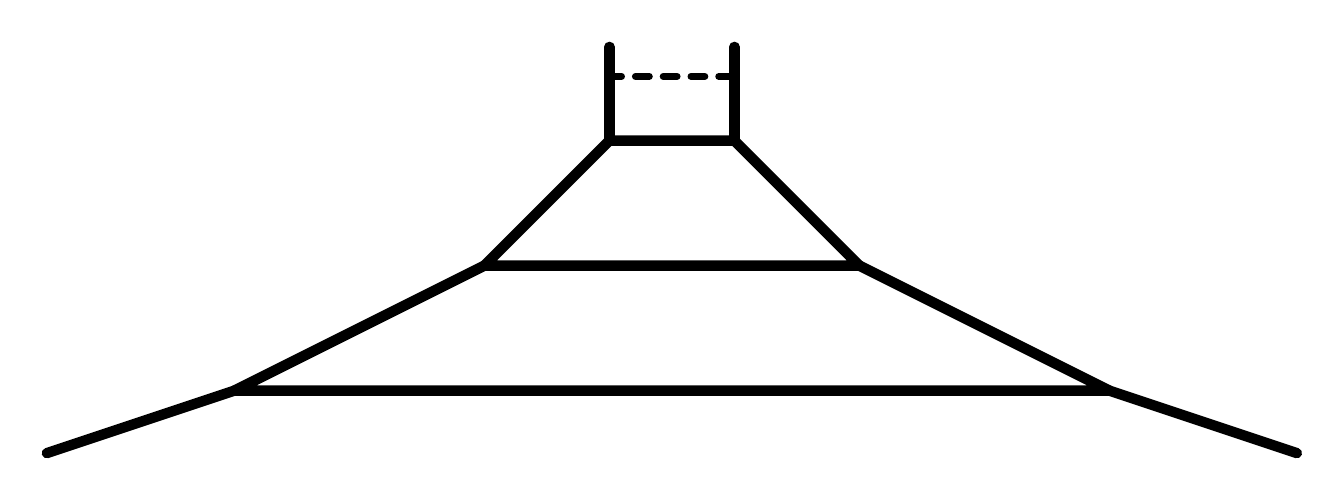} 
\caption{}
\label{}
\end{subfigure}
\caption{Webs for $SU(N)_N$ with $N=1,2$ and $3$.}
\label{web_SU(N)_N}
\end{figure}

As our first interesting application we consider pure $SU(N)$ theory with a CS level $\kappa = N$.
The corresponding 5-brane web, Fig.~\ref{web_SU(N)_N}, has a pair of parallel external NS5-branes.
The $N=1$ case, Fig.~\ref{web_SU(N)_N}a, is an empty theory.
The $N=2$ case, Fig.~\ref{web_SU(N)_N}b, is the third $SL(2,\mathbb{Z})$-inequivalent $SU(2)$ web.
It describes an $SU(2)$ theory with $\theta = 2\pi \sim 0$, which is equivalent to the $E_1$ theory \cite{Bergman:2013ala}.

We argued previously that the light state corresponding to the D-string between the external NS5-branes
decouples from the interacting fixed point theory.
An instructive way to explain this is to embed the $SU(N)_N$ web inside a larger web
with an extra internal D5-brane and no parallel external NS5-branes, Fig.~\ref{web_SU(N+1)_{N-1}}a.
This describes an $SU(N+1)_{N-1}$ theory at a generic point on its Coulomb branch.
A continuation beyond infinite coupling for this theory leads to an $SU(N)_N\times U(1)_{-1}$ theory, Fig.~\ref{web_SU(N+1)_{N-1}}b,
where the $U(1)$ gauge group factor is associated with the extra rectangular face.
The gauge multiplets that become massive correspond to the open fundamental string in the $U(1)$ face.
The shifted CS levels of the unbroken gauge groups are a result of integrating out the massive gauginos.
>From the point of view of the $SU(N)$ theory this corresponds to two fermions in the fundamental representation,
so the CS level is shifted by $1$ to $N$.
>From the $U(1)$ point of view there are $2N$ fermions, so the resulting CS level is $-1$.
It is apparent from the web that this deformation generates a flow to an interacting fixed point associated to $SU(N)_N$
plus an IR free $SU(2)$ theory. The D-string that becomes massless is just the $W$-boson of this $SU(2)$
(it carries $U(1)$ electric charge due to the CS term),
and is therefore completely decoupled from the dynamics of the $SU(N)_N$ fixed point theory.

\begin{figure}[h]
\center
\begin{subfigure}[]{0.3\textwidth}
\center
\includegraphics[width=\textwidth]{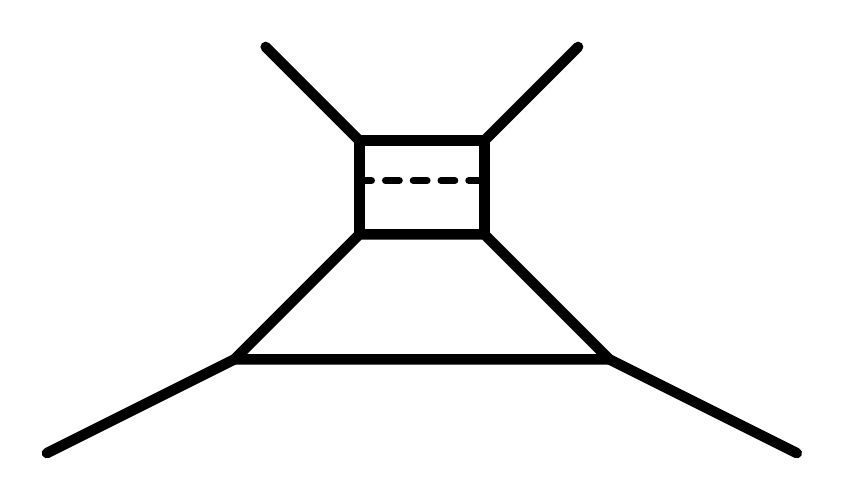} 
\caption{}
\label{}
\end{subfigure}
\hspace{2cm}
\begin{subfigure}[]{0.2\textwidth}
\center
\includegraphics[width=\textwidth]{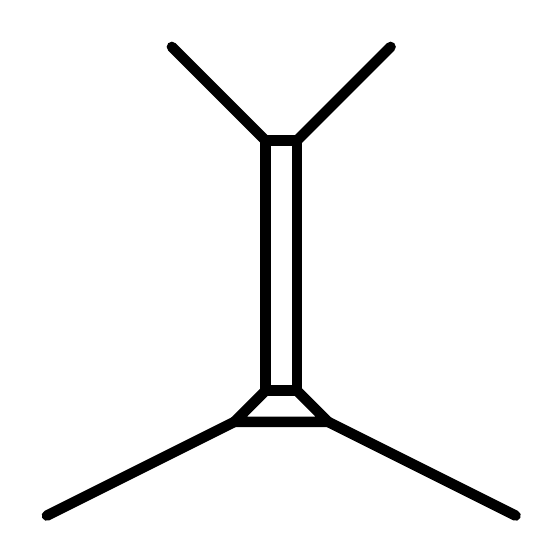} 
\caption{}
\label{}
\end{subfigure}
\caption{Embedding $SU(N)_N\times U(1)_{-1} \subset SU(N+1)_{N-1}$.}
\label{web_SU(N+1)_{N-1}}
\end{figure} 

As we will now show, the $SU(N)_N$ fixed point theory exhibits a non-perturbatively enhanced global symmetry $E_1 = SU(2)$
for all $N$. 

\subsection{The $SU(N)_N$ superconformal index}

The full global symmetry of the $SU(N)_N$ theory will be revealed by examining its superconformal index,
and specifically the coefficient of the $x^2$ term, which gets contributions from the conserved current multiplets.

The perturbative contribution has the form
\be
I_{pert}^{SU(N)} = 1 + x^2 + O(x^3) \,.
\ee
This is, of course, independent of the CS level.
The perturbative index exhibits only the $U(1)_T$ symmetry.
For $N=1$ the theory is empty, and therefore $I_{pert}^{SU(1)} = 1$.

The instanton contribution is given by a sum of $k$-instanton partition functions (\ref{insum}).
Let us first concentrate on the 1-instanton contribution.
For CS level $\kappa$ this is given by (see Appendix A)
\be
Z_{1}^{U(N)} = \frac{1}{2 \pi i} \oint \frac{(1-x^2)\, u^{\kappa + N - 1} \, du}{(1-x y)(1-\frac{x}{y})
\prod^{N}_{j=1} (u - x e^{i\alpha_j})(u - \frac{e^{i\alpha_j}}{x})} \,,
\label{eq:SUNlNintg}
\ee
where we have projected out the overall $U(1)$ by setting $\sum_i \alpha_i = 0$, and included
the sign factor $(-1)^\kappa$. 
This can be evaluated using the residue theorem. For the ``$SU(1)_1$" theory 
this gives
\be
Z_1^{U(1)_1} = \mbox{} - \frac{x^2}{(1-x y)(1-\frac{x}{y})} \,.
\label{U(1)_inst}
\ee
We interpret this as the contribution of the decoupled D-string state.
For the $``SU(2)_2"$ theory we find
\be
Z_1^{U(2)_2} = \frac{x^4\left(e^{2i\alpha} + 1 + e^{-2i\alpha} - x^2\right)}{(1 - x y)
(1 - \frac{x}{y})(1 - x^2 e^{2i\alpha})(1 - x^2 e^{-2i{\alpha}}) }\,.
\label{SU(2)_2_index}
\ee 

As we claimed previously, these partition functions are not invariant under $x \rightarrow 1/x$,
and therefore do not respect the superconformal symmetry.
We need to remove the contribution of the decoupled state (\ref{U(1)_inst}).
For the ``$SU(1)_1$" theory this obviously gives $Z_1^{SU(1)_1} = 0$, as it should for an empty theory.
For $SU(2)_2$ we get
\be
Z_1^{SU(2)_2} = Z_1^{U(2)_2} + \frac{x^2}{(1-xy)(1-\frac{x}{y})}  = 
\frac{x^2 (1+x^2)}{(1 - x y)(1 - \frac{x}{y})(1 - x^2 e^{2i\alpha})(1 - x^2 e^{-2i\alpha})} \,,
\ee
which is $x\rightarrow 1/x$ invariant.
In fact it is the same as the 1-instanton partition function of the $SU(2)_0$ theory,
which has an enhanced $E_1$ global symmetry.
The extra conserved currents come from the (subtracted) decoupled state. 
Again, this is what we expect, since this theory is really the $SU(2)$ theory with $\theta = 2\pi \sim 0$
\cite{Bergman:2013ala}.

Getting the explicit result for $SU(N)_N$ with $N>2$ is a bit cumbersome and not very illuminating. 
The lack of $x\rightarrow 1/x$ invariance of (\ref{eq:SUNlNintg}) for $\kappa = N$ is due to the existence
of an additional pole at $u\rightarrow \infty$ in this case, as compared with $\kappa < N$
(for $\kappa = -N$ there is an additional pole at $u=0$).
Since we assume that $x\ll 1$, the transformation $x\rightarrow 1/x$ implies that we
should change our pole prescription, and sum the residues of the poles {\em outside} the unit circle.
The contributions of all the other poles respect the symmetry since they come in pairs that
are interchanged under $x\rightarrow 1/x$. But the extra pole violates this symmetry.
The change in the partition function is related to the residue at the extra pole:
\bea
Z_{1}^{U(N)_{\pm N}} [x] - Z_{1}^{U(N)_{\pm N}} \left[\frac{1}{x}\right] &=& Res[u=0] 
- Res[u \rightarrow \infty] \nonumber\\
&=& \frac{1-x^2}{(1-xy)(1-\frac{x}{y})}  \,.
\label{DeltaZ_1}
\eea
Note that this is independent of $N$. 
Indeed it is the same for $N=1$.
Therefore by subtracting the $N=1$ partition function, thereby removing the decoupled state,
superconformal invariance is restored. 
Thus the 1-instanton partition function for $SU(N)_N$ is given by
\be
Z_{1}^{SU(N)_N} = Z_{1}^{U(N)_N} + \frac{x^2}{(1-xy)(1-\frac{x}{y})} \,,
\label{Z1_SU(N)_N}
\ee 
generalizing the $N=1,2$ cases above.

One can easily see from (\ref{eq:SUNlNintg}) that the leading contribution in the first term on the RHS
of (\ref{Z1_SU(N)_N}) is ${\cal O}(x^{2N})$.
The extra conserved current comes from the second term.
Adding the contribution of the anti-instanton, the full index up to $k=1$ has the form
\be
I^{SU(N)_N} = 1 + x^2 \left(q + 1 + \frac{1}{q}\right) + O(x^3) \,.
\ee
The coefficient of $x^2$ is precisely the adjoint character of $SU(2)$, which implies
that the global symmetry is enhanced from $U(1)_T$ to $E_1 = SU(2)$ for all $N$.
The same holds for the $SU(N)_{-N}$ theory.

In fact one can show that the full index can be expressed in terms of $SU(2)$ characters. 
Since $[{\cal Z}_{inst}(x,y,\alpha,q)]^* = {\cal Z}_{inst}(x,y,-\alpha,1/q)$,
the general expression for the index (\ref{eq:index}) can be written in this case as
\bea
I_{SU(N)_{\pm N}}(x,y,q) &=& \int [{\cal D}\alpha]\, PE[f_{vector}] \, \mathcal{Z}_{inst} (x,y,\alpha,q) \mathcal{Z}_{inst} (x,y,-\alpha,\frac{1}{q}) 
\nonumber \\ 
&=& \int [{\cal D}\alpha]\, PE[f_{vector}] \, \mathcal{Z}_{inst} (x,y,-\alpha,q) \mathcal{Z}_{inst} (x,y,\alpha,\frac{1}{q}) \,, 
\eea
where in the second line we changed the integration variables from $\alpha \rightarrow -\alpha$ and used the invariance of 
$f_{vector}$ and the Haar measure.
Therefore the full index is invariant under $q\rightarrow 1/q$, implying that it can be expressed in terms of
$SU(2)$ characters spanned by $q$.

\subsubsection{Comments on higher instantons}

Let us make a few observations about the contributions from higher instanton number. 
Our result for $k=1$ suggests that the full instanton contribution for $SU(N)_N$ is given by
\be  
\mathcal{Z}_{inst}^{SU(N)_N} = PE\left[\frac{q\, x^2}{(1-xy)(1-\frac{x}{y})}\right] 
\mathcal{Z}_{inst}^{U(N)_N} \,,
\label{eq:SUNlNfPF}
\ee
where the plethystic exponential removes the decoupled state contribution from the full partition function.
This is consistent with what we found for $SU(2)$ by comparing with $Sp(1)$ in section~\ref{sec:SU(N)vsU(N)}.
Let us test this at the 2-instanton level.
In this case there are two integration variables $u_1,u_2$, and we find
\bea
Z_{2}^{U(N)_N} [x]  &-&  Z_{2}^{U(N)_N} \left[\frac{1}{x}\right]  = 
\mbox{} - Res[u_1\rightarrow \infty, u_2\rightarrow \infty] - Res[u_1=x u_2 y, u_2\rightarrow \infty] \nonumber \\  
 &-&  Res[u_1=\frac{x u_2}{y}, u_2 \rightarrow \infty] -  \sum^N_{i=1} Res[u_1=x e^{i\alpha_i}, u_2\rightarrow \infty]  \nonumber \\
 &-& \sum_{i=1}^N Res[u_1\rightarrow \infty, u_2= xe^{i\alpha_i}] \nonumber \\
 &=&  \frac{1-x^2}{(1-xy)(1-\frac{x}{y})} Z_{1}^{U(N)_N}
- \frac{x (1-x^2)(y+\frac{1}{y}-x-x^3)}{(1+xy)(1+\frac{x}{y})(1-xy)^2(1-\frac{x}{y})^2} \,.
\label{DeltaZ_2}
\eea 
On the other hand, collecting the $q^2$ terms in (\ref{eq:SUNlNfPF}) we find that:
\be
Z_{2}^{SU(N)_N} = Z_{2}^{U(N)_N} + \Delta \,,
\ee
where
\be
\Delta = \frac{x^2}{(1-xy)(1-\frac{x}{y})} Z_{1}^{U(N)_N} + \frac{x^4 (1+x^2)}{(1+xy)(1+\frac{x}{y})(1-xy)^2(1-\frac{x}{y})^2} \,.
\ee
Using (\ref{DeltaZ_1}) one can easily show that $\Delta[x] - \Delta[\frac{1}{x}]$ is also given by (\ref{DeltaZ_2}),
and therefore that 
$Z_{2}^{SU(N)_N}$ is invariant under $x \rightarrow 1/x$. 
It would be interesting to prove this for all instanton numbers.\footnote{For the $SU(2)_2$ case a multi-instanton calculation and comparison with $SU(2)_0$ was done in \cite{Taki}.}

\section{The $SU(2)\times SU(2)$ quiver theories}

The simplest example of a non-trivial quiver theory in 5d is the $SU(2)\times SU(2)$ linear quiver,
which has a single matter hypermultiplet in the bifundamental representation.
We can think of this as the $\mathbb{Z}_2$ orbifold of an $Sp(2)$ theory with an antisymmetric hypermutiplet.
The global symmetry of the theory is $SU(2)_M \times U(1)_T \times U(1)'_T$,
where $SU(2)_M$ is the ``mesonic" symmetry associated to the bifundamental matter multiplet
(which, since it is real, can be decomposed into two half-hypermultiplets that are rotated by $SU(2)_M$), 
and the two $U(1)_T$'s
are the two topological symmetries associated with the instanton currents of the two $SU(2)$ factors.

\begin{figure}[h]
\center
\begin{subfigure}[]{0.35\textwidth}
\centering
\includegraphics[width=\textwidth]{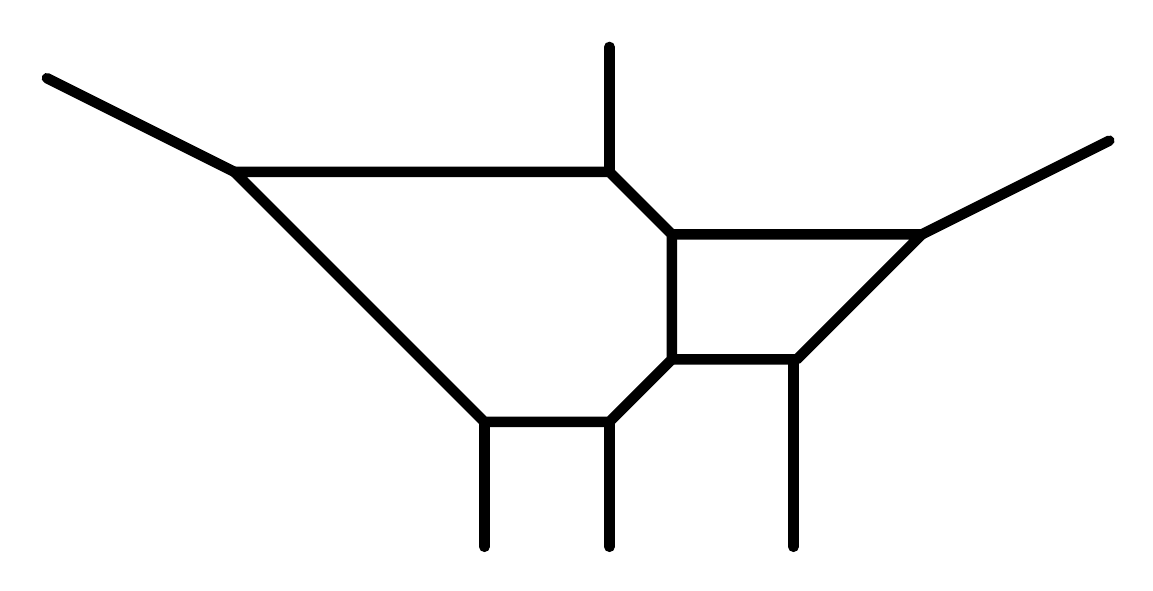} 
\caption{}
\label{}
\end{subfigure}
\begin{subfigure}[]{0.3\textwidth}
\centering
\includegraphics[width=\textwidth]{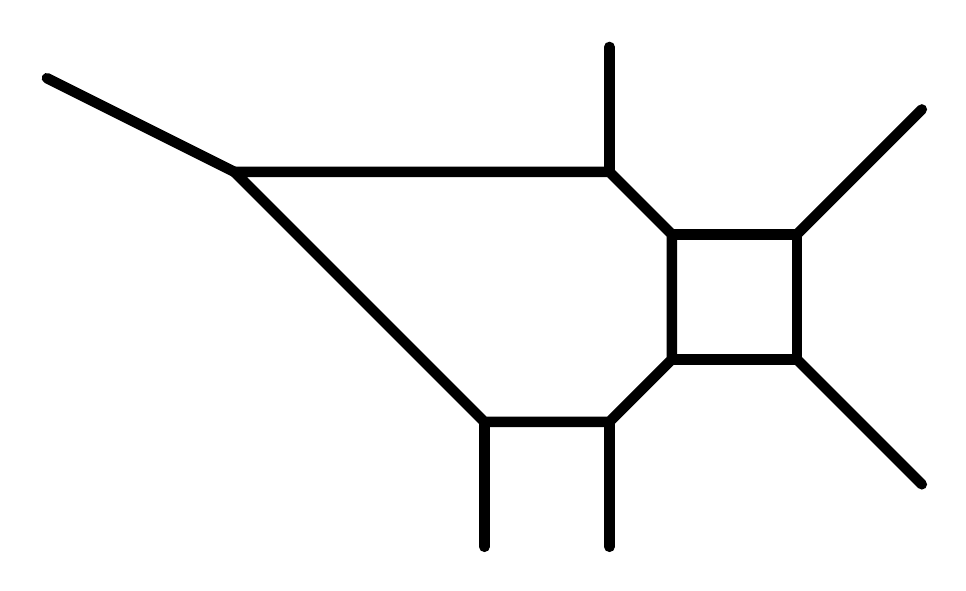} 
\caption{}
\label{}
\end{subfigure}
\hspace{1cm}
\begin{subfigure}[]{0.25\textwidth}
\centering
\includegraphics[width=\textwidth]{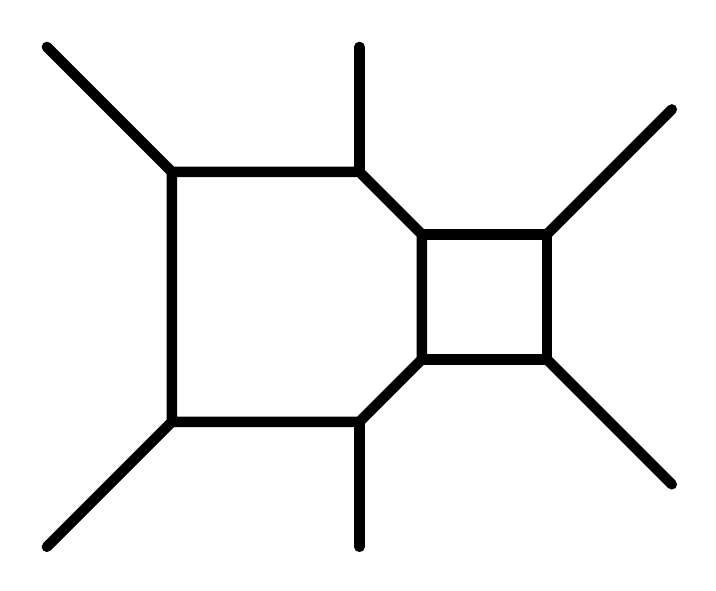} 
\caption{}
\label{}
\end{subfigure}
\caption{$SU(2)\times SU(2)$ linear quivers with $(\theta_1,\theta_2) = (0,0), (0,\pi)$ and $(\pi,\pi)$.}
\label{web_SU(2)xSU(2)}
\end{figure}

\begin{figure}[h]
\center
\includegraphics[height=0.2\textwidth]{webSU2pixSU2pi.pdf} 
\hspace{20pt}
\includegraphics[height=0.2\textwidth]{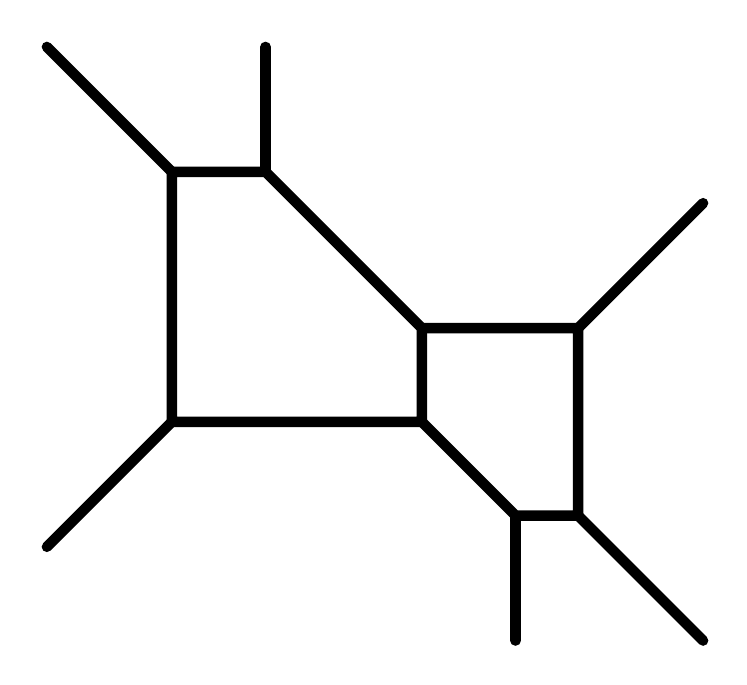} 
\hspace{20pt}
\includegraphics[height=0.2\textwidth]{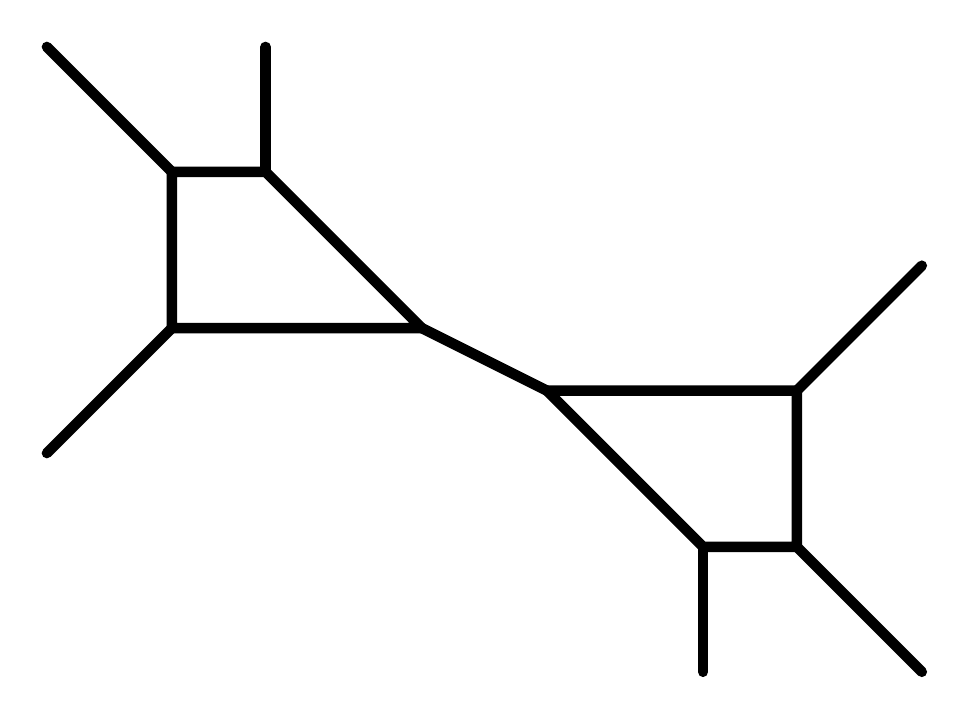} 
\caption{Step-by-step mass deformation (by ``flops") of the $SU(2)_\pi\times SU(2)_\pi$ theory.}
\label{web_SU(2)xSU(2)_mass_deformed}
\end{figure}

There are four $SU(2)\times SU(2)$ theories corresponding to the values of the two discrete $\theta$ parameters,
$(\theta_1,\theta_2)= (0,0)$, $(0,\pi)$, $(\pi,0)$, or $(\pi,\pi)$. The second and third theories are related
by exchanging the two gauge group factors. 
Note that the $\theta$ parameters are physical in this theory since the bifundamental multiplet contains
an even number of real massless fermions.
The corresponding 5-brane webs are shown in 
Fig.~\ref{web_SU(2)xSU(2)}.\footnote{As in the case of the $SU(2)$ theories, there are additional $SL(2,\mathbb{Z})$ 
inequivalent webs that correspond to the same theories. We have not included these.}
The correct identification of the $\theta$ parameters of each web is made clear by deforming the web in a way corresponding
to turning on a mass term for the bifundamental multiplet.
For a large enough mass this should reduce to two decoupled $SU(2)$ theories, with the corresponding values 
of $\theta$.
This is shown for the case of the $(\pi,\pi)$ theory in Fig.~\ref{web_SU(2)xSU(2)_mass_deformed}.

\subsection{Symmetry enhancement}
\label{section:SU(2)xSU(2)_enhancement}

The existence of parallel external NS5-branes in the webs for the $(0,0)$ and $(0,\pi)$ theories
suggests a non-perturbative enhancement of the global symmetry in those cases.
We will exhibit this explicitly in terms of the superconformal index.
But we can actually infer what the enhanced symmetry has to be by the group theoretic properties of the instantons.

Consider a single instanton of one of the $SU(2)$'s, say the $(1,0)$ instanton.
The analysis for the $(0,1)$ instanton is identical.
This will have the properties of an instanton of $SU(2)$ with two flavor hypermultiplets.
Recall that for this theory the perturbative global symmetry $SO(4)_F\times U(1)_T$ is enhanced
to $E_3 = SU(2) \times SU(3)$ at the fixed point.
Let us express the $SO(4)_F$ flavor symmetry as $SU(2)_F\times SU(2)'_F$.
The additional conserved currents are provided by the instanton (and anti-instanton),
which transforms as $({\bf 2},{\bf 1})_{+1}$ of the global symmetry.
>From the point of view of the quiver gauge theory one of the flavor $SU(2)$'s is identified with the second
gauge group, and the other with the global $SU(2)_M$ symmetry.
There are therefore two cases to consider.
In one case the instanton is charged under $SU(2)_M$ and leads to an enhancement of the global symmetry,
$SU(2)_M\times U(1)_T \rightarrow SU(3)$.
In the other case the instanton is charged under the second $SU(2)$ gauge group and does not lead to enhancement.
The two cases are in one-to-one correspondence with the $\theta$ parameter of the first $SU(2)$ gauge group.
For $\theta_1 = 0$ the $(1,0)$ instanton provides an additional conserved current,
and for $\theta_1 = \pi$ it does not.
The same conclusion holds also for the $(0,1)$ instanton and the value of $\theta_2$.
Therefore we assert that in the $(0,\pi)$ and $(\pi,0)$ theories, described by the web in Fig.~\ref{web_SU(2)xSU(2)}b
and its rotation by $\pi$, the global symmetry is enhanced to $SU(3)\times U(1)_T$,
where the $U(1)_T$ is associated to the gauge group that has $\theta = \pi$.
In the $(0,0)$ theory both instantons will contribute conserved currents, and we expect an enhancement to $SU(4)$.

Let us now verify these claims by examining the superconformal index.
For the sake of brevity, we present our results only to ${\cal O}(x^3)$, although they have been computed to ${\cal O}(x^7)$.
The perturbative part of the index is common to all four theories, and is given by
\bea
I_{pert}^{SU(2)\times SU(2)} & = & 1 + x^2 \left(\frac{1}{z^2} + 3 + z^2\right) 
+ x^3 \left(\frac{1}{y} + y\right) \left(\frac{1}{z^2} + 4 + z^2\right) + {\cal O}(x^4) \,,
\label{SU(2)xSU(2)_pert}
\eea
where $z$ is the fugacity associated to the Cartan subgroup of the $SU(2)_M$ symmetry.
The $x^2$ term clearly shows the classical global symmetry,
since $z^2 + 3 + z^{-2} = \chi_{\bf 3}(z) + 2$, corresponding to $SU(2)_M$ and the two $U(1)_T$'s.
Indeed all terms can be expressed in terms of characters of $SU(2)_M$.
Incorporating the instanton corrections requires us to deal with the issues described in section \ref{sec:issues}.
In particular, to avoid the problem associated with bifundamental matter in the $Sp(N)$ formalism we will
use the $U(N)$ formalism.
However this will require us to properly remove the $U(1)$ contributions.
Let us analyze the two theories in turn.

\subsubsection{The $(0,\pi)$ theory}

In the $U(N)$ formalism the $\theta$ parameter corresponds to the CS level.
In this case the $U(2)\times U(2)$ CS levels are $(\kappa_1,\kappa_2) = (1,0)$.
Note that due to the bifundamental matter multiplet, this is reversed relative to the identification in the $SU(2)$ theory, 
where $\theta = 0,\pi$ corresponds to $\kappa = 0,1$, respectively.

The 5-brane web, Fig.~\ref{web_SU(2)xSU(2)}b, has a pair of parallel external NS5-branes.
By analogy with the other cases of parallel NS5-branes, we must remove the contribution of the decoupled
D-string state. In this case 
\be
\mathcal{Z}_{inst}^{SU(2)_0\times SU(2)_\pi} = PE\left[\frac{q_1\, z\, x^2}{(1-x y)(1-\frac{x}{y})}\right] 
\mathcal{Z}_{inst}^{U(2)_1\times U(2)_0} \,,
\label{SU(2)xSU(2)_inst}
\ee 
where $q_1$ is the fugacity associated with the instanton number of the first $SU(2)$.
The $z$ dependence is due to the fermionic zero modes from the bifundamental hypermultiplet.
In addition, the contributions of instantons of the second $SU(2)$ are corrected by the sign factor
$(-1)^{k_2(\kappa + N_f/2)} = (-1)^{k_2}$.
As a consistency check of this formula, we have verified that all the instanton partition functions
we computed exhibit the full classical global symmetry, as well as  $x \rightarrow 1/x$ invariance.
They also agree with the $Sp(N)$ formalism, when that can be used.

The instanton partition functions for $U(2)\times U(2)$ have contributions from the two gauge multiplets
and from the bifundamental hypermultiplet. The contribution of the gauge multiplets is basically two copies 
of the gauge multiplet contribution for $U(2)$. 
The contribution of the bifundamental hypermultiplet is given in eq.~(\ref{U(N)_BF}) in Appendix A.
To ${\cal O}(x^3)$ there is only a contribution from the $(1,0)$ instanton.
The correction to the index to this order is given by
\bea
I_{(1,0)}^{SU(2)_0\times SU(2)_\pi} & = & x^2 \left(q_1 + \frac{1}{q_1}\right)\left(z + \frac{1}{z}\right) 
+ x^3 \left(y + \frac{1}{y}\right)\left(q_1 + \frac{1}{q_1}\right)\left(z + \frac{1}{z}\right) 
+ {\cal O}(x^4)\,. \nonumber \\
\label{(1,0)_inst}
\eea
Adding this to the perturbative contribution (\ref{SU(2)xSU(2)_pert}), the full index to this order can be expressed 
in terms of $SU(3)$ characters
\be
I^{SU(2)_0\times SU(2)_\pi}  = 1 + x^2 \left(1+\chi^{SU(3)}_{\bf 8}[q_1,z]\right) + x^3 \chi^{SU(2)}_{\bf 2}[y]
\left(2 + \chi^{SU(3)}_{\bf 8}[q_1,z]\right) + {\cal O}(x^4) \,.
\ee
Here the basic $SU(3)$ characters are given by $\chi_{\bf 3}[q_1,z] = {q_1^{1/3}}(z+\frac{1}{z})+q_1^{-2/3}$
and $\chi_{\overline{\bf 3}}[q_1,z] = {q_1^{-1/3}}(z+\frac{1}{z})+q_1^{2/3}$.
The characters for the other $SU(3)$ representations can easily be obtained from these.
This shows the enhanced $SU(3)$ global symmetry.
In particular the $x^2$ term exhibits the $SU(3)\times U(1)_T$ conserved current multiplets.

Other instanons contribute to higher order terms in the index.
For example the $(0,1)$ instanton enters only at ${\cal O}(x^4)$:
\bea
I_{(0,1)}^{SU(2)_0\times SU(2)_\pi} & = & x^4 + x^5\left(y + \frac{1}{y}\right) + {\cal O}(x^6) \,.
\label{(0,1)_inst}
\eea
As explained above, the instanton of one of the gauge $SU(2)$ factors is charged
either under the other gauge $SU(2)$ or under the global $SU(2)_M$.
This depends on the $\theta$ parameter associated with the instanton.
In the present case the $(0,1)$ instanton is charged under the first gauge $SU(2)$,
and can therefore only contribute when combined with the anti-instanton.
This is what we see in (\ref{(0,1)_inst}). The contribution begins at ${\cal O}(x^4)$,
and does not depend on the $(0,1)$ instanton fugacity. 
By contrast, the $(1,0)$ instanton contributes at ${\cal O}(x^2)$, and its contribution
depends on both the instanton fugacity $q_1$ and the $SU(2)_M$ fugacity $z$ (\ref{(1,0)_inst}), since it is gauge invariant.

We have extended the computation of the index to ${\cal O}(x^7)$, which includes contributions from 
(1,0), (0,1), (2,0), (1,1), (0,2), (3,0) and (1,2) instantons.
Other instantons with total instanton number 3, like the $(0,3)$ instanton, have partition functions that enter at this order,
but do not contribute to the index due to non-trivial gauge charges.
The final result, expressed in terms of $SU(3)$ characters, is given in Appendix B.

\subsubsection{The $(0,0)$ theory}

For the theory with $\theta$ parameters $(0,0)$ the $U(2)\times U(2)$ CS levels are either $(1,1)$ or $(1,-1)$.
The web in Fig.~\ref{web_SU(2)xSU(2)}a corresponds to $(\kappa_1,\kappa_2) = (1,1)$.
There is an inequivalent web that describes the same $SU(2)\times SU(2)$ theory, but which corresponds
to $U(2)\times U(2)$ CS levels $(1,-1)$. That web has two pairs of parallel external NS5-branes,
instead of three parallel NS5-branes.
This is analogous to the two webs that describe the $SU(2)$ theory with $\theta = 0$.
We will focus on the web shown in Fig.~\ref{web_SU(2)xSU(2)}a, namely on $U(2)_1\times U(2)_1$,
because the calculation turns out to be technically easier in this case.
Of course the final result should be the same in the other case, as it was for the two $SU(2)_0$ webs.
The removal of the decoupled states is achieved by \footnote{For $U(2)_1\times U(2)_{-1}$ we would instead need to
take
$$
\mathcal{Z}_{inst}^{SU(2)_0\times SU(2)_0} = PE\left[\frac{(q_1+q_2)\, z\, x^2}{(1-x y)(1-\frac{x}{y})}\right] 
\mathcal{Z}_{inst}^{U(2)_1\times U(2)_{-1}} \,,
$$
where the two terms in the numerator correspond to the D-strings between the parallel NS5-branes.}
\be
\mathcal{Z}_{inst}^{SU(2)_0\times SU(2)_0} = PE\left[\frac{x^2(q_1 z + \frac{q_2}{z} + q_1 q_2)}{(1-x y)(1-\frac{x}{y})}\right] 
\mathcal{Z}_{inst}^{U(2)_1\times U(2)_1} \,,
\label{SU(2)_0^2_partition}
\ee
where the three terms in the numerator correspond to the D-strings between any two of the three parallel external NS5-branes. 

As before, we have verified that all partition functions that we computed exhibit the complete classical symmetry,
including $x\rightarrow 1/x$, and that they agree with the $Sp(N)$ formalism when available.
We have also verified, at least to total instanton number 4, that the results using $(\kappa_1,\kappa_2) = (1,1)$
and $(1,-1)$ agree.

To ${\cal O}(x^3)$ there are contributions from the $(1,0)$, $(0,1)$ and $(1,1)$ instantons.
These combine to a correction given by
\bea
I_{(1,0)+(0,1)+(1,1)}^{SU(2)_0\times SU(2)_0} & = & 
x^2 \Big((z+\frac{1}{z})(q_1+\frac{1}{q_1}+q_2+\frac{1}{q_2}) + q_1 q_2+\frac{1}{q_1 q_2} \Big) \nonumber \\
&+& x^3 (y+\frac{1}{y})\Big((z+\frac{1}{z}) (q_1+\frac{1}{q_1} + q_2 +\frac{1}{q_2}) + q_1 q_2+\frac{1}{q_1 q_2}\Big) \nonumber \\
&+& {\cal O}(x^4) \,.
\eea
The $x^2$ term exhibits ten additional conserved currents. Adding to the five in the perturbative index 
(\ref{SU(2)xSU(2)_pert}), this gives the fifteen of the enhanced $SU(4)$ symmetry.
Indeed the full index can be expressed in terms of $SU(4)$ characters:
\be
I^{SU(2)_0\times SU(2)_0}  =  1 + x^2 \chi^{SU(4)}_{\bf 15}[q_1,q_2,z] 
+ x^3 \chi_{\bf 2}[y](1 + \chi^{SU(4)}_{\bf 15}[q_1,q_2,z])  + {\cal O}(x^4) \,,
\ee
where we have used that the character in the fundamental representation of $SU(4)$ is given by
$\chi_{\bf 4}[q_1,q_2,z] = (q_1/q_2)^{1/4} (z+\frac{1}{z})+(q_2/q_1)^{1/4} (\sqrt{q_1 q_2}+ \frac{1}{\sqrt{q_1 q_2}})$.
This shows the enhancement of the global symmetry at the fixed point from $SU(2)_M\times U(1)_T \times U(1)'_T$ to $SU(4)$.

We have carried out the computation of the superconformal index to ${\cal O}(x^7)$, which
requires including instantons with charges
(1,0), (0,1), (2,0), (0,2), (1,1), (3,0), (2,1), (1,2), (0,3), (2,2), (3,1), (1,3), (3,2), (2,3) and (3,3). 
The contributions of the $(2,3)$, $(3,2)$ and $(3,3)$ instantons were evaluated only using $(\kappa_1,\kappa_2) = (1,1)$.
In this case the computation is somewhat easier, in that these contributions come solely from the plethystic exponential factor 
in (\ref{SU(2)_0^2_partition}), and there was no need to carry out the lengthy computation of the $U(2)_1\times U(2)_1$
instanton partition functions.
The final result, which we present in Appendix B, can be expressed in terms of $SU(4)$ characters, confirming the enhancement.

\subsection{Duality}

\begin{figure}[h]
\center
\begin{subfigure}[]{0.2\textwidth}
\center
\includegraphics[width=\textwidth]{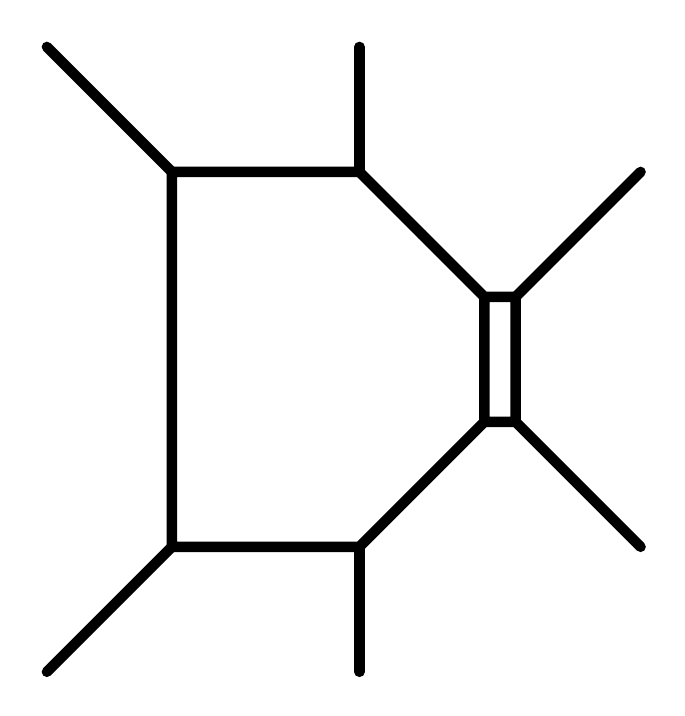} 
\caption{}
\label{}
\end{subfigure}
\hspace{3cm}
\begin{subfigure}[]{0.2\textwidth}
\center
\includegraphics[width=\textwidth]{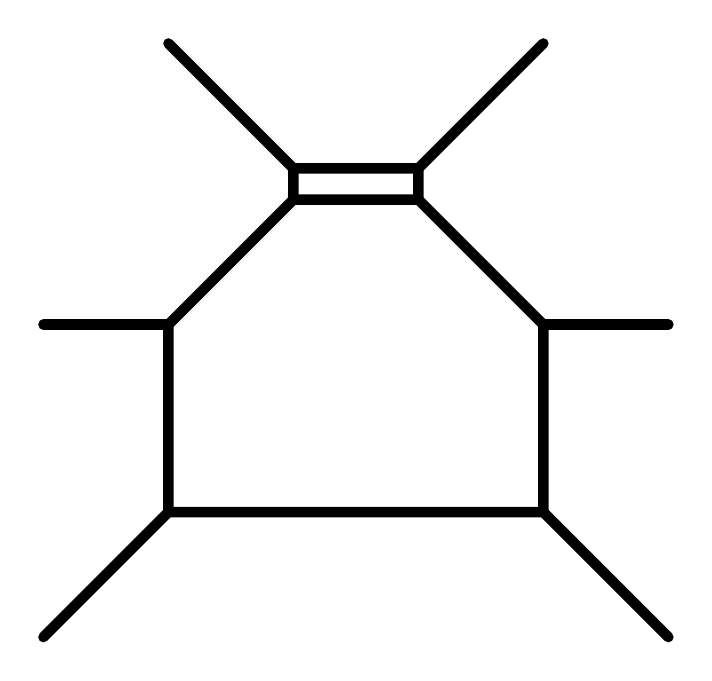} 
\caption{}
\label{}
\end{subfigure}
\caption{Duality: (a) $SU(2)\times SU(2)$ with $({\bf 2},{\bf 2})$, (b) $SU(3)$ with $N_f=2$.}
\label{web_SU(3)+2}
\end{figure}

Let us now concentrate on the $(\pi,\pi)$ theory, Fig.~\ref{web_SU(2)xSU(2)}c.
The singularity on the Coulomb branch is clearly visible in the web, Fig.~\ref{web_SU(3)+2}a.
However at this point we have already gone beyond infinite bare coupling for one of the $SU(2)$'s.
As realized in \cite{Aharony:1997ju}, the theory is now more appropriately described in terms of the S-dual web,
Fig.~\ref{web_SU(3)+2}b, which gives $SU(3)$ with two matter multiplets in the fundamental representation.

\begin{figure}[h]
\center
\includegraphics[height=0.15\textwidth]{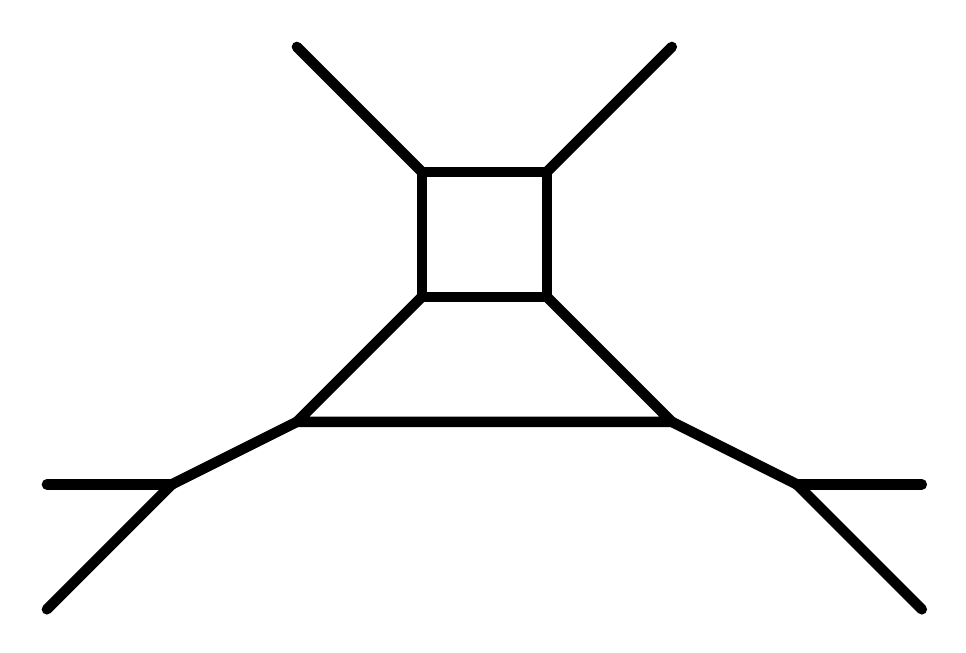} 
\caption{Mass deformation of $SU(3) + 2$ with $m_{1,2}<0$..}
\label{web_SU(3)+2_mass}
\end{figure}

In general, the 5d $SU(3)$ theory can have a CS term.
However the CS level of this theory is zero,
since the web, or more precisely the external 5-branes of the web, are unchanged under a $\pi$ rotation,
namely under charge conjugation.
Another way to see this is to deform the web by moving the external D5-branes to infinity,
corresponding to giving the flavors an infinite mass.
This leaves a pure $SU(3)$ theory with a CS level
\be
\kappa = \kappa_0 + \frac{1}{2}\sum_{i=1}^2 \mbox{sign}(m_i) \,,
\ee
where $\kappa_0$ is the bare CS level of the $SU(3) + 2$ theory.
We can then read off $\kappa$ by comparing the resulting pure $SU(3)$ sub-web with Fig.~\ref{web_SU(3)},
though we may need to use $SL(2,\mathbb{Z})$.
For example, moving both D5-branes down we get the web in Fig.~\ref{web_SU(3)+2_mass}.
The $SU(3)$ part of the web is related by the T-transformation of $SL(2,\mathbb{Z})$ to the
web in Fig.~\ref{web_SU(3)}b, which corresponds to $\kappa = 1$. Therefore $\kappa_0=0$.
The basic duality proposal is then
$$
\mbox{Duality 1:} \;\; SU(2)_\pi\times SU(2)_\pi + (\funda,\funda)  \longleftrightarrow SU(3)_0 + 2\, \funda \,.
$$

The global symmetries of the two theories agree.
The $SU(3) + 2$ theory has a $U(2)\sim SU(2)_M\times U(1)_B$ flavor symmetry, and an additional $U(1)_T$
topological symmetry. Evidently the $SU(2)_M$ part should agree with that of the $SU(2)\times SU(2)$ theory,
and the baryonic and instantonic charges of the $SU(3)$ theory should map to the two instantonic charges
of the $SU(2)\times SU(2)$ theory. 
Denoting the former as $B$ and $I$, and the latter as $I_1$ and $I_2$,
it is reasonable to guess that $I \propto I_1 + I_2$ and $B\propto I_1 - I_2$.
But in order to determine the precise map, as well as to confirm the lack of symmetry enhancement in both theories,
we will need the superconformal indices.

The discrete symmetries of the two theories also agree.
In particular, the quiver theory has a $\mathbb{Z}_2$ symmetry that exchanges the two gauge groups.
This is the quantum symmetry in the orbifold construction.
There is no such symmetry on the $SU(3)$ side.
On the other hand the $SU(3)$ theory has a $\mathbb{Z}_2$ symmetry with no analog in the $SU(2)\times SU(2)$
theory, charge conjugation. All the fields in the quiver theory are real, 
so charge conjugation acts trivially.
Clearly the duality should relate these two.\footnote{We thank Nathan Seiberg
for pointing this out to us.} This is consistent with our guess about the charge map above,
since on the quiver side $I_1 \leftrightarrow I_2$, and on the $SU(3)$ side $B\rightarrow -B$.
Note that it is crucial that the CS level of the $SU(3)$ theory vanishes, since the 5d CS term is odd 
under charge conjugation.

\subsubsection{Comparing indices}

Let us then compare the superconformal indices of the two theories.
We begin with the perturbative contributions.
The perturbative index of the $SU(2)\times SU(2)$ theory was already given, to ${\cal O}(x^3)$, in (\ref{SU(2)xSU(2)_pert}).
We reproduce it here to ${\cal O}(x^4)$, since that is where it differs from the $SU(3)$ theory:
\bea
I_{pert}^{SU(2)^2}  &=&  1 + x^2 \left(\frac{1}{z^2} + 3 + z^2\right) 
+ x^3 \left(\frac{1}{y} + y\right) \left(\frac{1}{z^2} + 4 + z^2\right) \nonumber \\
&+&  x^4 \left((\frac{1}{y^2} + y^2)(\frac{1}{z^2} + 4 + z^2) + \frac{1}{z^4} + \frac{3}{z^2} + 7 + 3 z^2 + z^4 \right) 
+ {\cal O}(x^5) \,.
\eea
For $SU(3)$ with $N_f = 2$ we find 
\bea
I_{pert}^{SU(3)+2}  &=&  1 + x^2 \left(\frac{1}{z^2} + 3 + z^2\right) + x^3 \left(\frac{1}{y} + y\right) 
\left(\frac{1}{z^2} + 4 + z^2\right) \nonumber \\
& + & x^4 \left( (\frac{1}{y^2} + y^2) (\frac{1}{z^2} + 4 + z^2) + \frac{1}{z^4} + \frac{3}{z^2} + 9 + 3 z^2 + z^4 \right) 
+ {\cal O}(x^5) \,.
\eea
The perturbative indices agree to ${\cal O}(x^3)$, which is consistent with the agreement of the global symmetries.
However they start to differ at ${\cal O}(x^4)$.

If the duality is correct the difference should be accounted for by instanton corrections.
To ${\cal O}(x^4)$ only the 1-instanton contributes in the $SU(3)$ theory, and only the $(1,0)$,
$(0,1)$ and $(1,1)$ instantons contribute in the $SU(2)\times SU(2)$ theory.
The corrections to the respective indices are given by
\bea
I_{1}^{SU(3)_0+2}  =  x^3 \left(\frac{1}{q} + q\right)\left(\frac{1}{z} + z\right) + x^4 \left(\frac{1}{q} + q\right)
\left(\frac{1}{y} + y\right)\left(\frac{1}{z} + z\right)  + {\cal O}(x^5) \,,
\eea
where $q$ is the $SU(3)$ instanton fugacity, and
\bea
I_{(1,0)+(0,1)+(1,1)}^{SU(2)_\pi^2}  &=& x^3 \left(q_1 q_2 + \frac{1}{q_1 q_2}\right)\left(z+\frac{1}{z}\right) \nonumber \\
&+& x^4 \left[ 2 + \left(y+\frac{1}{y}\right)\left(q_1 q_2 + \frac{1}{q_1 q_2}\right)\left(z+\frac{1}{z}\right) \right] + {\cal O}(x^5) \,,
\eea
where $q_1, q_2$ are the instanton fugacities associated with the two $SU(2)$ factors.
The contributions of the $(1,0)$ and $(0,1)$ instantons can be computed in either the $U(N)$ or $Sp(N)$ formalism.
In this case there is no $U(1)$ factor, since there is no decoupled state to remove.
These instantons carry gauge charge, and therefore contribute only
in instanton-anti-instanton combinations starting with the ``2" at ${\cal O}(x^4)$.
The contribution of the $(1,1)$ instanton was computed in the $U(N)$ formalism, using $U(2)_0\times U(2)_0$,
due to the problem related to the bifundamental matter.
This instanton contributes from ${\cal O}(x^3)$.

Comparing the full indices to ${\cal O}(x^4)$ we find a perfect agreement if we identify $q=q_1 q_2$,
namely the $SU(3)$ instanton charge is given by the sum the two instanton charges
of the quiver theory, $I = I_1 + I_2$.

To see baryonic states in the $SU(3)+2$ theory we need to go to at least ${\cal O}(x^5)$.\footnote{Since there are only two flavors
the simplest baryonic operator $\epsilon^{\alpha\beta\gamma} Q^a_\alpha Q^b_\beta Q^c_\gamma$ vanishes.
The lowest dimension non-trivial baryonic operator is schematically 
$B^c \sim \epsilon^{\alpha\beta\gamma} \epsilon_{ab} Q^a_\alpha Q^b_\beta \partial_+ \partial_- Q^c_\gamma$, and variants where some of the derivatives are replaced by contractions with the gaugino, for which $2(R+j_1)=5$.}
This requires taking into account several higher instanton number contributions on both sides.
We have done the computation to ${\cal O}(x^7)$, which includes the contributions of the 1-instanton and 2-instanton
of the $SU(3)$ theory, and the (1,0), (0,1), (1,1), (2,0), (0,2), (2,1), (1,2) and (2,2) instantons of the
$SU(2)\times SU(2)$ theory. Setting $q=q_1 q_2$ the indices agree to ${\cal O}(x^4)$.
At ${\cal O}(x^5)$ we find 
\bea
I^{SU(2)_\pi^2} &=& \cdots \mbox{} +
x^5 \bigg\{ \chi_{\bf 4}[y] (3 + \chi_{\bf 3}[z]) + \chi_{\bf 2}[y] (\chi_{\bf 5}[z] + 5\chi_{\bf 3}[z] + 6) \\
&-& \left(\frac{q_1}{q_2} + \frac{q_2}{q_1}\right)\chi_{\bf 2}[z] 
+ \left(q_1 q_2 + \frac{1}{q_1 q_2}\right) \left(\chi_{\bf 3}[y] \chi_{\bf 2}[z] + \chi_{\bf 4}[z] + \chi_{\bf 2}[z]\right) \bigg\}
+ {\cal O}(x^6)\,, \nonumber
\eea
and 
\bea
I^{SU(3)_0+2} &=&  \cdots \mbox{} +
x^5 \bigg\{ \chi_{\bf 4}[y] (3 + \chi_{\bf 3}[z]) + \chi_{\bf 2}[y] (\chi_{\bf 5}[z] + 5\chi_{\bf 3}[z] + 6) \\
&-& \left(b^3 + \frac{1}{b^3}\right)\chi_{\bf 2}[z] 
+ \left(q + \frac{1}{q}\right) \left(\chi_{\bf 3}[y] \chi_{\bf 2}[z] + \chi_{\bf 4}[z] + \chi_{\bf 2}[z]\right) \bigg\}
+ {\cal O}(x^6) \,. \nonumber
\eea
Setting $q=q_1 q_2$ we then find perfect agreement if $b^3 = q_1/q_2$,
which means that the charges are related as $B = 3(I_1 - I_2)$.
The agreement extends to ${\cal O}(x^7)$ (See Appendix B).
This completes our derivation of the charge map between the two theories.

\section{Generalizations}

In this section we will consider four simple generalizations of the $SU(2)\times SU(2)$ quiver theory,
gotten by increasing the ranks, adding flavor, or adding another $SU(2)$ factor.
We will focus just on the theories that have potential gauge theory duals,
namely on those whose rotated 5-brane webs describe a Lagrangian gauge theory.
These are the analogs of the $SU(2)_\pi \times SU(2)_\pi$ theory.
The other cases are also interesting from the point of view of enhanced global symmetries,
but we leave them out for the sake of compactness.

\subsection{Higher ranks}
\label{higher_rank}

The most natural higher-rank generalization of the $SU(2)\times SU(2)$ quiver is an $Sp(N)\times Sp(N)$ quiver,
corresponding to $N$ D4-branes in the Type I' $\mathbb{Z}_2$ orbifold.
More generally one can consider $Sp(N+M)\times Sp(N)$, corresponding to 
$N$ whole D4-branes and $M$ fractional D4-branes.
The global symmetry is again $SU(2)_M\times U(1)_T \times U(1)'_T$.
Here we will concentrate on the two simplest generalizations, $Sp(2)\times SU(2)$ and $Sp(2)\times Sp(2)$.
We will comment on higher $N$ generalizations in the conclusions.

\begin{figure}[h]
\center
\begin{subfigure}[]{0.25\textwidth}
\center
\includegraphics[width=\textwidth]{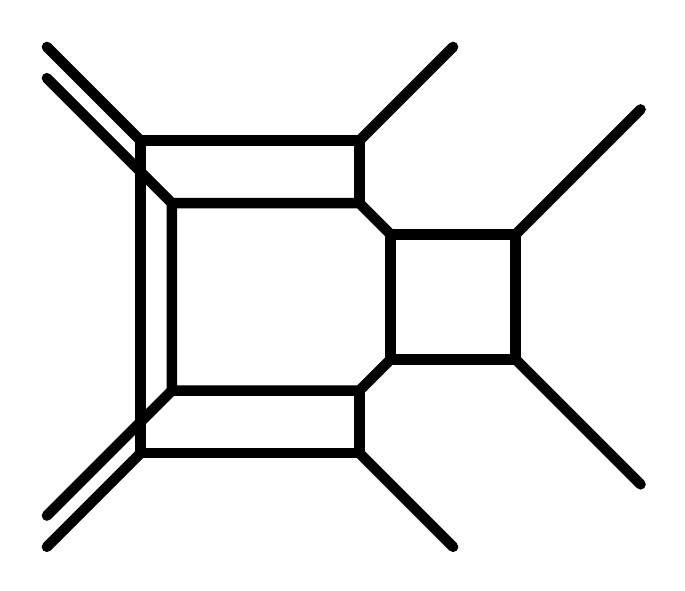} 
\caption{}
\label{}
\end{subfigure}
\hspace{3cm}
\begin{subfigure}[]{0.25\textwidth}
\center
\includegraphics[width=\textwidth]{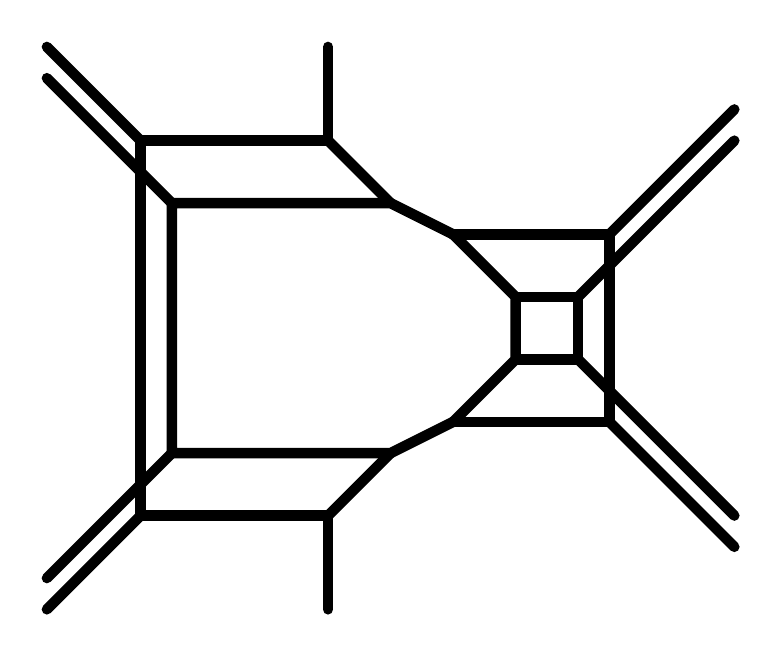} 
\caption{}
\label{}
\end{subfigure}
\caption{5-brane webs for (a) $Sp(2)\times SU(2)$ quiver, (b) $Sp(2)\times Sp(2)$ quiver.}
\label{web_higher_rank}
\end{figure}

The 5-brane webs for the theories we are interested in are shown in 
Fig.~\ref{web_higher_rank}.
Note that the intersections of 5-branes in these webs cannot break.
Each pair of coincident external 5-branes should be considered as ending on the same 7-brane.
This implies that they must connect to different D5-branes, as required by the {\em s}-rule.
This is also consistent with the dimensions of the Coulomb and Higgs branches of these theories (see \cite{Bergman:2012kr}).
For example, the web for $Sp(2)\times SU(2)$, Fig.~\ref{web_higher_rank}a, has three independent faces
corresponding to the three Coulomb moduli.

\begin{figure}[h]
\center
\begin{subfigure}[]{0.4\textwidth}
\center
\includegraphics[width=\textwidth]{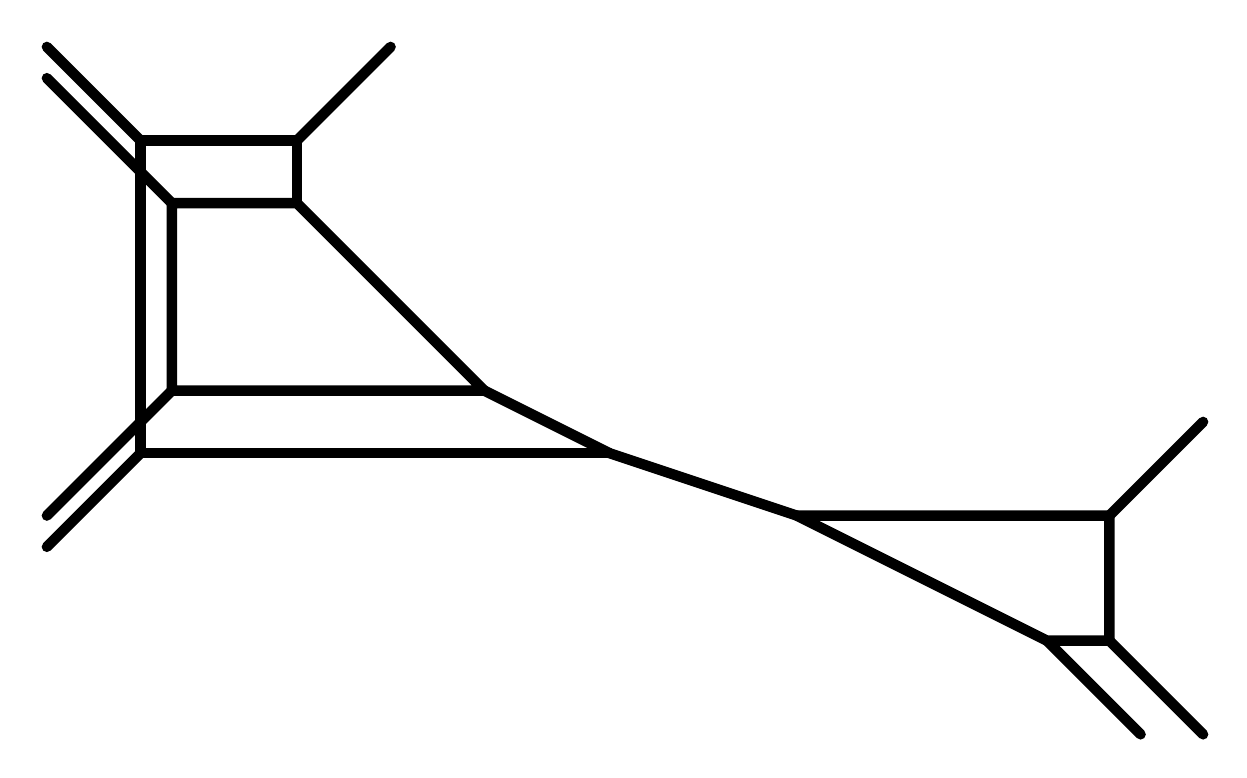} 
\caption{}
\label{}
\end{subfigure}
\hspace{1cm}
\begin{subfigure}[]{0.5\textwidth}
\center
\includegraphics[width=\textwidth]{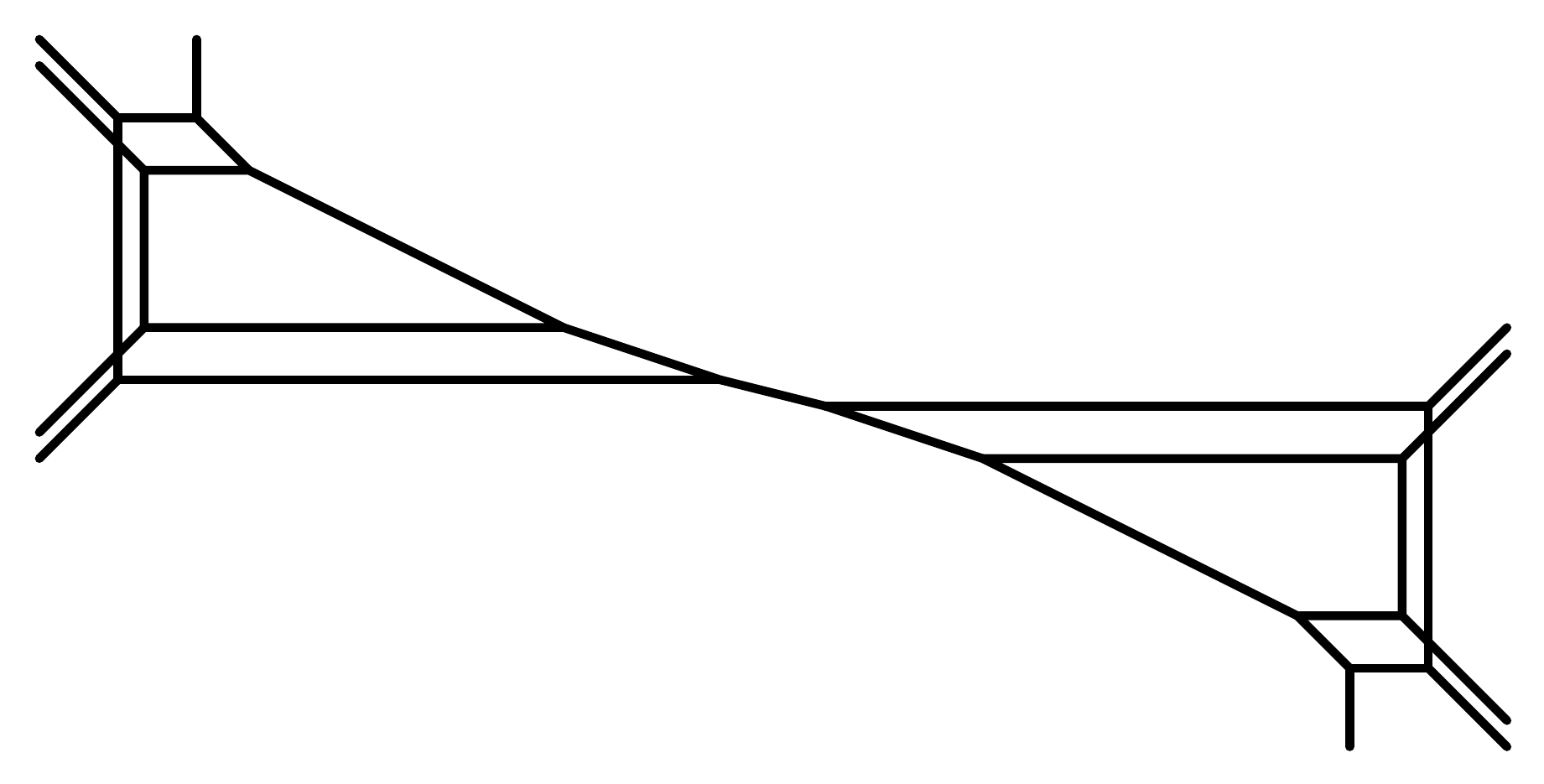} 
\caption{}
\label{}
\end{subfigure}
\caption{Mass deformation of (a) $Sp(2)\times SU(2)$ quiver, (b) $Sp(2)\times Sp(2)$ quiver.}
\label{web_higher_rank_deformed}
\end{figure}

The first question to address concerns the values of the $\theta$ parameters $(\theta_1,\theta_2)$.
As before we can determine these by mass-deforming the web into two sub-webs
corresponding to the two gauge group factors.
The deformations of the webs of Fig.~\ref{web_higher_rank} are shown in Fig.~\ref{web_higher_rank_deformed}.
For the $SU(2)$ part in the $Sp(2)\times SU(2)$ theory clearly $\theta = 0$, since the $SU(2)$ sub-web is 
related by $SL(2,\mathbb{Z})$ to the $SU(2)_0$ web of Fig.~\ref{web_SU(2)}c.
For the $Sp(2)$ sub-webs the situation is less obvious.
Clearly the two sub-webs in the $Sp(2)\times Sp(2)$ case have the same $\theta$,
and the $Sp(2)$ sub-web in the $Sp(2)\times SU(2)$ case has a different $\theta$ from that.
We claim that in the former case $\theta = \pi$ and in the latter case $\theta = 0$. 
This can be understood by going on the Coulomb branch of the $Sp(2)$ factor so that it is broken to $SU(2) \times U(1)$. 
The adjoint ${\bf 10}$ representation of $Sp(2)$ decomposes on the Coulomb branch as
${\bf 3} + 2\cdot {\bf 2} + 3\cdot {\bf 1}$.
The $SU(2)$ part corresponds to the internal face in the $Sp(2)$ sub-web (Fig.~\ref{web_higher_rank_deformed}).
One can see that the $\theta$ parameter for this $SU(2)$ is $\pi$ in the $Sp(2)\times SU(2)$ theory,
and 0 in the $Sp(2)\times Sp(2)$ theory.
On the other hand the $SU(2)$ $\theta$ parameter on the Coulomb branch is the opposite of the $Sp(2)$ $\theta$
parameter at the origin, because the effective theory on the Coulomb branch involves integrating
out two vector multiplets of opposite mass in the ${\bf 2}$ representation of $SU(2)$.
%
%
Therefore $(\theta_1,\theta_2) = (0,0)$ for our $Sp(2)\times SU(2)$ theory,
and $(\theta_1,\theta_2) = (\pi,\pi)$ for our $Sp(2)\times Sp(2)$ theory.

Therefore we do not expect an enhancement of the global symmetry in the $Sp(2)\times Sp(2)$ theory.
On the other hand, for the $Sp(2)\times SU(2)$ theory we expect some enhancement.
This is also hinted by the parallel external 5-branes.
Following the same logic as in section \ref{section:SU(2)xSU(2)_enhancement}, we can determine the nature of the enhancement.
The $(0,1)$ instanton has the properties of an instanton of $SU(2)$ with four flavors.
In this theory there is
an enhancement of the perturbative global symmetry $SO(8)_F \times U(1)_T$
to $E_5 = SO(10)$. 
The $SO(8)_F$ part contains both the $Sp(2)=SO(5)$ gauge symmetry as well as the $SU(2)_M$
global symmetry of the quiver theory. Indeed $SO(5)\times SU(2)$ is a maximal subgroup of $SO(8)$.
The extra currents in $SO(10)$ provided by the instanton transform in one of the two spinor representations
of $SO(8)$,  ${\bf 8}_s$ or ${\bf 8}_c$.
In terms of $Sp(2)\times SU(2)_M \times U(1)_T$ this decomposes either as $({\bf 5},{\bf 1})_{+1} + ({\bf 1},{\bf 3})_{+1}$
or as $({\bf 4},{\bf 2})_{+1}$. 
This reflects the choice of the $\theta$ parameter for the $SU(2)$ part, $\theta_2$.
Only the former case contains a gauge-invariant component, the $({\bf 1},{\bf 3})_{+1}$,
which (together with its complex conjugate) should lead to an enhancement of 
$SU(2)_M \times U(1)_T \rightarrow Sp(2)$. This corresponds to the choice of $\theta _2 =0$.
We will confirm our assertion about $(\theta_1,\theta_2)$ for both theories, and exhibit
the above enhancement, in the index computation.

\begin{figure}[h]
\center
\begin{subfigure}[]{0.2\textwidth}
\center
\includegraphics[width=\textwidth]{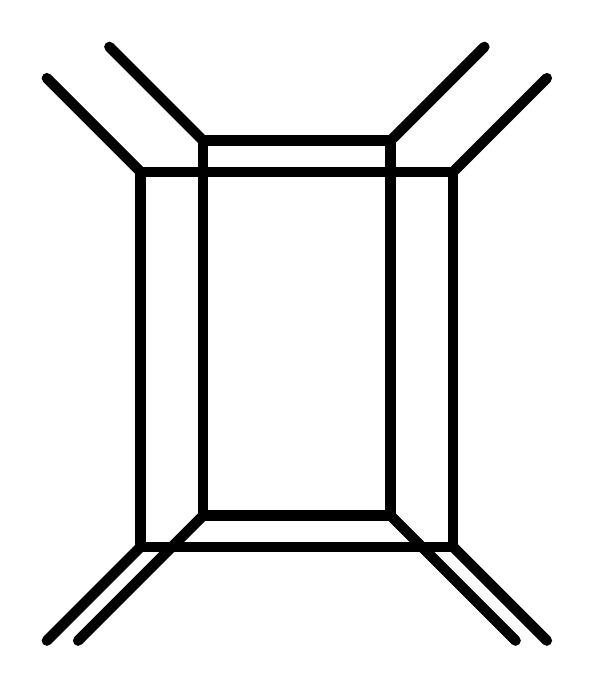} 
\caption{}
\label{}
\end{subfigure}
\hspace{2cm}
\begin{subfigure}[]{0.22\textwidth}
\center
\includegraphics[width=\textwidth]{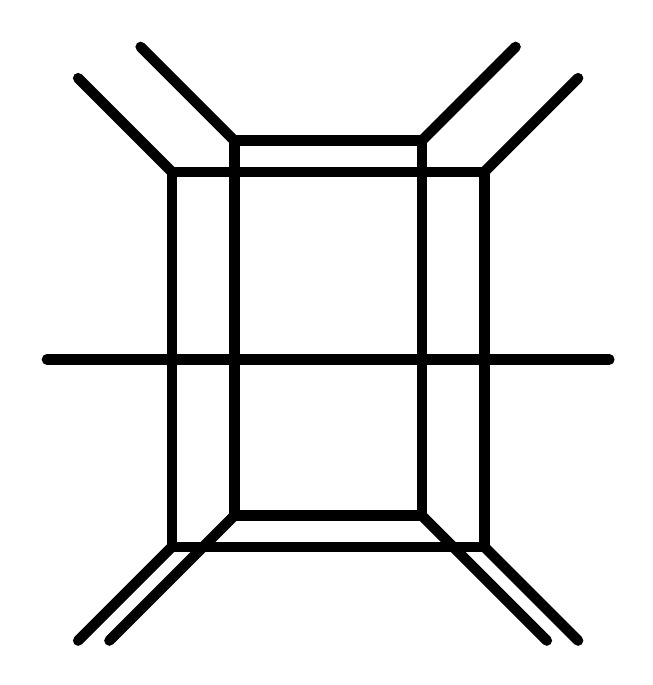} 
\caption{}
\label{}
\end{subfigure}
\caption{Webs on the Higgs branch (a) $SU(4) + 2$, (b) $SU(5) + 2$.}
\label{web_Higgs}
\end{figure}

As before, the continuation past infinite coupling instructs us to view the webs sideways.
This suggests that the dual gauge groups for $Sp(2)\times SU(2)$ and
$Sp(2)\times Sp(2)$ are $SU(4)$ and $SU(5)$, respectively.
But what is the matter content?
For the $SU(5)$ theory, Fig.~\ref{web_higher_rank}b viewed sideways, one might conclude that 
the two external D5-branes give two matter multiplets in the fundamental representation of $SU(5)$,
as in the $SU(3)$ theory previously.
For the $SU(4)$ theory, Fig.~\ref{web_higher_rank}a, the situation is less clear, since instead
of external D5-branes the web has an additional (1,1)5-brane and an additional (1,-1)5-brane.
Actually, in both cases the two matter multiplets are in the rank 2 antisymmetric 
tensor representation.\footnote{This is the other natural generalization of the $SU(3) + 2$ theory, 
since for $SU(3)$ the fundamental and rank 2 antisymmetric representations are equivalent.}
We can understand this indirectly by going to a generic point on the Higgs branch,
shown in Fig.~\ref{web_Higgs}.
This clearly shows the unbroken symmetry in both cases to be $SU(2)\times SU(2)$,
consistent with the pattern of symmetry breaking for two antisymmetric hypermultiplets.\footnote{$SU(4)$ with two antisymmetrics,
namely ${\bf 6}$'s, is equivalent to $SO(6)$ with two vectors. This breaks $SO(6)$ to $SO(4)\sim SU(2)\times SU(2)$.}
By contrast, two fundamental hypermutiplets would generically leave an unbroken symmetry $SU(3)$
in the $SU(4)$ case, and $SU(4)$ in the $SU(5)$ case.

\begin{figure}[h]
\center
\includegraphics[height=0.15\textwidth]{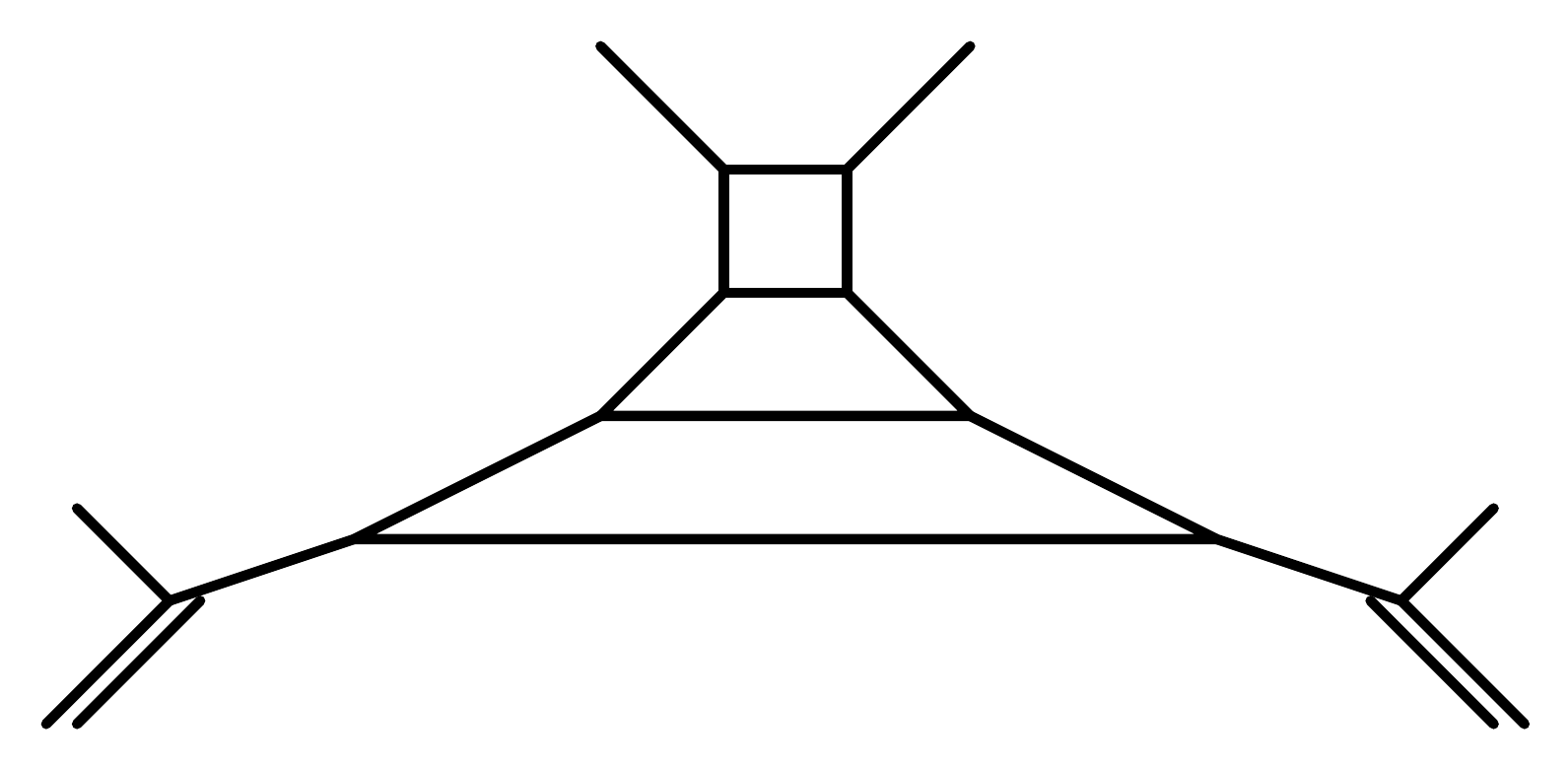} 
\caption{Mass-deformed $SU(4) + 2$}
\label{web_SU(4)_deformed}
\end{figure}

Finally, let us determine the CS levels of the dual theories.
For the $SU(5) + 2$ theory clearly $\kappa_0 = 0$ since the web is invariant under a $\pi$ rotation.
For the $SU(4) + 2$ theory $\kappa_0 \neq 0$, and we can find its value by considering
a mass deformation. 
Turning on negative masses for the matter fields in the $SU(4)$ theory
corresponds to moving the middle external 5-branes in Fig.~\ref{web_higher_rank}a to the left.
Turning the resulting web on its side yields the web in Fig.~\ref{web_SU(4)_deformed}.
The effective CS level in this case is $\kappa = 2$. 
In 5d, the contribution to $\kappa$ of a matter multiplet in a given representation is scaled up
from the contribution of the fundamental representation by the cubic index of that representation.
The cubic index of the antisymmetric representation of $SU(N)$ is $N-4$.
So for $SU(4)$ we find that $\kappa = \kappa_0 = 2$.
Our two new duality conjectures are therefore:
\begin{eqnarray*}
\mbox{Duality 2:} \;\; Sp(2)_0\times SU(2)_0 + (\funda,\funda) & \longleftrightarrow & SU(4)_2 + 2\, \asymm \\[5pt]
\mbox{Duality 3:} \;\; Sp(2)_\pi\times Sp(2)_\pi + (\funda,\funda) & \longleftrightarrow & SU(5)_0 + 2\, \asymm \,.
\end{eqnarray*}

In Duality 3 both the classical global $SU(2)_M \times U(1)^2_T$ symmetry and the discrete $\mathbb{Z}_2$ symmetry agree.
In Duality 2 the classical global symmetries do not agree.
Since $SU(4)\sim SO(6)$, and the antisymmetric representation is real, the flavor symmetry of the 
$SU(4) + 2$ theory is enhanced from $U(2)_F$ to $Sp(2)_F$. But as we argued above, we expect the classical $SU(2)_M$ 
symmetry of the $Sp(2)\times SU(2)$ theory to be enhanced, together with the $U(1)_T$ associated
to the $SU(2)$ instantons, to $Sp(2)$ as well.
The discrete symmetries also match: there is no exchange symmetry in the quiver theory,
and charge conjugation is violated by the non-zero CS level in the $SU(4)_2$ theory.

\subsubsection{Comparing indices}

We begin with the $Sp(2)\times Sp(2)$ and $SU(5)+2A$ theories.
In this case the perturbative indices begin to differ at ${\cal O}(x^6)$: 
\bea
I_{pert}^{Sp(2)\times Sp(2)}  &=&  1 + x^2 \left(\frac{1}{z^2} + 3 + z^2\right) 
+ x^3 \left(\frac{1}{y} + y\right) \left(\frac{1}{z^2} + 4 + z^2\right)  \\
& +&  x^4 \left[ \left(\frac{1}{y^2} + y^2\right) \left(\frac{1}{z^2} + 4 + z^2\right) + \frac{2}{z^4} + \frac{6}{z^2} + 14 + 6 z^2 + 2 z^4 \right]
\nonumber \\
 & + & x^5 \left(\frac{1}{y} + y\right)\left[\left(\frac{1}{y^2} + y^2\right)\left(\frac{1}{z^2} + 4 + z^2\right) + \frac{2}{z^4} + \frac{10}{z^2} + 18 + 10 z^2 + 2 z^4 \right] \nonumber \\
  & + & x^6 \left[ 3\left(\frac{1}{y^2} + y^2\right) \left(\frac{1}{z^4}+ \frac{5}{z^2} + 11 + 5 z^2 + z^4\right) 
  + \left(\frac{1}{y^4} + y^4\right) \left(\frac{1}{z^2} + 4 + z^2\right) \right. \nonumber \\
 & & \qquad \qquad \mbox{} +  \left. \frac{2}{z^6} + \frac{10}{z^4}+\frac{\bf 32}{z^2} + {\bf 58} + {\bf 32} z^2 + 10 z^4 + 2z^6  \right]  \nonumber \\
 & + & x^7 \left(\frac{1}{y} + y\right) \bigg[\left(\frac{1}{y^4}+1 + y^4\right)\left(\frac{1}{z^2} + 4 + z^2\right) \nonumber \\ 
 & & \qquad\qquad \mbox{} +  3\left(\frac{1}{y^2} + y^2\right) \left(\frac{1}{z^4}+\frac{6}{z^2}+ 12 + 6z^2 +z^4\right)+\frac{3}{z^6}+ \frac{19}{z^4} + \frac{\bf 49}{z^2} \nonumber \\
 & & \qquad\qquad \mbox{} + {\bf 74} + {\bf 49} z^2 + 19z^4 + 3z^6 \bigg] + O(x^8) \nonumber
\eea
and
\bea
I_{pert}^{SU(5)+2A} & = & 
1 + x^2 \left(\frac{1}{z^2} + 3 + z^2\right) 
+ x^3 \left(\frac{1}{y} + y\right) \left(\frac{1}{z^2} + 4 + z^2\right) \\
& +&  x^4 \left[ \left(\frac{1}{y^2} + y^2\right) \left(\frac{1}{z^2} + 4 + z^2\right) + \frac{2}{z^4} + \frac{6}{z^2} + 14 + 6 z^2 + 2 z^4 \right]
\nonumber \\
 & + & x^5 \left(\frac{1}{y} + y\right)\left[\left(\frac{1}{y^2} + y^2\right)\left(\frac{1}{z^2} + 4 + z^2\right) + \frac{2}{z^4} + \frac{10}{z^2} + 18 + 10 z^2 + 2 z^4 \right] \nonumber \\
  & + & x^6 \left[ 3\left(\frac{1}{y^2} + y^2\right) \left(\frac{1}{z^4}+ \frac{5}{z^2} + 11 + 5 z^2 + z^4\right) 
  + \left(\frac{1}{y^4} + y^4\right) \left(\frac{1}{z^2} + 4 + z^2\right) \right. \nonumber \\
  & + & \left. \frac{2}{z^6} + \frac{10}{z^4}+\frac{\bf 34}{z^2} + {\bf 62} + {\bf 34} z^2 + 10 z^4 + 2z^6  \right]  \nonumber \\ 
& + & x^7 \bigg\{\left(\frac{1}{y} + y\right) \bigg[\left(\frac{1}{y^4}+1 + y^4\right)\left(\frac{1}{z^2} + 4 + z^2\right)  \nonumber  \\
& &  \mbox{} +  3\left(\frac{1}{y^2} + y^2\right) \left(\frac{1}{z^4}+\frac{6}{z^2}+ 12 + 6z^2 +z^4\right)
+\frac{3}{z^6} + \frac{19}{z^4} + \frac{\bf 51}{z^2} \nonumber \\ 
& & \mbox{} +  {\bf 78} + {\bf 51} z^2 + 19z^4 + 3z^6 \bigg]
- \left({\bf b^5+\frac{1}{b^5}}\right)\left({\bf z + \frac{1}{z}}\right)\left({\bf z^2 + \frac{1}{z^2}}\right)\bigg\} + O(x^8) \,, \nonumber
\eea
where we have indicated in boldface the differing terms.
Note that baryonic states in the $SU(5)+2A$ theory begin to contribute at ${\cal O}(x^7)$, corresponding
to an operator of the form 
$\epsilon^{\alpha_1 \cdots \alpha_5} \epsilon^{\beta_1 \cdots \beta_5} 
A^1_{\alpha_1\alpha_2} A^1_{\alpha_3\alpha_3} A^2_{\beta_1 \beta_2} A^2_{\beta_3 \beta_4} \partial_+ \partial_- A^a_{\alpha_5 \beta_5}$, or variants with different flavor numbers, derivative placement and replacement of some derivatives with gaugino contractions. 

The computation of the instanton corrections is made complicated by the issues discussed in section~\ref{AS_BF_issues}.
We are only able to compute instanton partition functions for $(k_1,0)$ and $(0,k_2)$ instantons in the $Sp(2)\times Sp(2)$ theory,
since these can be treated as instantons of an $Sp(2)$ theory with four flavors. 
The $(1,0)$ and $(0,1)$ instantons contribute
\bea
I_{(1,0)+(0,1)}^{Sp(2)_\pi^2} &=& 2x^6 \left(\frac{1}{z} + z\right)^2 \\
&+& x^7 \left(2\left(\frac{1}{z} + z\right)^2 \left(\frac{1}{y} + y\right) 
- \left(\frac{q_1}{q_2} + \frac{q_2}{q_1}\right)\left(\frac{1}{z} + z\right)\left(\frac{1}{z^2} + z^2\right)\right) + O(x^8) \,. \nonumber
\label{eq:fiouspf}
\eea
For $k_1>1$ or $k_2>1$ the contributions begin at higher orders.
As expected, these instantons contribute only in instanton-anti-instanton combinations, since for $\theta = \pi$
they are charged under the other gauge group.
The ${\cal O}(x^6)$ term precisely makes up the difference between the perturbative indices at ${\cal O}(x^6)$.
Then the ${\cal O}(x^7)$ term precisely makes up the difference at ${\cal O}(x^7)$ if we identify
$b^5 = q_1/q_2$, which means that the charges in Duality 3 are related as $B = 5(I_1 - I_2)$.

We are not able to compute the contributions of instantons in the quiver theory with both $k_1,k_2\neq 0$,
or of instantons in the $SU(5)$ theory, so we cannot confirm the second part of the charge map 
$I = I_1 + I_2$.
However it is reassuring that the superconformal indices agree to ${\cal O}(x^7)$ with
both of these contributions omitted
(see Appendix B for the complete expression).

\medskip

Let us now turn to the $Sp(2)\times SU(2)$ and $SU(4)+2A$ theories, and to Duality 2.
Not surprisingly, in this case the perturbative indices begin to differ already at ${\cal O}(x^2)$,
since the classical global symmetries are different.
The perturbative index of the $Sp(2)\times SU(2)$ theory is given by
\bea
\label{Sp(2)xSU(2)_pert}
I_{pert}^{Sp(2)\times SU(2)}   =   1 + x^2 \left(\frac{1}{z^2} + 3 + z^2\right) + x^3 \left(y + \frac{1}{y}\right)\left(\frac{1}{z^2} + 4 + z^2\right)
+ {\cal O}(x^4) \,,
\eea  
and that of the $SU(4)+2A$ theory is given by
\bea
\label{SU(4)+2_pert}
I_{pert}^{SU(4) + 2A}   &=&   1 + x^2 \left[2 + \left(\frac{1}{z^2} + 1 + z^2\right)\left(\frac{1}{l^2} + 1 + l^2\right) \right] \nonumber \\
& & \mbox{} +  x^3 \left(y + \frac{1}{y}\right) \left[3 + \left(\frac{1}{z^2} + 1 + z^2\right)\left(\frac{1}{l^2} + 1 + l^2\right) \right]
+ {\cal O}(x^4) \,.
\eea
where $z$ and $l$ span the $Sp(2)_F$ flavor symmetry.

As in the previous case, we are only able to compute the corrections due to $(k_1,0)$ and $(0,k_2)$ instantons
in the quiver theory.
As we will now see, the latter are sufficient to reproduce the perturbative index of the $SU(4)$ theory.
The contribution of the $(0,1)$ instanton is given by
\bea
I_{(0,1)}^{Sp(2)_0\times SU(2)_0}  & = &  x^2 \left(q_2 + \frac{1}{q_2}\right)\left(\frac{1}{z^2} + 1 + z^2\right) \nonumber \\
&& \mbox{} + x^3 \left(q_2 + \frac{1}{q_2}\right)\left(y + \frac{1}{y}\right)\left(\frac{1}{z^2} + 1 + z^2\right) 
+ {\cal O}(x^4) \,.
\eea  
Adding this to (\ref{Sp(2)xSU(2)_pert}) we reproduce (\ref{SU(4)+2_pert}) upon identifying $q_2=l^2$.
This confirms Duality 2 to this order, and in particular shows that the global symmetry of the quiver theory
is indeed enhanced from $SU(2)_M\times U(1)_T$ to $Sp(2)$.
The $(0,2)$ and $(0,3)$ instanton contributions are given by
\bea
I_{(0,2)}^{Sp(2)_0\times SU(2)_0}  & = &  x^4 \left(\frac{1}{q_2^2} + q_2^2\right)
\left(\frac{1}{z^2} + 1 + z^2\right)\left(\frac{1}{z^2} + z^2\right) 
\nonumber \\
& + & x^5 \left(\frac{1}{q_2^2} + q_2^2\right)\left(y + \frac{1}{y}\right)\left(\frac{1}{z^2} + 1 + z^2\right)^2 \nonumber \\  
& + & x^6 \bigg\{ \left(\frac{1}{q_2} + q_2\right) \left(\frac{1}{z^6} + \frac{2}{z^4} + \frac{5}{z^2} + 5 + 5z^2 + 2z^4 + z^6 \right)  \nonumber \\ 
& & \qquad \mbox{} +  \left(\frac{1}{q_2^2} + q_2^2\right)\left(\frac{1}{z^2} + 1 + z^2\right) 
  \bigg[ \left(y^2 + \frac{1}{y^2}\right)\left(\frac{2}{z^2} + 1 + 2z^2\right) \nonumber \\
& & \qquad\qquad \mbox{} + \frac{1}{z^4} + \frac{5}{z^2} + 4  +   5z^2 + z^4 \bigg] \bigg\} \nonumber \\
&+& x^7 \left(y + \frac{1}{y}\right)\left(\frac{1}{z^2} + 1 + z^2\right)\bigg\{ \left(\frac{1}{q_2} + q_2\right)
\left( \frac{2}{z^4} + \frac{3}{z^2} + 7 + 3z^2 + 2z^4\right)  \nonumber \\ 
&  & \qquad \mbox{} + \left(\frac{1}{q_2^2} + q_2^2\right)\bigg[ 2\left(y^2 + \frac{1}{y^2}\right)\left(\frac{1}{z^2} + 1 + z^2\right) \nonumber \\
&& \qquad\qquad \mbox{} + \frac{2}{z^4} + \frac{9}{z^2} + 9 + 9z^2 + 2z^4 \bigg] \bigg\} + {\cal O}(x^8) \,,
\eea  
and
\bea
I_{(0,3)}^{Sp(2)_0\times SU(2)_0}  & = &  x^6 \left(\frac{1}{q_2^3} + q_2^3\right)
\left(\frac{1}{z^6} + \frac{1}{z^4} + \frac{2}{z^2} + 2 + 2z^2 + z^4 + z^6\right) \nonumber \\
 & + & x^7 \left(\frac{1}{q_2^3} + q_2^3\right)\left(y + \frac{1}{y}\right)\left(\frac{1}{z^6} + \frac{2}{z^4} 
 + \frac{4}{z^2} + 4 + 4z^2 + 2z^4 + z^6\right) \nonumber \\
& +& {\cal O}(x^8) \,.
\eea  
Setting $q_2=l^2$ these lead to a complete agreement with the the perturbative index of $SU(4)+2A$ to ${\cal O}(x^7)$
(see Appendix B for the complete expression in terms of $Sp(2)_F$ characters).
Note that the $\theta$ parameter of $Sp(2)$ does not enter into these computations, since they do not include $Sp(2)$ instantons.

Completing the duality map requires the $Sp(2)$ instantons on the quiver side and the $SU(4)$ instantons on the other side.
In particular the $Sp(2)$ instanton number should be related to the $SU(4)$ instanton number.
However, due to the issues that we raised, we are not able to compute the contributions of the latter.
In particular this means that our computation so far has been insensitive to the $SU(4)$ CS level.
In principle, we can compute the contributions of pure $Sp(2)$ instantons in the quiver theory, but 
we do not have anything to compare them with.

\subsection{Extra flavor}
\label{extra_flavor}

\begin{figure}[h]
\center
\begin{subfigure}[]{0.25\textwidth}
\center
\includegraphics[width=\textwidth]{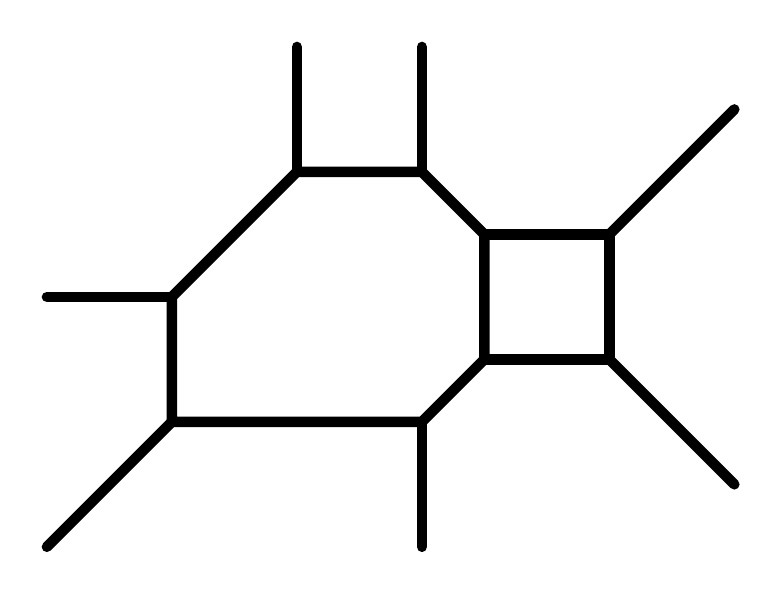}
\caption{}
\label{}
\end{subfigure}
\hspace{2cm}
\begin{subfigure}[]{0.25\textwidth}
\center
\includegraphics[width=\textwidth]{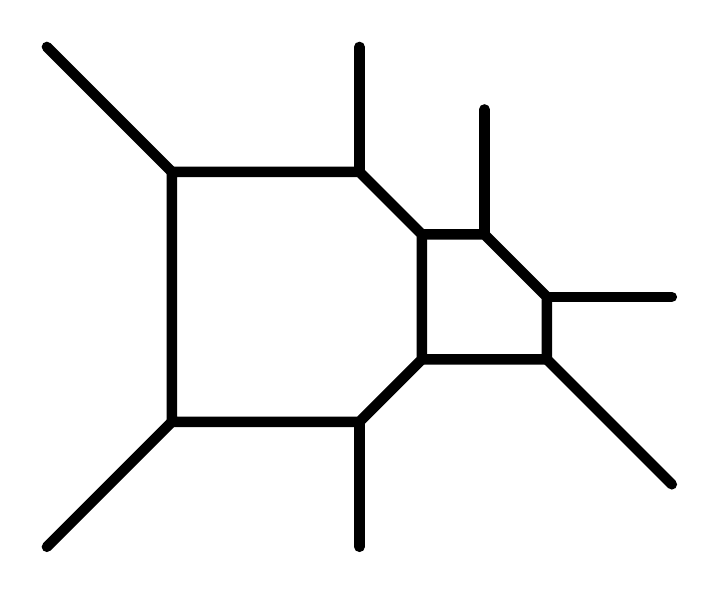} 
\caption{}
\label{}
\end{subfigure}
\caption{$SU(2)\times SU(2)$ quiver with $N_f=1$.}
\label{web_SU(2)xSU(2)+1}
\end{figure}

Going back to the $SU(2)\times SU(2)$ quiver, let
us now add a single matter multiplet in the fundamental representation of one of the $SU(2)$ factors.
The 5-brane webs for the two possibilities are shown in Fig.~\ref{web_SU(2)xSU(2)+1}.
The $\theta$ parameter of the flavored $SU(2)$ is irrelevant.
For the flavorless $SU(2)$ we keep $\theta = \pi$.

Viewing these webs sideways we are led to the proposal that the dual theory is $SU(3)$ with three 
fundamental matter hypermultiplets.
Because there is an odd number of flavors, there is an anomaly unless we shift the quantization
condition of the CS level by one-half. This is the 5d analog of the parity anomaly in 3d.
The two possibilities in Fig.~\ref{web_SU(2)xSU(2)+1}a,b correspond to $\kappa_0 = \pm 1/2$.
So our conjecture in this case is:
$$
\mbox{Duality 4:} \;\; SU(2)\times SU(2)_\pi + (\funda,\funda) + (\funda,{\bf 1})  \longleftrightarrow SU(3)_{\pm\frac{1}{2}} + 3\, \funda \,.
$$

The classical global symmetries of the two theories do not agree.
The quiver theory has $SU(2)_M \times U(1)_T \times U(1)'_T \times U(1)_F$,
whereas the $SU(3)$ theory has $U(3)_F\times U(1)_T$.
But as we are now accustomed to, the existence of parallel external NS5-branes suggests that there
is an instanton-led enhancement of the global symmetry in the quiver theory.
In particular, the instanton of the flavored $SU(2)$ sees effectively three flavors, and therefore,
if not for the other $SU(2)$ gauge group, would enhance the global symmetry from 
$SO(6)_F\times U(1)_T = SU(4)_F \times U(1)_T$
to $E_4 = SU(5)$. The $SO(6)_F$ contains the $SU(2)_M\times U(1)_F$ part of the global symmetry
of the quiver theory, together with the unflavored $SU(2)$ gauge group, as a maximal subgroup.
The instanton transforms as a ${\bf 4}$ of $SO(6)_F$, which decomposes
as $({\bf 2},{\bf 1}) + ({\bf 1},{\bf 2})$. We see that we have the correct gauge-invariant states to enhance
$SU(2)_M \times U(1)_T \rightarrow SU(3)$, making the global symmetries agree.
We will confirm this below using the superconformal index.

The discrete symmetries also agree.
There is no extra $\mathbb{Z}_2$ in either theory.

\subsubsection{Comparing indices}

We begin as usual with the perturbative indices of the two theories.
We present the results to ${\cal O}(x^3)$, since all the information is contained to that order.
The final result to ${\cal O}(x^5)$ is given in Appendix B.
For the $SU(2)\times SU(2)+1$ theory
\bea
I_{pert}^{SU(2)^2+1} & = & 1 + x^2 \left(\frac{1}{z^2} + 4 + z^2\right) 
+ x^3 \left(\frac{1}{y} + y\right)\left(\frac{1}{z^2} + 5 + z^2\right) + {\cal O}(x^4) \,,
\label{eq:sspfpert} \\ \nonumber 
\eea
and for the $SU(3)+3$ theory
\bea
I_{pert}^{SU(3)+3} & = & 1 + x^2 \left[\frac{1}{z^2} + 4 + z^2 + \left(\frac{1}{z} + z\right)\left(\frac{1}{p} + p\right)\right] \label{eq:iwfd} \nonumber \\
& + & x^3 \left[ \left(\frac{1}{y} + y\right)\left[\frac{1}{z^2} + 5 + z^2 + \left(\frac{1}{z} + z\right)\left(\frac{1}{p} + p\right)\right] + b^3 p + \frac{1}{b^3 p} \right]
\nonumber \\ 
&+& {\cal O}(x^4) \,,
\eea
where we have assumed the following decomposition for the $U(3)_F$ flavor symmetry:
\be
\begin{pmatrix}
   b z & 0 & 0\\
  0 & \frac{b}{z} & 0\\  
  0 & 0 & b p\\  
\end{pmatrix} \,.
\ee
We will also denote the flavor fugacity of the $SU(2)\times SU(2) +1$ theory as $l$.
Naturally it does not appear in the perturbative index, since all flavored states are necessarily 
charged under the flavored $SU(2)$.

As expected, the perturbative indices differ already at ${\cal O}(x^2)$.
The $SU(2)\times SU(2)+1$ index exhibits the classical global symmetry $SU(2)_M\times U(1)_T^2 \times U(1)_F$,
whereas the $SU(3)+3$ index exhibits $U(3)_F\times U(1)_T$.
The classical quiver theory is missing four conserved currents.

We turn next to the instanton contributions.
We use the $U(N)$ formalism to compute instanton partition functions in the quiver theory
due to the complication related to the bifundamental hypermultiplet.
We must therefore remove the decoupled state associated with the parallel external NS5-branes.
Taking into account the different fermionic zero modes of the decoupled D-string, this is achieved by
\be
\mathcal{Z}_{inst}^{SU(2)\times SU(2)_\pi +1} 
= PE\left[\frac{q_1 x^2}{z \sqrt{l}(1-x y)(1-\frac{x}{y})}\right] \mathcal{Z}_{inst}^{U(2)_{-1/2}\times U(2)_0 + 1}  \,,
\label{eq:susupcor}
\ee
where $q_1$ is the instanton fugacity for the flavored $SU(2)$, which we took to be the first $SU(2)$.
We have also taken $\kappa_1 = -1/2$ for the first $U(2)$ on the RHS.
For $\kappa_1 = +1/2$ the flavor factor $z\sqrt{l}$ in the plethystic exponential would appear in the numerator, as explained 
below eq.~(\ref{eq:suspftf}) in section~\ref{sec:SU(N)vsU(N)}. 

To ${\cal O}(x^3)$ there are contributions from the $(1,0)$, $(0,1)$ and $(1,1)$ instantons in the quiver theory:
\bea
I_{(1,0)+(0,1)+(1,1)}^{SU(2)\times SU(2)_\pi +1} &=& x^2 \left(\frac{1}{z} + z\right)\left(\frac{\sqrt{l}}{q_1} + \frac{q_1}{\sqrt{l}}\right) \nonumber \\
&+& x^3 \bigg\{
\left[\left(\frac{1}{y} + y\right)\left(\frac{1}{z} + z\right)\left(\frac{\sqrt{l}}{q_1} 
+ \frac{q_1}{\sqrt{l}}\right) + \left(\frac{1}{q_2} + q_2\right)\left(\frac{1}{l} + l\right)\right] \nonumber \\
&& \qquad\qquad \mbox{} + \left(\frac{1}{\sqrt{l}q_1 q_2} + \sqrt{l}q_1 q_2\right)\left(\frac{1}{z} + z\right) \bigg\} + {\cal O}(x^4) \,.
\eea
Note that only the $(1,0)$ instanton contributes at ${\cal O}(x^2)$.
This is for the same reason as in the $SU(2)_0\times SU(2)_\pi$ theory, namely that only this instanton is gauge invariant.
The four extra conserved currents leading to the enhanced $SU(3)$ global symmetry are clearly visible in the ${\cal O}(x^2)$ term,
once we identify $\sqrt{l}/q_1 = p$.

To this order there is also a contribution from the $SU(3)$ instanton:
\bea
I_{1}^{SU(3)_{1/2} +3} & = & x^3 \left[ q \sqrt{\frac{p}{b}} + \frac{1}{q} \sqrt{\frac{b}{p}} 
+ \left(\frac{1}{z} + z\right)\left(\frac{q}  {\sqrt{b p}} + \frac{\sqrt{b p}}{q}\right) \right] + {\cal O}(x^4) \,,
\label{eq:iwtmf} 
\eea
where $q$ is the $SU(3)$ instanton fugacity, and we have chosen $\kappa=+\frac{1}{2}$ (the choice $\kappa=-\frac{1}{2}$ differs by taking $q\rightarrow \frac{1}{q}$).
Taking all the ${\cal O}(x^3)$ contributions, and setting $\sqrt{l}/q_1 = p$, we find a complete agreement if we also identify
$b^3 = \sqrt{l}\, q_1/q_2$ and $q/\sqrt{bp} = q_1 q_2\sqrt{l}$.

We have extended the comparison to ${\cal O}(x^5)$, which requires including also the contributions from the $(0,2)$ and $(1,2)$ instantons.
We find a complete agreement of the indices of the two theories.
The result is shown in Appendix B.

\subsection{Extra node}

\begin{figure}[h]
\center
\begin{subfigure}[]{0.3\textwidth}
\center
\includegraphics[width=\textwidth]{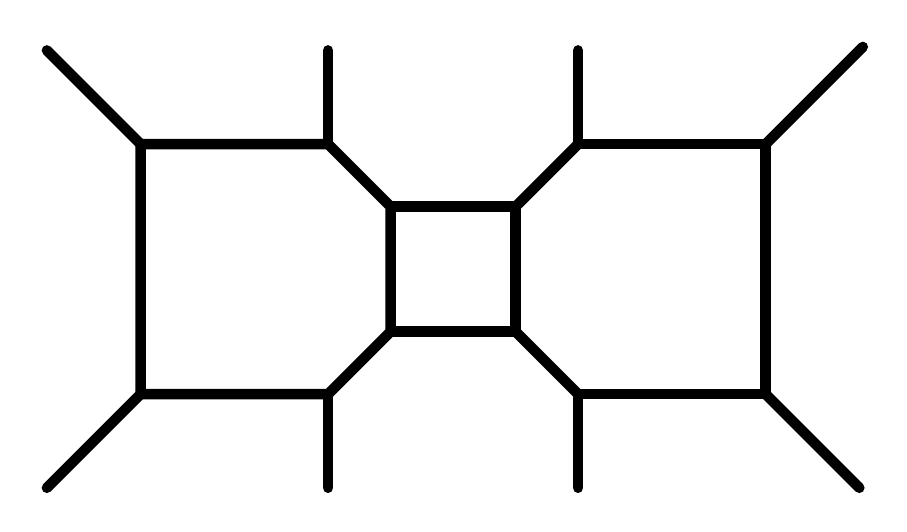}
\caption{}
\label{}
\end{subfigure}
\hspace{2cm}
\begin{subfigure}[]{0.35\textwidth}
\center
\includegraphics[width=\textwidth]{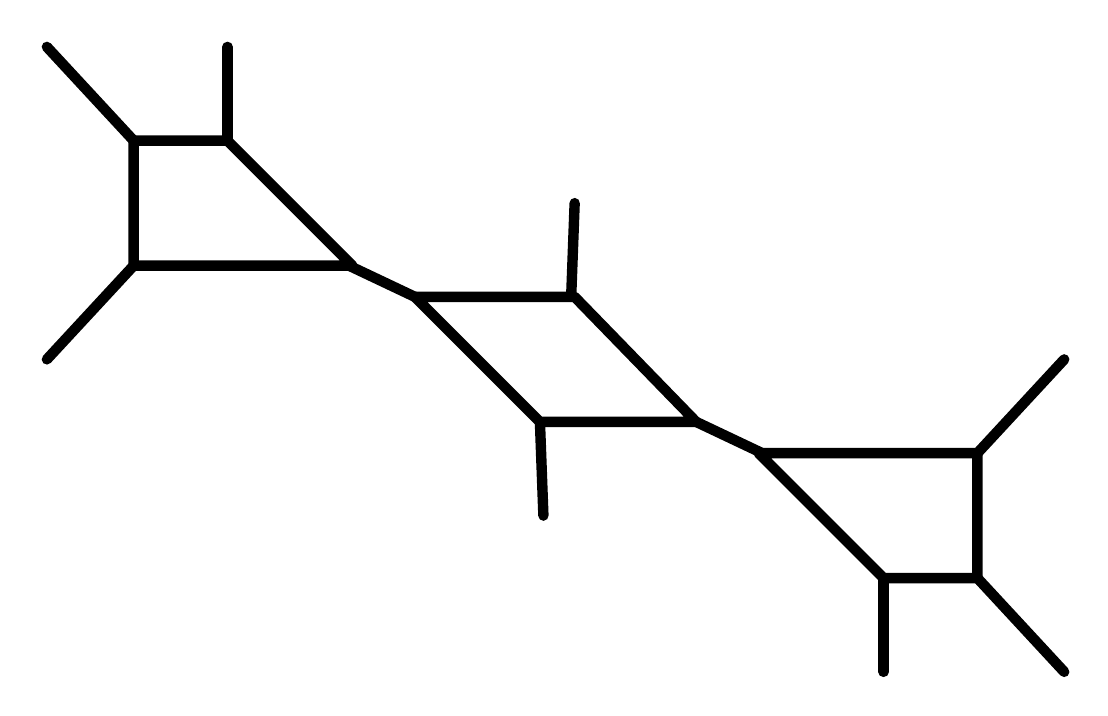} 
\caption{}
\label{}
\end{subfigure}
\caption{$SU(2)_\pi\times SU(2)_0\times SU(2)_\pi$ linear quiver}
\label{web_SU(2)^3}
\end{figure}

As our final generalization, let us add another $SU(2)$ node to the quiver.
The theory is thus $SU(2)\times SU(2)\times SU(2)$, with matter mutiplets
in $({\bf 2},{\bf 2},{\bf 1}) + ({\bf 1},{\bf 2},{\bf 2})$.
We will consider the theory whose 5-brane web realization is shown in Fig.~\ref{web_SU(2)^3}a.
The $\theta$ parameters are easily extracted by mass deforming the web, as in Fig.~\ref{web_SU(2)^3}b.
Clearly $(\theta_1,\theta_2,\theta_3) = (\pi,0,\pi)$.

\begin{figure}[h]
\center
\begin{subfigure}[]{0.15\textwidth}
\center
\includegraphics[width=\textwidth]{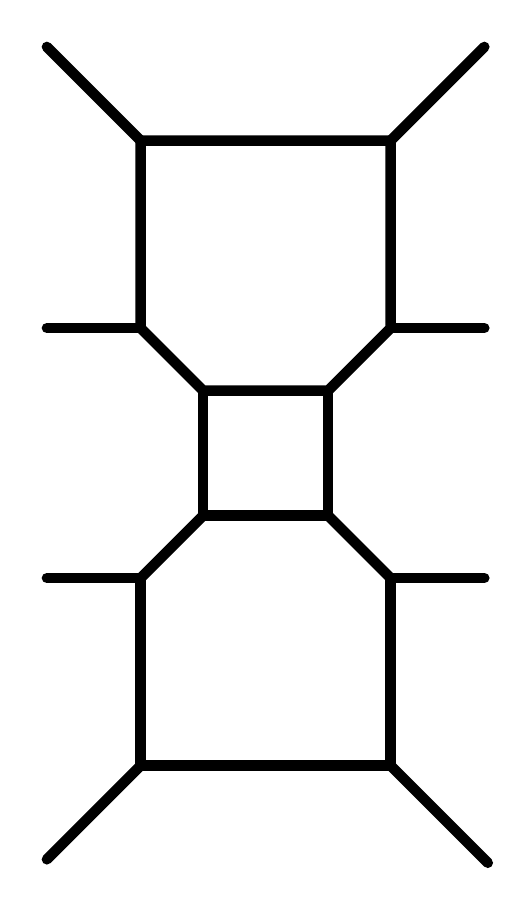}
\caption{}
\label{}
\end{subfigure}
\hspace{3cm}
\begin{subfigure}[]{0.15\textwidth}
\center
\includegraphics[width=\textwidth]{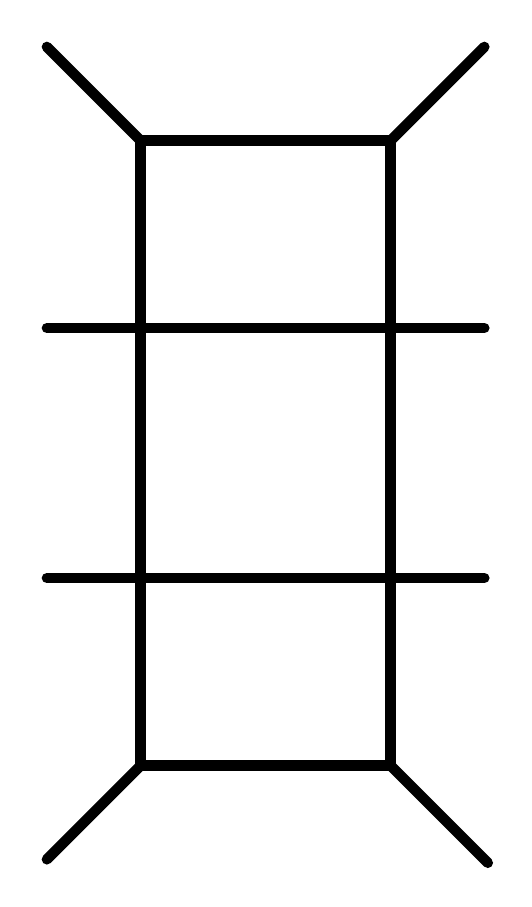} 
\caption{}
\label{}
\end{subfigure}
\caption{$SU(4) + 4$ (a) Coulomb branch, (b) Higgs branch}
\label{web_SU(4)+4}
\end{figure}

Our proposal for the dual theory, Fig.~\ref{web_SU(4)+4}a, is $SU(4)$ with four matter multiplets.
But again we must raise the question of the representations of the matter fields: are they fundamentals,
antisymmetrics, or something else?
Going to the Higgs branch, Fig.~\ref{web_SU(4)+4}b, we see that the unbroken gauge group is $SU(2)$.
This is the correct result for four fundamental hypermutiplets.
On the other hand for four antisymmetrics we would get $SO(2)$.
Our conjecture is therefore
$$
\mbox{Duality 5:} \;\; SU(2)_\pi\times SU(2)_0\times SU(2)_\pi + (\funda,\funda,1) + (1,\funda,\funda)  
\longleftrightarrow SU(4)_{0} + 4\, \funda \,.
$$
The CS level clearly vanishes, since charge conjugation is respected.
The discrete symmetries agree.

This is another example where the global symmetry of the quiver theory should be enhanced.
The classical symmetry is $SU(2)_M\times SU(2)'_M\times U(1)_T\times U(1)'_T\times U(1)''_T$, whereas
that of the $SU(4) + 4$ theory is $U(4)_F\times U(1)_T$.
There are three types of instantons in this theory.
The $(1,0,0)$ and $(0,0,1)$ instantons do not induce enhancement, since $\theta = \pi$ for
the first and last $SU(2)$ factors.
The $(0,1,0)$ instanton is similar to the case discussed in section \ref{higher_rank},
since it sees effectively four flavors. In this case the quiver theory realizes a maximal subgroup
of $SO(8)_F$ given by $SO(4)_M\times SO(4)_{gauge}$, where
$SO(4)_M\sim SU(2)_M\times SU(2)'_M$, and $SO(4)_{gauge}\sim SU(2)_\pi\times SU(2)_\pi$.
There are two inequivalent decompositions of the ${\bf 8}_s$, corresponding to the extra currents in $E_5= SO(10)$:
$({\bf 2},{\bf 2}) + ({\bf 2'},{\bf 2'})$ or $({\bf 4},{\bf 1}) + ({\bf 1},{\bf 4})$.
The latter is the relevant one, since $\theta_2=0$, and leads to the required enhancement of $SO(4)_M\times U(1)_T \rightarrow SU(4)$.
We will also see this below in the superconformal index.

\subsubsection{Comparing indices}

We will need to go to ${\cal O}(x^4)$ to see baryonic states in the $SU(4)+4$ theory.
The perturbative index of the quiver theory to this order is given by
\bea
I_{pert}^{SU(2)^3}  & = &  1 + x^2 \left(\frac{1}{z^2} + \frac{1}{z^{\prime 2}} + 5 + z^{\prime 2} + z^2 \right) 
+ x^3 \left(y+\frac{1}{y}\right)\left(\frac{1}{z^2} + \frac{1}{z^{\prime 2}} + 6 + z^{\prime 2} + z^2 \right) \nonumber \\  
& + & x^4 \bigg[ \left(z^{\prime 2}+ \frac{1}{z^{\prime 2}}\right)\left(z^2 + \frac{1}{z^2}\right) 
+ \left(y^2 + \frac{1}{y^2}\right)\left(z^{\prime 2}+ \frac{1}{z^{\prime 2}}+z^2 + \frac{1}{z^2}+6\right) 
\nonumber \\  
&& \qquad\qquad \mbox{} +  \frac{1}{z^4} + \frac{5}{z^2} + 5 z^2 + z^4 
+ \frac{1}{z^{\prime 4}} + \frac{5}{z^{\prime 2}} + 5 z^{\prime 2} + z^{\prime 4} + 16 \bigg] + {\cal O}(x^5) \,, \nonumber \\  
\eea    
where $z$ and $z'$ are the fugacities associated to $SU(2)_M$ and $SU(2)_M'$, respectively. 
The adjoint characters of $SU(2)_M\times SU(2)_M' \times U(1)^3_T$ are clearly visible in the $x^2$ term. 
With the privilege of hindsight, we will use the following decomposition of the global $U(4)_F$ symmetry of the $SU(4)+4$ theory:
\be
\begin{pmatrix}
  \frac{bz}{h} & 0 & 0 & 0\\
  0 & \frac{b}{zh} & 0 & 0\\  
  0 & 0 & b z' h & 0\\  
  0 & 0 &0  & \frac{bh}{z'}
\end{pmatrix}  \,.
\ee
In terms of these, the perturbative index of the $SU(4)+4$ theory is given by
\bea
I_{pert}^{SU(4)+4}  & = &  1 + x^2 \left[\frac{1}{z^2} + \frac{1}{z^{\prime 2}} + 5 + z^{\prime 2} + z^2 
+ \left(h^2 +\frac{1}{h^2}\right)\left(z'+\frac{1}{z'}\right)\left(z+\frac{1}{z}\right) \right] \nonumber \\  
& + & x^3 \left(y+\frac{1}{y}\right)\left[\frac{1}{z^2} + \frac{1}{z^{\prime 2}} + 6 + z^{\prime 2} + z^2 
+ \left(h^2 +\frac{1}{h^2}\right)\left(z'+\frac{1}{z'}\right)\left(z+\frac{1}{z}\right) \right] \nonumber \\  
& + & x^4 \bigg\{ 2\left(z^{\prime 2}+ \frac{1}{z^{\prime 2}}\right)\left(z^2 + \frac{1}{z^2}\right) 
+ \left(y^2 + \frac{1}{y^2}+1\right)\bigg[z^{\prime 2}+ \frac{1}{z^{\prime 2}}+z^2 + \frac{1}{z^2}+6 \nonumber \\ 
&  & \mbox{} + \left(h^2 +\frac{1}{h^2}\right)\left(z'+\frac{1}{z'}\right)\left(z+\frac{1}{z}\right) \bigg]
+ \frac{1}{z^4} + \frac{6}{z^2} + 6 z^2 + z^4 + \frac{1}{z^{\prime 4}} + \frac{6}{z^{\prime 2}} + 6 z^{\prime 2} \nonumber \\
&& \mbox{} + z^{\prime 4} +  17 + \left(h^4 +\frac{1}{h^4}\right)
\left[1+\left(z^{\prime 2}+1+ \frac{1}{z^{\prime 2}}\right)\left(z^2 +1+ \frac{1}{z^2}\right)\right] \nonumber \\
& & \mbox {} + b^4 +\frac{1}{b^4} 
+  \left(h^2 +\frac{1}{h^2}\right)\left[\left(z^{\prime 3}+\frac{1}{z^{\prime 3}}\right)\left(z+\frac{1}{z}\right) 
+ \left(z'+\frac{1}{z'}\right)\left(z^3+\frac{1}{z^3}\right) \right. \nonumber \\
&& \qquad \left. \mbox{} + 6\left(z'+\frac{1}{z'}\right)\left(z+\frac{1}{z}\right) \right] \bigg\} + {\cal O}(x^5) \,.
\eea 
The $x^2$ term contains in fact the adjoint character of $SU(4)_F$.

Turning our attention to the instanton contributions, we will again need to include a correction factor for the quiver theory.
By now, we can simply read it off from the 5-brane web:
\be
\mathcal{Z}_{inst}^{SU(2)_\pi \times SU(2)_0 \times SU(2)_\pi} 
= PE\left[\frac{q_2\, x^2(z z' + \frac{1}{z z'})}{(1-x y)(1-\frac{x}{y})}\right] 
\mathcal{Z}_{inst}^{U(2)_0 \times U(2)_1 \times U(2)_0} \,.
\label{eq:susuffcor}
\ee
The dependence on the two bifundamental fugacities can be understood from the dependence of the mass 
of the D-string between the parallel NS5-branes on the masses of the two bifundamental matter multiplets.
This is basically the same as the $SU(2) +4$ case in eq.~(\ref{eq:SU(2)+4}).

To ${\cal O}(x^3)$ there is only a contribution from the $(0,1,0)$ instanton of the quiver theory:
\bea
I_{(0,1,0)}^{SU(2)^3}   &=&   x^2 \left(q_2+\frac{1}{q_2}\right)\left(z'+\frac{1}{z'}\right)\left(z+\frac{1}{z}\right) \nonumber \\
&+& x^3 \left(y+\frac{1}{y}\right)\left(q_2+\frac{1}{q_2}\right)\left(z'+\frac{1}{z'}\right)\left(z+\frac{1}{z}\right) + {\cal O}(x^4) \,.
\eea
Adding this to the perturbative result reproduces the $SU(4)+4$ index to this order if we identify $q_2=h^2$.
So indeed, it appears that the global symmetry is enhanced to $SU(4)_F \times U(1)_T^2$ in the quiver theory.

To complete the charge map we need to go to ${\cal O}(x^4)$, which is where the baryonic states enter in the $SU(4)+4$ theory.
To this order there are contributions from the $(0,1,0)$, $(1,0,0)$, $(0,0,1)$, $(0,2,0)$, $(1,0,1)$, $(1,1,1)$ and $(1,2,1)$ instantons.
Their sum gives
\bea
I_{many}^{SU(2)^3} &=& \cdots \mbox{} + x^4 \bigg\{ \left(q_2+\frac{1}{q_2}\right)\left(z'+\frac{1}{z'}\right)\left(z+\frac{1}{z}\right)\left( \frac{1}{z^2} + \frac{1}{c^2} + \frac{1}{y^2} + 5 + y^2 + c^2 + z^2 \right) \nonumber \\  
& & \mbox{} +  \left(z' +\frac{1}{z'}\right)^2 \left(z+\frac{1}{z}\right)^2 + 3 + \frac{q_1}{q_3}+ \frac{q_3}{q_1} \nonumber \\
& & \mbox{} + \left(q_2^2 +\frac{1}{q_2^2}\right)\left[1+\left(z^{\prime 2}+1+ \frac{1}{z^{\prime 2}}\right)\left(z^2 +1+ \frac{1}{z^2}\right) \right] \nonumber \\
& & \mbox{} + q_1 q_3 +\frac{1}{q_1 q_3} + \left(q_1 q_2 q_3 +\frac{1}{q_1 q_2 q_3}\right)\left(z +\frac{1}{z}\right)\left(z' +\frac{1}{z'}\right) 
+ q_1 q_2^2 q_3 +\frac{1}{q_1 q_2^2 q_3} \bigg\} \nonumber \\
&+& {\cal O}(x^5) \,.
\eea
Other low number instantons, like the $(1,1,0)$ instanton, do not contribute to this order since they are gauge-charged.
For the $SU(4)+4$ theory, only the 1-instanton contributes to this order:
\bea
I_{1}^{SU(4)_0+4}  & = &  x^4 \left(q +\frac{1}{q}\right)\left[h^2+\frac{1}{h^2}+\left(z'+ \frac{1}{z'}\right)\left(z + \frac{1}{z}\right) \right]
+ {\cal O}(x^5) \,.
\eea
Including these contributions we find a complete agreement between the two theories if we also
identify $b^4 = q_1/q_3$ and $q = q_1 q_2 q_3$, in addition to $q_2 = h^2$.
The calculation was actually carried out to ${\cal O}(x^5)$, and the indices agree with the above identifications.
The explicit result in terms of $SU(4)_F$ characters is given in Appendix B.

\section{Conclusions}

In this paper we set out to explore properties of 5d superconformal field theories that can be described as UV fixed
points of ${\cal N}=1$ supersymmetric gauge theories.
Our main tools have been the 5-brane web constructions of \cite{Aharony:1997ju},
and the computation of 5d superconformal indices via localization pioneered in \cite{KKL}.

We have uncovered several new cases of non-perturbatively enhanced global symmetries,
analogous to the exceptional global symmetries of the $SU(2) + N_f$ theories.
In particular we have shown that in the ${\cal N}=1$ $SU(N)$ gauge theory with CS level $\kappa = \pm N$,
which sits on the borderline of well-defined fixed points in terms of the relation between 
$\kappa$ and $N$ \cite{Intriligator:1997pq}, the global topological $U(1)_T$ symmetry is enhanced to $SU(2)$ at the fixed point.
Unlike the $SU(2) + N_f$ examples, we do not have a stringy description of this enhancement.

This result can be generalized to $SU(N)_\kappa$ with $N_f$ hypermultiplets in the fundamental representation 
such that $N_f + 2|\kappa| = 2N$. 
For $0 < |\kappa| < N$, one combination of the topological and baryonic $U(1)$'s, either the symmetric or antisymmetric
combination, depending on the sign of $\kappa$, is enhanced to $SU(2)$.
For $\kappa =0$ and $N_f = 2N$, both combinations are enhanced to $SU(2)$'s. 

Our analysis suggests more generally that any theory described by a 5-brane web
with external parallel NS5-branes, or more generally 5-branes that are not D5-branes,
will exhibit a non-perturbatively enhanced global symmetry.
Indeed we have shown this in a number of other examples, starting with the $SU(2)\times SU(2)$
linear quiver theory with $(\theta_1,\theta_2) = (0,\pi)$ and $(0,0)$.
The former exhibits enhancement from $SU(2)_M\times U(1)_T^2$ to $SU(3)\times U(1)_T$,
and the latter to $SU(4)$.
There are several generalizations of this that one can explore.
For example, our studies indicate that the $SU(2)^n$ linear quiver theory should exhibit 
enhanced global symmetries $SU(2n)$, $SU(2n-1)$ and $SU(2n-2)$ for the cases
$(\theta_1,\ldots,\theta_n) = (0,\ldots,0)$, $(\pi,0,\ldots,0)$ and
$(\pi,0,\ldots,0,\pi)$, respectively.

\medskip

We have also formulated a number of duality conjectures via a ``continuation past infinite coupling"
as suggested by the 5-brane web construction, generalizing the one between the $SU(2)\times SU(2)$  
theory and $SU(3)$ with $N_f = 2$, made in \cite{Aharony:1997ju}.
In each case we showed that the superconformal index for the two theories was equal to 
a reasonably high order in an expansion for small $x$.
This provides very solid evidence for the conjectured duailties.

There are a number of further directions to explore.
The dualities we proposed are just the simplest generalizations of the one for $SU(2)\times SU(2)$.
What is the more general relation for rank $N$ and $N_f$ flavors?

The natural guess for the dual of the $Sp(N)\times Sp(N)$ theory is $SU(2N+1)$ with two antisymmetrics
and the natural guess for $Sp(N+1)\times Sp(N)$ is $SU(2N+2)$ with two antisymmetrics.
What about $Sp(N+M)\times Sp(N)$ for $M>1$?
At large $N$ these theories are dual to Massive IIA supergravity on $AdS_6 \times S^4/{\mathbb{Z}_2}$
with and without vector structure, and with some additional fluxes corresponding to fractional branes \cite{Bergman:2012kr}.
Can one understand the field theory dualities from the point of view of these backgrounds?

Adding $N_f$ flavors to the $SU(2)\times SU(2)$ theory, 
the natural guess for the dual is $SU(3)$ with $N_f+2$ flavors.
For the $SU(3)$ theory a fixed point exists only if $N_f + 2\kappa_0 \leq 4$ \cite{Intriligator:1997pq}.
This should also be observed on the $SU(2)\times SU(2)$ side.
More generally, one might guess that the dual of $Sp(N)\times Sp(N)$ with $N_f$ flavors
is $SU(2N+1)$ with two antisymmetrics and $N_f$ fundamentals.

One can also ask about the generalization to more nodes, $SU(2)^n$.
In this case the 5-brane web suggests that the dual is $SU(n+1)$ with $2n-2$ fundamentals.
This is closely related to the duality studied in \cite{Bao:2011rc}.
One can go on and combine the different generalizations.
The problem is that as we increase the ranks, number of flavors and number of nodes,
the index computations become more and more cumbersome.
Perhaps more efficient techniques can be found.

We haven't said anything about $SO(N)$ theories.
5-brane webs provide a very nice way to realize gauge theories with $SU(N)$ and $Sp(N)$ gauge groups,
but to realize $SO(N)$ we would need to add orientifold planes.
It would be interesting to study the continuation past infinite coupling in these configurations,
and then to compare with computations of the superconformal indices,
which should be possible for $SO(N)$.

Another interesting direction to explore are relations of 5d dualities to 4d dualities,
along the lines of the relations between 4d and 3d dualities of \cite{Aharony:2013dha}.

\section*{Acknowledgements}

We are grateful to Ofer Aharony, Davide Gaiotto, Noppadol Mekareeya, Shlomo Razamat and Nathan Seiberg for useful comments, 
and to Hee-Cheol Kim and Kimyeong Lee for helpful correspondence.
O.B. would like to thank the High Energy Physics group at the University of Oviedo for their hospitality.
O.B. is supported in part by the Israel Science Foundation under grants no. 392/09, and 352/13,
the US-Israel Binational Science Foundation under grants no. 2008-072, and 2012-041,
the German-Israeli Foundation for Scientific Research and Development under grant no. 1156-124.7/2011,
and by the Technion V.P.R Fund.
G.Z. is supported in part by Israel Science Foundation under grant no. 392/09.
D.R-G is supported by the Ram\'on y Cajal fellowship RyC-2011-07593, as well as by the Spansih Ministry of Science and Education grant FPA2012-35043-C02-02.

\appendix

\section{Instanton partition functions}

In this appendix we collect all the relevant expressions for the instanton partition functions that we use
in computing the instanton contributions to the superconformal indices.
Some of these results are taken from \cite{KKL}, and others where lifted from the 4d results of \cite{Nekrasov:2004vw,Shad,Shad2}.
Each of these is given as a contour integral over the Cartan subgroup of the dual gauge group, which depends on the gauge group and instanton number.
We will express everything in terms of fugacities.
The fugacities associated with the gauge group will be denoted by $s_i \equiv e^{i\alpha_i}$, and those of the dual gauge group by $u_a$.
We will also denote by $f_n = e^{im_n}$ the fugacities associated with flavor matter fields (in fundamental representations),
and by $z$ the fugacity associated with either bifundamental or antisymmetric matter fields.
We assume that $|x|<<1$, so the relevant poles are the ones at $u_a \propto x^{positive}$.
Generically, the instanton partition function is given by 
\bea
Z_{inst}  =  \oint [du] \, z_{G}[u_a]  z_{M}[u_a] \,,
\label{generic_inst}
\eea
where  $[du]$ denotes the Haar measure of the dual gauge group,
$z_{G}$ denotes the contribution of the gauge multiplet, and $z_{M}$ denotes
the contribution of matter multiplets.

\subsection{$SU(N)$}

Strictly speaking, the instanton partition functions for $SU(N)$ are really computed for $U(N)$, and then 
one sets $\prod_{i=1}^N s_i=1$.
As we mentioned in section~\ref{sec:SU(N)vsU(N)}, one often needs to include an additional factor to remove remnants of the overall $U(1)$ dependence.

For $k$ instantons in $U(N)$ the dual gauge group is $U(k)$. 
The Haar measure for $U(k)$ is given by
\be
[du] = \frac{1}{k!} \prod^{k}_{a=1} \frac{du_a}{u_a} \,
\prod^{k}_{a<b} \left(\frac{u_b}{u_a}+\frac{u_a}{u_b}-2\right) \,.\label{U(N)_inst_Haar}
\ee
The contribution of the $U(N)$ gauge multiplet is 
\bea
z_{G}^k[u_a]  &=& \prod^{k}_{a=1} \frac{(1-x^2) u^{\kappa}_a} {(1-x y)(1-\frac{x}{y})\prod^N_{i=1} (x+\frac{1}{x}-\frac{u_a}{s_i} -\frac{s_i}{u_a})} \nonumber \\  
&& \qquad \qquad \mbox{} \times \prod^{k}_{a<b} \frac{
(\frac{u_b}{u_a}+\frac{u_a}{u_b}-x^2 
-\frac{1}{x^2})}{(\frac{u_b}{u_a}+\frac{u_a}{u_b}-x y - 
\frac{1}{xy})(\frac{u_b}{u_a}+\frac{u_a}{u_b}-\frac{x}{y}-\frac{y}{x})} \,,
\label{U(N)_inst_gauge}
\eea
where $\kappa$ is the $U(N)$ CS level.
For $N_f$ matter multiplets in the fundamental representation, the matter contribution is 
\bea
z_{F}^k[u_a] =  \prod^{k}_{a=1} \prod^{N_f}_{n=1} \left(\sqrt{u_a f_n}-\frac{1}{\sqrt{u_a f_n}}\right) \,. 
\eea

The computation of the contour integral gets quite involved for $k>1$.
There are several poles to consider, some of which end up summing to a vanishing contribution.
This problem was solved in \cite{Nekrasov:2002qd}.
The poles can be classified by $N$ Young diagrams with a total of $k$ boxes.
For each set of $N$ Young diagrams there is a corresponding set of poles.
Each box 
corresponds to one $u_a$. 
The first box in each of the $N$ diagrams corresponds to one of the $N$ basic poles at $u=x s_i$.
If a box representing $u_b$ appears below 
a box representing $u_a$ their poles are related by $u_b = x u_a/y$.
If it appears to the right of the $u_a$ box the poles are related as
$u_b = x y u_a$. 
Each set of Young diagrams actually gives $k!$ equivalent sets of poles, corresponding 
to permutations of $\{u_a\}$, which cancels the $k!$ in (\ref{U(N)_inst_Haar}).

For example, for 1 instanton there are $N$ possibilities for the position of
the 1-box diagram in the set, which correspond to the $N$ poles at $u = x s_i$. 
For 2 instantons, there are three types of sets, the first containing two 1-box diagrams, 
and the two others containing the two possible 2-box diagrams.
The former has $N(N-1)/2$ possibilities, corresponding to poles at
$u_1 = x s_i, u_2 = x s_j$. 
The latter has $2N$ possibilities, 
corresponding to poles of the form $u_1= x y u_2, u_2 = x s_i$ and $u_1= x u_2/y, u_2 = x s_i$.

The contribution of matter in the antisymmetric representation is a bit more involved and is given by 
(lifting the 4d results from \cite{Shad})
\bea
z_{A}^k[u_a] &=& (-1)^N \prod^k_{a=1} \frac{\prod^N_{i=1} \left(\sqrt{u_a s_i z} 
- \frac{1}{\sqrt{u_a s_i z}}\right)}{u^2_a z + \frac{1}{u^2_a z} - x - \frac{1}{x}}  \nonumber \\
&& \qquad\qquad \mbox{} \times
\prod^k_{a<b} \frac{(u_a u_b + \frac{1}{u_a u_b} - z y - \frac{1}{z y})(u_a u_b + \frac{1}{u_a u_b} 
- \frac{z}{y} - \frac{y}{z})}{(u_a u_b + \frac{1}{u_a u_b} - z x - \frac{1}{z x})(u_a u_b + \frac{1}{u_a u_b} 
- \frac{z}{x} - \frac{x}{z})} \,. \label{ASSU}
\eea
We did not actually use this, due to the problems mentioned in section~\ref{AS_BF_issues}.

For quiver theories, the partition functions for di-group instantons will have a contribution 
from bifundamental matter.
A single bifundamental of $U(N_1)\times U(N_2)$ contributes to the $(k_1,k_2)$ instanton
partition function integrand the factor (lifting this time from \cite{Shad2}):
\bea
z^{k_1, k_2}_{BF}[u,u'] &=& \prod_{a,j=1}^{k_1,N_2}
\left( \sqrt{\frac{u_a z}{s'_{j}}} - \sqrt{\frac{s'_{j}}{u_a z}} \right) 
\prod_{b,i=1}^{k_2,N_1} 
\left( \sqrt{\frac{u'_{b}}{z s_i}} - \sqrt{\frac{s_i z}{u'_{b}}} \right) 
\prod_{a,b}^{k_1,k_2}
\frac{\frac{u_a z}{u'_b}+\frac{u'_b}{u_a z} - y - \frac{1}{y}}
{\frac{u_a z}{u'_b}+\frac{u'_b}{u_a z} - x - \frac{1}{x}} \,,
\nonumber \\
\label{U(N)_BF}
\eea
where $s_i, s'_j$ are the fugacities associated with the two gauge groups $U(N_1), U(N_2)$,
and $u_a, u'_{b}$ are the fugacities associated with the two dual gauge groups
$U(k_1), U(k_2)$. 
This contributes additional poles to the integral, and one must give a pole prescription. 
We follow \cite{KKL}, and ignore these poles.

\subsection{$Sp(N)$}

For $k$ $Sp(N)$ instantons the dual gauge group is $O(k)$.
The instanton partition function consists of two parts associated with the two disconnected
components of $O(k)$, $O(k)_+ = SO(k)$ and $O(k)_-$.
There are two possible combinations corresponding to the two possible values of the discrete
$\theta$ parameter \cite{Bergman:2013ala}:
\be
Z_k^{Sp(N)} = \left\{
\begin{array}{ll}
\frac{1}{2}(Z_{k}^{+} + Z_{k}^{-}) & \theta = 0 \\[5pt]
\frac{(-1)^k}{2}(Z_{k}^{+} - Z_{k}^{-}) & \theta = \pi \,.
\end{array} \right.
\label{theta_inst}
\ee

The even and odd $k$ cases are qualitatively different, so we will present them separately.
The Haar measures in the different cases are given by
\be
[du] = \left\{
\begin{array}{ll}
\frac{(-1)^n}{2^n n!} \prod_{a=1}^n \frac{(u_a + \frac{1}{u_a} - 2)du_a}{u_a}\,
 \prod^{n}_{a<b} \left(u_a+\frac{1}{u_a}-u_b-\frac{1}{u_b}\right)^2 & k=2n+1, O_+ \\
\frac{1}{2^{n-1} n!} \prod_{a=1}^n \frac{du_a}{u_a}\,
 \prod^{n}_{a<b} \left(u_a+\frac{1}{u_a}-u_b-\frac{1}{u_b}\right)^2 & k=2n, O_+ \\
 \frac{1}{2^n n!} \prod_{a=1}^n \frac{(u_a + \frac{1}{u_a} + 2)du_a}{u_a}\,
  \prod^{n}_{a<b} \left(u_a+\frac{1}{u_a}-u_b-\frac{1}{u_b}\right)^2 & k=2n+1, O_- \\
 \frac{(-1)^n}{2^n n!} \prod_{a=1}^n \frac{(u^2_a + \frac{1}{u^2_a} - 2)du_a}{u_a}\,
  \prod^{n}_{a<b} \left(u_a+\frac{1}{u_a}-u_b-\frac{1}{u_b}\right)^2 & k=2n+2, O_- \\
\end{array} \right.
\ee
The contributions of the gauge multiplet in the different cases are given by 
\bea
&& z_{G+}^{2n+1}[u_a] =  \frac{(-1)^n x (1-x^2)^n}
{(1-x y)^{n+1}(1-\frac{x}{y})^{n+1} \prod^{N}_{i=1} (x+\frac{1}{x}-s_i-\frac{1}{s_i})} \nonumber\\
&& 
\prod^{n}_{a=1}   \frac{(u_a+\frac{1}{u_a}-x^2
-\frac{1}{x^2})}{(u_a+\frac{1}{u_a}-x y-\frac{1}{x y})(u_a+\frac{1}{u_a}-\frac{y}{x}-\frac{x}{y})
(u^2_a+\frac{1}{u^2_a}-x y-\frac{1}{x y})(u^2_a+\frac{1}{u^2_a}-\frac{y}{x}-\frac{x}{y})} \nonumber \\
&& 
 \prod^{n}_{a=1} \prod_{i=1}^N 
\frac{1}{(u_a+\frac{1}{u_a}-x s_i-\frac{1}{x s_i})(u_a+\frac{1}{u_a}-\frac{s_i}{x}-\frac{x}{s_i})} \\
&& 
\prod^{n}_{a<b} 
\frac{(u_a u_b + \frac{1}{u_a u_b} - x^2 - \frac{1}{x^2})(\frac{u_a}{u_b} + \frac{u_b}{u_a} - x^2 - \frac{1}{x^2})}
{(u_a u_b + \frac{1}{u_a u_b} - x y-\frac{1}{x y})(\frac{u_a}{u_b} + \frac{u_b}{u_a} - x y-\frac{1}{x y})
(u_a u_b + \frac{1}{u_a u_b} - \frac{y}{x} - \frac{x}{y})(\frac{u_a}{u_b} + \frac{u_b}{u_a} - \frac{y}{x} - \frac{x}{y})}
\nonumber
\eea
\bea
&& z_{G+}^{2n}[u_a] = 
 \prod^{n}_{a=1}  \left[\frac{(1-x^2)}
{(1-x y)(1-\frac{x}{y})(u^2_a+\frac{1}{u^2_a}-x y-\frac{1}{x y})(u^2_a+\frac{1}{u^2_a}-\frac{y}{x}-\frac{x}{y})}
\right. \nonumber\\
&& \qquad\qquad\qquad \qquad \left. \mbox{} \times
\prod^N_{i=1}\frac{1}{(u_a+\frac{1}{u_a}-x s_i-\frac{1}{x s_i})(u_a+\frac{1}{u_a}-\frac{s_i}{x}-\frac{x}{s_i})}\right] \\
&& 
\prod^{n}_{a<b} \frac{(u_a u_b + \frac{1}{u_a u_b} - x^2 - \frac{1}{x^2})
(\frac{u_a}{u_b} + \frac{u_b}{u_a} - x^2 - \frac{1}{x^2})}{(u_a u_b + \frac{1}{u_a u_b} - x y-\frac{1}{x y})
(\frac{u_a}{u_b} + \frac{u_b}{u_a} - x y-\frac{1}{x y})(u_a u_b + \frac{1}{u_a u_b} 
- \frac{y}{x} - \frac{x}{y})(\frac{u_a}{u_b} + \frac{u_b}{u_a} - \frac{y}{x} - \frac{x}{y})} \nonumber
\eea
\bea
&& z_{G-}^{2n+1}[u_a] = \frac{x(1-x^2)^n}
{(1-x y)^{n+1}(1-\frac{x}{y})^{n+1} \prod^{N}_{i=1} (x+\frac{1}{x}+s_i+\frac{1}{s_i})} \nonumber \\
&& \prod^{n}_{a=1}  \left[\frac{ (u_i+\frac{1}{u_i}+x^2+\frac{1}{x^2})}{ (u_i+\frac{1}{u_i}+x y+\frac{1}{x y})(u_i+\frac{1}{u_i}+\frac{y}{x}+\frac{x}{y})(u^2_i+\frac{1}{u^2_i}-x y-\frac{1}{x y})(u^2_i+\frac{1}{u^2_i}-\frac{y}{x}-\frac{x}{y}) } \right. 
 \nonumber \\
 & & \qquad \qquad \qquad \qquad \left. \mbox{} \times
     \prod^N_{i=1}\frac{1}{(u_a+\frac{1}{u_a}-x s_i-\frac{1}{x s_i})(u_a+\frac{1}{u_a}-\frac{s_i}{x}-\frac{x}{s_i})}  \right] \\
 && \prod^{n}_{a<b} \frac{(u_a u_b + \frac{1}{u_a u_b} - x^2 - \frac{1}{x^2})(\frac{u_a}{u_b} 
 + \frac{u_b}{u_a} - x^2 - \frac{1}{x^2})}{(u_a u_b + \frac{1}{u_a u_b} - x y-\frac{1}{x y})(\frac{u_a}{u_b} 
 + \frac{u_b}{u_a} - x y-\frac{1}{x y})(u_a u_b + \frac{1}{u_a u_b} - \frac{y}{x} - \frac{x}{y})(\frac{u_a}{u_b} 
 + \frac{u_b}{u_a} - \frac{y}{x} - \frac{x}{y})}
 \nonumber
\eea
\bea  
&& z_{G-}^{2n}[u_a] = \frac{(-1)^{n-1} x^2(1+x^2)(1-x^2)^{n-1}}{(1-x y)^{n}(1-\frac{x}{y})^{n} (1-x^2 y^2)(1-\frac{x^2}{y^2}) \prod^{N}_{i=1} (x^2+\frac{1}{x^2}-s^2_i-\frac{1}{s^2_i})} \nonumber \\
&& \prod^{n-1}_{a=1}  \left[ \frac{(u^2_a+\frac{1}{u^2_a}-x^4-\frac{1}{x^4})}
{(u^2_a+\frac{1}{u^2_a}-x y-\frac{1}{x y})(u^2_a+\frac{1}{u^2_a}-\frac{y}{x}-\frac{x}{y})
(u^2_a+\frac{1}{u^2_a}-x^2 y^2-\frac{1}{x^2 y^2})(u^2_a+\frac{1}{u^2_a}-\frac{y^2}{x^2}-\frac{x^2}{y^2})} 
\right. \nonumber \\  
& &\qquad\qquad\qquad\qquad  \left. \mbox{} \times \prod^N_{i=1}
\frac{1}{(u_a+\frac{1}{u_a}-x s_i-\frac{1}{x s_i})(u_i+\frac{1}{u_a}-\frac{s_i}{x}-\frac{x}{s_i})}  \right]  \\
& & \prod^{n-1}_{a<b} 
\frac{(u_a u_b + \frac{1}{u_a u_b} - x^2 - \frac{1}{x^2})
(\frac{u_a}{u_b} + \frac{u_b}{u_a} - x^2 - \frac{1}{x^2})}
{(u_a u_b + \frac{1}{u_a u_b} - x y-\frac{1}{x y})
(\frac{u_a}{u_b} + \frac{u_b}{u_a} - x y-\frac{1}{x y})
(u_a u_b + \frac{1}{u_a u_b} - \frac{y}{x} - \frac{x}{y})
(\frac{u_a}{u_b} + \frac{u_b}{u_a} - \frac{y}{x} - \frac{x}{y})} \nonumber 
\eea

The contribution of matter multiplets in the fundamental representation of $Sp(N)$ is given by: 
\be
z_F^k[u_a] = 
\prod^{N_f}_{r=1} \prod_{a=1}^n (f_r+\frac{1}{f_r}-u_a-\frac{1}{u_a}) \times
\left\{
\begin{array}{ll}
\prod^{N_f}_{r=1} (\sqrt{f_r} \mp \frac{1}{\sqrt{f_r}}) & k=2n+1, O_{\pm} \\
1 & k=2n, O_+ \\
\prod^{N_f}_{r=1} ({f_r}-\frac{1}{{f_r}}) & k=2n+2, O_- 
\end{array} \right.
\ee
Note that in the presence of matter multiplets in the fundamental representation
the effect of the $\theta$ parameter can be absorbed into the sign of the mass.
Replacing $f_r \rightarrow 1/f_r$ for an odd number of flavors has the effect
of exchanging the two cases in (\ref{theta_inst}).

The contribution of a matter multiplet in the rank 2 antisymmetric representation
of $Sp(N)$ is given for the $O(k)_+$ part by:
\bea
&& z^{2n+\chi}_{A+}[u_a]  =   \left[\frac{\prod^N_{i=1}( z + \frac{1}{z} - s_i - \frac{1}{s_i})}{z + \frac{1}{ z} - x - \frac{1}{x}} \prod^n_{a=1}  \frac{(u_a+\frac{1}{u_a} - y z -\frac{1}{y z})
(u_a+\frac{1}{u_a} - \frac{y}{z} -\frac{z}{y})}
{(u_a+\frac{1}{u_a} - x z-\frac{1}{x z})(u_a+\frac{1}{u_a} - \frac{x}{z} -\frac{z}{x})} \right]^{\chi} \nonumber \\
& &  \prod^n_{a=1} \frac{(z + \frac{1}{ z} - y - \frac{1}{y})
\prod^N_{i=1} (u_a+\frac{1}{u_a} - z s_i -\frac{1}{z s_i})(u_a+\frac{1}{u_a} - \frac{z}{s_i} -\frac{s_i}{z})}
{(z + \frac{1}{ z} - x - \frac{1}{x})(u_a^2 + \frac{1}{u_a^2} - x z - \frac{1}{x z})
(u_a^2 + \frac{1}{u_a^2} - \frac{x}{z} - \frac{z}{x})} \label{ASSPplus} \\
& &   \prod^n_{a<b} \frac{(u_a u_b + \frac{1}{u_a u_b} - z y - \frac{1}{z y})
(u_a u_b + \frac{1}{u_a u_b} - \frac{z}{y} - \frac{y}{z})
(\frac{u_a}{u_b} + \frac{u_b}{u_a} - z y - \frac{1}{z y})
(\frac{u_a}{u_b} + \frac{u_b}{u_a} - \frac{z}{y} - \frac{y}{z})}
{(u_a u_b + \frac{1}{u_a u_b} - z x - \frac{1}{z x})
(u_a u_b + \frac{1}{u_a u_b} - \frac{z}{x} - \frac{x}{z})
(\frac{u_a}{u_b} + \frac{u_b}{u_a} - z x - \frac{1}{z x})(\frac{u_a}{u_b} + \frac{u_b}{u_a} - \frac{z}{x} - \frac{x}{z})}
\nonumber
\eea
where $\chi = 0,1$ correspond to $k=2n, 2n+1$, respectively.
For the $O(k)_-$ part, the contributions of the antisymmetric multiplet are given by:
\bea
&& z^{2n+1}_{A-}[u_a]  = \frac{\prod^N_{i=1}( z + \frac{1}{z} + s_i + \frac{1}{s_i})}{z + \frac{1}{ z} - x - \frac{1}{x}} 
\prod^n_{a=1}  \frac{(u_a+\frac{1}{u_a} + y z +\frac{1}{y z})(u_a+\frac{1}{u_a} + \frac{y}{z} +\frac{z}{y})}
{(u_a+\frac{1}{u_a} + x z+\frac{1}{x z})(u_a+\frac{1}{u_a} + \frac{x}{z} +\frac{z}{x})} \nonumber \\
& & \prod^n_{a=1} \frac{(z + \frac{1}{ z} - y - \frac{1}{y})
\prod^N_{l=1} (u_a+\frac{1}{u_a} - z s_i -\frac{1}{z s_i})
(u_a+\frac{1}{u_a} - \frac{z}{s_i} -\frac{s_i}{z})}
{(z + \frac{1}{ z} - x - \frac{1}{x})(u_a^2 + \frac{1}{u_a^2} - x z - \frac{1}{x z})
(u_a^2 + \frac{1}{u_a^2} - \frac{x}{z} - \frac{z}{x})} \label{ASSPminuso} \\
& &  \prod^n_{a<b} \frac{(u_a u_b + \frac{1}{u_a u_b} - z y - \frac{1}{z y})
(u_a u_b + \frac{1}{u_a u_b} - \frac{z}{y} - \frac{y}{z})
(\frac{u_a}{u_b} + \frac{u_b}{u_a} - z y - \frac{1}{z y})
(\frac{u_a}{u_b} + \frac{u_b}{u_a} - \frac{z}{y} - \frac{y}{z})}
{(u_a u_b + \frac{1}{u_a u_b} - z x - \frac{1}{z x})(u_a u_b + \frac{1}{u_a u_b} - \frac{z}{x} - \frac{x}{z})
(\frac{u_a}{u_b} + \frac{u_b}{u_a} - z x - \frac{1}{z x})(\frac{u_a}{u_b} + \frac{u_b}{u_a} - \frac{z}{x} - \frac{x}{z})} 
\nonumber
\eea
for $k=2n+1$, and
\bea
&& z^{2n}_{A-}[u_a]  = 
\frac{(z + \frac{1}{z} + y + \frac{1}{y}) (z + \frac{1}{z} - y - \frac{1}{y})^{n-1}
\prod^N_{i=1}( z^2 + \frac{1}{z^2} - s^2_i - \frac{1}{s^2_i})}
{(z + \frac{1}{z} + x + \frac{1}{x})(z + \frac{1}{z} - x - \frac{1}{x})^{n+1}} 
\label{ASSPminuse} \\
& & \prod^{n-1}_{a=1}  \frac{(u_a^2 + \frac{1}{u_a^2} - z^2 y^2 - \frac{1}{z^2 y^2})
(u_a^2 + \frac{1}{u_a^2} - \frac{z^2}{y^2} - \frac{y^2}{z^2}) 
\prod^N_{i=1} (u_a + \frac{1}{u_a} - z s_i - \frac{1}{z s_i})(u_a + \frac{1}{u_a} - \frac{z}{s_i} - \frac{s_i}{z})}
{(u_a^2 + \frac{1}{u_a^2} - z^2 x^2 - \frac{1}{z^2 x^2})(u_a^2 + \frac{1}{u_a^2} - \frac{z^2}{x^2} - \frac{x^2}{z^2})
(u_a^2 + \frac{1}{u_a^2} - z x - \frac{1}{z x})(u_a^2 + \frac{1}{u_a^2} - \frac{z}{x} - \frac{x}{z})}
\nonumber \\
&& \prod^{n-1}_{a<b} \frac{(u_a u_b + \frac{1}{u_a u_b} - z y - \frac{1}{z y})
(u_a u_b + \frac{1}{u_a u_b} - \frac{z}{y} - \frac{y}{z})(\frac{u_a}{u_b} + \frac{u_b}{u_a} - z y - \frac{1}{z y})
(\frac{u_a}{u_b} + \frac{u_b}{u_a} - \frac{z}{y} - \frac{y}{z})}{(u_a u_b + \frac{1}{u_a u_b} - z x - \frac{1}{z x})
(u_a u_b + \frac{1}{u_a u_b} - \frac{z}{x} - \frac{x}{z})(\frac{u_a}{u_b} + \frac{u_b}{u_a} - z x - \frac{1}{z x})
(\frac{u_a}{u_b} + \frac{u_b}{u_a} - \frac{z}{x} - \frac{x}{z})} \nonumber
\eea
for $k=2n$.

For an $Sp(N_1) \times Sp(N_2)$ quiver theory the di-group instanton partition functions will
get a contribution from the bifundamental matter multiplet.
In this case there are many expressions since there are four disconnected components of
$O(k_1)\times O(k_2)$,
and even and odd $k$'s are different.
These can be evaluated using the methods and results in \cite{Shad, KKL}.

The results for the $(+,+)$ component can also be lifted from the result from the 4d expression in \cite{HKS}.
These are given by
\bea
&& z^{k_1,k_2}_{BF++}[u,u'] = 
\left[ \prod^{N_1}_{i=1}(z+\frac{1}{z} - s_i -\frac{1}{s_i}) 
\prod^{n_1}_{a=1} \frac{(z+\frac{1}{z} - y u_a -\frac{1}{y u_a})(z+\frac{1}{z} - \frac{y}{u_a} -\frac{u_a}{y})}
{(z+\frac{1}{z} - x u_a-\frac{1}{x u_a})(z+\frac{1}{z} - \frac{x}{u_a} -\frac{u_a}{x})} \right]^{\chi_2} 
\nonumber \\
&& \left[\prod^{N_2}_{j=1} (z+\frac{1}{z} - s'_j -\frac{1}{s'_j}) 
\prod^{n_2}_{b=1} \frac{(z+\frac{1}{z} - y u'_b -\frac{1}{y u'_b})
(z+\frac{1}{z} - \frac{y}{u'_b} -\frac{u'_b}{y})}
{(z+\frac{1}{z} - x u'_b-\frac{1}{x u'_b})(z+\frac{1}{z} - \frac{x}{u'_b} -\frac{u'_b}{x})} \right]^{\chi_1} 
\left[\frac{z+\frac{1}{z} - y-\frac{1}{y}}{z+\frac{1}{z} - x-\frac{1}{x}}\right]^{\chi_1 \chi_2} 
\nonumber\\
 && \prod_{a,b=1}^{n_1,n_2} 
 \frac{(z+\frac{1}{z} - y u_a u'_b -\frac{1}{y u_a u'_b})(z+\frac{1}{z} - \frac{y u'_b}{u_a} -\frac{u_a}{y u'_b})
 (z+\frac{1}{z} - \frac{y u_a}{u'_b}  -\frac{u'_b}{y u_a})(z+\frac{1}{z} - \frac{y}{u_a u'_b} -\frac{u_a u'_b}{y})}
 {(z+\frac{1}{z} - x u_a u'_b -\frac{1}{x u_a u'_b})(z+\frac{1}{z} - \frac{x u'_b}{u_a} -\frac{u_a}{x u'_b})
 (z+\frac{1}{z} - \frac{x u_a}{u'_b}  -\frac{u'_b}{x u_a})(z+\frac{1}{z} - \frac{x}{u_a u'_b} -\frac{u_a u'_b}{x})} 
 \nonumber \\
&&  \prod_{a,j=1}^{n_1,N_2} (z+\frac{1}{z} - u_a s'_j-\frac{1}{u_a s'_j})(z+\frac{1}{z} - \frac{u_a}{s'_j}-\frac{s'_j}{u_a})
 \prod_{b,i=1}^{n_2,N_1}
 (z+\frac{1}{z} - u'_b s_i-\frac{1}{u'_b s_i})(z+\frac{1}{z} - \frac{u'_b}{s_i}-\frac{s_i}{u'_b}) ,
 \nonumber \\
 \eea
where $k_1 = 2n_1 + \chi_1, k_2 = 2n_2 + \chi_2$, where $\chi_{1, 2}$ being either 0 or 1.
For the $(+,-)$ component we separate the odd and even $k_2$ cases:
\bea
&& z^{k_1, 2n_2+1}_{BF+-}[u,u'] =  
\left[\prod^{N_2}_{j=1} (z+\frac{1}{z} - s'_j -\frac{1}{s'_j}) \prod^{n_2}_{b=1} 
\frac{(z+\frac{1}{z} - y u'_b -\frac{1}{y u'_b})(z+\frac{1}{z} - \frac{y}{u'_b} -\frac{u'_b}{y})}
{(z+\frac{1}{z} - x u'_b-\frac{1}{x u'_b})(z+\frac{1}{z} - \frac{x}{u'_b} -\frac{u'_b}{x})} \right]^{\chi_1} \nonumber\\
&& \left[\frac{z+\frac{1}{z} + y+\frac{1}{y}}{z+\frac{1}{z} + x+\frac{1}{x}}\right]^{\chi_1} 
 \prod^{N_1}_{i=1}(z+\frac{1}{z} + s_i +\frac{1}{s_i}) \prod^{n_1}_{a=1} 
\frac{(z+\frac{1}{z} + y u_a +\frac{1}{y u_a})(z+\frac{1}{z} + \frac{y}{u_a} +\frac{u_a}{y})}
{(z+\frac{1}{z} + x u_a+\frac{1}{x u_a})(z+\frac{1}{z} + \frac{x}{u_a} +\frac{u_a}{x})} \\
&&  \prod_{a,b}^{n_1,n_2}
\frac{(z+\frac{1}{z} - y u_a u'_b -\frac{1}{y u_a u'_b})(z+\frac{1}{z} - \frac{y u'_b}{u_a} -\frac{u_a}{y u'_b})
(z+\frac{1}{z} - \frac{y u_a}{u'_b}  -\frac{u'_b}{y u_a})(z+\frac{1}{z} - \frac{y}{u_a u'_b} -\frac{u_a u'_b}{y})}
{(z+\frac{1}{z} - x u_a u'_b -\frac{1}{x u_a u'_b})(z+\frac{1}{z} - \frac{x u'_b}{u_a} -\frac{u_a}{x u'_b})
(z+\frac{1}{z} - \frac{x u_a}{u'_b}  -\frac{u'_b}{x u_a})(z+\frac{1}{z} - \frac{x}{u_a u'_b} -\frac{u_a u'_b}{x})}
\nonumber \\
&& \prod_{a,j}^{n_1,N_2} (z+\frac{1}{z} - u_a s'_j-\frac{1}{u_a s'_j})(z+\frac{1}{z} - \frac{u_a}{s'_j}-\frac{s'_j}{u_a})
\prod_{b,i}^{n_2,N_1} (z+\frac{1}{z} - u'_b s_i-\frac{1}{u'_b s_i})(z+\frac{1}{z} - \frac{u'_b}{s_i}-\frac{s_i}{u'_b})
\nonumber
\eea
and
\bea
&& z^{k_1, 2n_2}_{BF+-}[u,u'] =  
\left[\prod^{N_2}_{j=1} (z+\frac{1}{z} - s'_j -\frac{1}{s'_j}) \prod^{n_2-1}_{b=1} 
\frac{(z+\frac{1}{z} - y u'_b -\frac{1}{y u'_b})(z+\frac{1}{z} - \frac{y}{u'_b} -\frac{u'_b}{y})}
{(z+\frac{1}{z} - x u'_b-\frac{1}{x u'_b})(z+\frac{1}{z} - \frac{x}{u'_b} -\frac{u'_b}{x})} \right]^{\chi_1} \nonumber\\
&&   \left[\frac{z^2+\frac{1}{z^2} - y^2-\frac{1}{y^2}}{z^2+\frac{1}{z^2} - x^2-\frac{1}{x^2}} \right]^{\chi_1} 
\prod^{N_1}_{i=1} (z^2+\frac{1}{z^2} - s^2_i-\frac{1}{s^2_i})
\prod^{n_1}_{a=1} \frac{(z^2+\frac{1}{z^2} - y^2 u^2_a -\frac{1}{y^2 u^2_a})(z^2+\frac{1}{z^2} - \frac{y^2}{u^2_a} -\frac{u^2_a}{y^2})}
{(z^2+\frac{1}{z^2} - x^2 u^2_a-\frac{1}{x^2 u^2_a})(z^2+\frac{1}{z^2} - \frac{x^2}{u^2_a} -\frac{u^2_a}{x^2})} \nonumber \\
&& \prod_{a,b}^{n_1,n_2-1}
\frac{(z+\frac{1}{z} - y u_a u'_b -\frac{1}{y u_a u'_b})(z+\frac{1}{z} - \frac{y u'_b}{u_a} -\frac{u_a}{y u'_b})
(z+\frac{1}{z} - \frac{y u_a}{u'_b}  -\frac{u'_b}{y u_a})(z+\frac{1}{z} - \frac{y}{u_a u'_b} -\frac{u_a u'_b}{y})}
{(z+\frac{1}{z} - x u_a u'_b -\frac{1}{x u_a u'_b})(z+\frac{1}{z} - \frac{x u'_b}{u_a} -\frac{u_a}{x u'_b})
(z+\frac{1}{z} - \frac{x u_a}{u'_b}  -\frac{u'_b}{x u_a})(z+\frac{1}{z} - \frac{x}{u_a u'_b} -\frac{u_a u'_b}{x})}
\nonumber \\
&& \prod_{a,j}^{n_1,N_2} (z+\frac{1}{z} - u_a s'_j-\frac{1}{u_a s'_j})(z+\frac{1}{z} - \frac{u_a}{s'_j}-\frac{s'_j}{u_a})
\prod_{b,i}^{n_2-1,N_1} (z+\frac{1}{z} - u'_b s_i-\frac{1}{u'_b s_i})(z+\frac{1}{z} - \frac{u'_b}{s_i}-\frac{s_i}{u'_b}) ,
\nonumber
\eea
and similarly for the $(-,+)$ component.
Finally, for the $(-,-)$ component the contributions are given by
\bea
&& z^{2n_1+1, 2n_2+1}_{BF--}[u,u'] = 
\prod^{N_2}_{j=1} (z+\frac{1}{z} + s'_j +\frac{1}{s'_j}) \prod^{n_2}_{b=1} \frac{(z+\frac{1}{z} + y u'_b +\frac{1}{y u'_b})
(z+\frac{1}{z} + \frac{y}{u'_b} +\frac{u'_b}{y})}{(z+\frac{1}{z} + x u'_b+\frac{1}{x u'_b})(z+\frac{1}{z} + \frac{x}{u'_b} +\frac{u'_b}{x})}
\nonumber \\
&&  \left[\frac{z+\frac{1}{z} - y-\frac{1}{y}}{z+\frac{1}{z} - x-\frac{1}{x}}\right] \,
\prod^{N_1}_{i=1}(z+\frac{1}{z} + s_i +\frac{1}{s_i}) \prod^{n_1}_{a=1} \frac{(z+\frac{1}{z} + y u_a +\frac{1}{y u_a})
(z+\frac{1}{z} + \frac{y}{u_a} +\frac{u_a}{y})}{(z+\frac{1}{z} + x u_a+\frac{1}{x u_a})(z+\frac{1}{z} + \frac{x}{u_a} +\frac{u_a}{x})}
 \\
&& \prod_{a,b}^{n_1,n_2}
\frac{(z+\frac{1}{z} - y u_a u'_b -\frac{1}{y u_a u'_b})(z+\frac{1}{z} - \frac{y u'_b}{u_a} -\frac{u_a}{y u'_b})
(z+\frac{1}{z} - \frac{y u_a}{u'_b}  -\frac{u'_b}{y u_a})(z+\frac{1}{z} - \frac{y}{u_a u'_b} -\frac{u_a u'_b}{y})}
{(z+\frac{1}{z} - x u_a u'_b -\frac{1}{x u_a u'_b})(z+\frac{1}{z} - \frac{x u'_b}{u_a} -\frac{u_a}{x u'_b})
(z+\frac{1}{z} - \frac{x u_a}{u'_b}  -\frac{u'_b}{x u_a})(z+\frac{1}{z} - \frac{x}{u_a u'_b} -\frac{u_a u'_b}{x})}
\nonumber \\
&& \prod_{a,j}^{n_1,N_2} (z+\frac{1}{z} - u_a s'_j-\frac{1}{u_a s'_j})(z+\frac{1}{z} - \frac{u_a}{s'_j}-\frac{s'_j}{u_a})
\prod_{b,i}^{n_2,N_1} (z+\frac{1}{z} - u'_b s_i-\frac{1}{u'_b s_i})(z+\frac{1}{z} - \frac{u'_b}{s_i}-\frac{s_i}{u'_b}) ,
\nonumber
\eea
\bea
&& z^{2n_1+1, 2n_2}_{BF--}[u,u'] = 
\prod^{N_2}_{j=1} (z+\frac{1}{z} + s'_j +\frac{1}{s'_j}) \prod^{n_2-1}_{b=1} \frac{(z+\frac{1}{z} + y u'_b +\frac{1}{y u'_b})
(z+\frac{1}{z} + \frac{y}{u'_b} +\frac{u'_b}{y})}{(z+\frac{1}{z} + x u'_b+\frac{1}{x u'_b})(z+\frac{1}{z} + \frac{x}{u'_b} +\frac{u'_b}{x})} \\
&&  \left[ \frac{z^2+\frac{1}{z^2} - y^2-\frac{1}{y^2}}{z^2+\frac{1}{z^2} - x^2-\frac{1}{x^2}}   \right] \,
\prod^{N_1}_{i=1} (z^2+\frac{1}{z^2} - s^2_i -\frac{1}{s^2_i})
 \prod^{n_1}_{a=1} \frac{(z^2+\frac{1}{z^2} - y^2 u^2_a -\frac{1}{y^2 u^2_a})(z^2+\frac{1}{z^2} - \frac{y^2}{u^2_a} -\frac{u^2_a}{y^2})}
 {(z+\frac{1}{z} - x^2 u^2_a-\frac{1}{x^2 u^2_a})(z+\frac{1}{z} - \frac{x^2}{u^2_a} -\frac{u^2_a}{x^2})} 
  \nonumber \\
&& \prod_{a,b}^{n_1,n_2-1}
\frac{(z+\frac{1}{z} - y u_a u'_b -\frac{1}{y u_a u'_b})(z+\frac{1}{z} - \frac{y u'_b}{u_a} -\frac{u_a}{y u'_b})
(z+\frac{1}{z} - \frac{y u_a}{u'_b}  -\frac{u'_b}{y u_a})(z+\frac{1}{z} - \frac{y}{u_a u'_b} -\frac{u_a u'_b}{y})}
{(z+\frac{1}{z} - x u_a u'_b -\frac{1}{x u_a u'_b})(z+\frac{1}{z} - \frac{x u'_b}{u_a} -\frac{u_a}{x u'_b})
(z+\frac{1}{z} - \frac{x u_a}{u'_b}  -\frac{u'_b}{x u_a})(z+\frac{1}{z} - \frac{x}{u_a u'_b} -\frac{u_a u'_b}{x})}
\nonumber \\
&& \prod_{a,j}^{n_1,N_2} (z+\frac{1}{z} - u_a s'_j-\frac{1}{u_a s'_j})(z+\frac{1}{z} - \frac{u_a}{s'_j}-\frac{s'_j}{u_a})
\prod_{b,i}^{n_2-1,N_1} (z+\frac{1}{z} - u'_b s_i-\frac{1}{u'_b s_i})(z+\frac{1}{z} - \frac{u'_b}{s_i}-\frac{s_i}{u'_b}) ,
\nonumber
\eea
and similarly for $(k_1,k_2)=(2n_1,2n_2+1)$, and
\bea
&& z^{2n_1, 2n_2}_{BF--}[u,u'] = 
\prod^{N_2}_{j=1} (z^2+\frac{1}{z^2} - s^{\prime 2}_j -\frac{1}{s^{\prime ^2}_j})
\prod^{n_2-1}_{b=1} 
  \frac{(z^2+\frac{1}{z^2} - y^2 u^{\prime 2}_b -\frac{1}{y^2 u^{\prime 2}_b})
 (z^2+\frac{1}{z^2} - \frac{y^2}{u^{\prime 2}_b} -\frac{u^{\prime 2}_b}{y^2})}
 {(z^2+\frac{1}{z^2} - x^2 u^{\prime 2}_b-\frac{1}{x^2 u^{\prime 2}_b})(z^2+\frac{1}{z^2} - \frac{x^2}{u^{\prime 2}_b} -\frac{u^{\prime 2}_b}{x^2})}
\nonumber \\
&&  \left[ \frac{z^2+\frac{1}{z^2} - y^2-\frac{1}{y^2}}{z^2+\frac{1}{z^2} - x^2-\frac{1}{x^2}}   \right] \,
\prod^{N_1}_{i=1} (z^2+\frac{1}{z^2} - s^2_i -\frac{1}{s^2_i})
 \prod^{n_1-1}_{a=1} \frac{(z^2+\frac{1}{z^2} - y^2 u^2_a -\frac{1}{y^2 u^2_a})(z^2+\frac{1}{z^2} - \frac{y^2}{u^2_a} -\frac{u^2_a}{y^2})}
 {(z+\frac{1}{z} - x^2 u^2_a-\frac{1}{x^2 u^2_a})(z+\frac{1}{z} - \frac{x^2}{u^2_a} -\frac{u^2_a}{x^2})} 
  \nonumber \\
&& \prod_{a,b}^{n_1-1,n_2-1}
\frac{(z+\frac{1}{z} - y u_a u'_b -\frac{1}{y u_a u'_b})(z+\frac{1}{z} - \frac{y u'_b}{u_a} -\frac{u_a}{y u'_b})
(z+\frac{1}{z} - \frac{y u_a}{u'_b}  -\frac{u'_b}{y u_a})(z+\frac{1}{z} - \frac{y}{u_a u'_b} -\frac{u_a u'_b}{y})}
{(z+\frac{1}{z} - x u_a u'_b -\frac{1}{x u_a u'_b})(z+\frac{1}{z} - \frac{x u'_b}{u_a} -\frac{u_a}{x u'_b})
(z+\frac{1}{z} - \frac{x u_a}{u'_b}  -\frac{u'_b}{x u_a})(z+\frac{1}{z} - \frac{x}{u_a u'_b} -\frac{u_a u'_b}{x})}
\nonumber \\
&& \prod_{a,j}^{n_1-1,N_2} (z+\frac{1}{z} - u_a s'_j-\frac{1}{u_a s'_j})(z+\frac{1}{z} - \frac{u_a}{s'_j}-\frac{s'_j}{u_a})
\prod_{b,i}^{n_2-1,N_1} (z+\frac{1}{z} - u'_b s_i-\frac{1}{u'_b s_i})(z+\frac{1}{z} - \frac{u'_b}{s_i}-\frac{s_i}{u'_b}) .
\nonumber
\eea

As explained in section~\ref{AS_BF_issues}, there is a problem in evaluating the contribution of bifundamentals to
the partition functions of di-group instantons.
For the $SU(2)\times SU(2)$ theories we have the option of using the $U(N)$ formalism, which we do.
In this formalism the $SU(2)$ $\theta$ parameters correspond to the $U(2)$ CS levels.

For the contributions of single group instantons there is no problem,
since one can treat them as $Sp(N)$ instantons with flavors in the fundamental representation.
The second $Sp(N)$ gauge group is embedded in the flavor symmetry in a way determined by the $\theta$ parameter.

In the presence of a sufficiently large number of fundamentals (for example, if $N_f>2$ for $SU(2)$),
there is a problem related to the existence of parallel external NS5-branes.
This manifests itself in the instanton partition function by the appearance of extra poles at zero or infinity.
As shown in section 4.1, this also leads to a lack of invariance under $x \rightarrow 1/x$. 
A pole prescription must be chosen for these additional poles.
The difference between including them and excluding them corresponds to the correction associated with the decoupled state.
We have chosen to include them.
%

\section{Explicit superconformal indices} 

Here we present the explicit expressions for the superconformal indices of all the theories discussed in the paper,
to the highest order in $x$ that we computed.

\medskip

\subsection{{$SU(2)\times SU(2)$}}

There are three inequivalent cases corresponding to the values of the two $\theta$ parameters.
For $(\theta_1,\theta_2)=(0,\pi)$:
\bea
I^{SU(2)_0\times SU(2)_\pi}  & = & 1 + x^2 \left(1+\chi^{SU(3)}_{\bf 8}\right) + x^3 \chi_{\bf 2}[y]
\left(2 + \chi^{SU(3)}_{\bf 8}\right) \nonumber\\
&+& x^4 \Big(\chi_{\bf 3}[y]\left(2 + \chi^{SU(3)}_{\bf 8}\right) + \chi^{SU(3)}_{\bf 27} + \chi^{SU(3)}_{\bf 8}+1\Big) \nonumber \\  
& + & x^5 \Big(\chi_{\bf 4}[y]\left(2 + \chi^{SU(3)}_{\bf 8}\right) 
+ \chi_{\bf 2}[y] \big(\chi^{SU(3)}_{\bf 27} + \chi^{SU(3)}_{\bf 10}
+ \chi^{SU(3)}_{\overline{\bf 10}} + 4\chi^{SU(3)}_{\bf 8}+2\big)\Big) \nonumber \\  
& + & x^6 \Big(\chi_{\bf 5}[y](2 + \chi^{SU(3)}_{\bf 8}) + \chi_{\bf 3}[y] (2\chi^{SU(3)}_{\bf 27} + \chi^{SU(3)}_{\bf 10} 
+ \chi^{SU(3)}_{\overline{\bf 10}}
+ 7\chi^{SU(3)}_{\bf 8}+6) \nonumber \\
&& \qquad\qquad \mbox{} +\chi^{SU(3)}_{\bf 64}+\chi^{SU(3)}_{\bf 27}
+ \chi^{SU(3)}_{\bf 10} + \chi^{SU(3)}_{\overline{\bf 10}}+ 4\chi^{SU(3)}_{\bf 8}
+2 \nonumber \\
&& \qquad\qquad \mbox{} - q_1^{2/3} q_2^2 \, \chi^{SU(3)}_{\bf 3} - q_1^{-2/3} q_2^{-2} \, \chi^{SU(3)}_{\overline{\bf 3}}\Big) \nonumber \\
&+& x^7 \Big(\chi_{\bf 6}[y](2 + \chi^{SU(3)}_{\bf 8})  +   \chi_{\bf 4}[y] (2\chi^{SU(3)}_{\bf 27} + 2\chi^{SU(3)}_{\bf 10} 
+ 2\chi^{SU(3)}_{\overline{\bf 10}} + 10\chi^{SU(3)}_{\bf 8}+7) \nonumber \\
&& \qquad \mbox{} + \chi_{\bf 2}[y](\chi^{SU(3)}_{\bf 64} + \chi^{SU(3)}_{\bf 35} + \chi^{SU(3)}_{\overline{\bf 35}} 
+   5\chi^{SU(3)}_{\bf 27} + 2\chi^{SU(3)}_{\bf 10}
+ 2\chi^{SU(3)}_{\overline{\bf 10}} \nonumber \\
&& \qquad \mbox{} + 9\chi^{SU(3)}_{\bf 8} + 7  - q_1^{2/3} q_2^2 \, \chi^{SU(3)}_{\bf 3}
   -  q_1^{-2/3} q_2^{-2} \, \chi^{SU(3)}_{\overline{\bf 3}})\Big)  \nonumber \\
 & + & O(x^8),
\eea
where $q_2$ is the fugacity associated with the instanton number of the second $SU(2)$.
This shows enhancement of the global symmetry to $SU(3)\times U(1)_T$.
The fugacity of the remaining global $U(1)_T$ symmetry is given by $q_1^{2/3} q_2^2$.

\medskip 

\noindent For $(\theta_1,\theta_2)=(0,0)$:
\bea
I^{SU(2)_0\times SU(2)_0} & = & 1 + x^2 \chi^{SU(4)}_{\bf 15} + x^3 \chi_{\bf 2}[y](1 + \chi^{SU(4)}_{\bf 15}) \nonumber \\
&+& x^4 \Big(\chi_{\bf 3}[y](1 + \chi^{SU(4)}_{\bf 15}) + \chi^{SU(4)}_{\bf 20} + \chi^{SU(4)}_{\bf 84}\Big) \nonumber \\  
& + & x^5 \Big(\chi_{\bf 4}[y](1 + \chi^{SU(4)}_{\bf 15}) + \chi_{\bf 2}[y] (\chi^{SU(4)}_{\bf 20} + \chi^{SU(4)}_{\bf 84} 
+ 2\chi^{SU(4)}_{\bf 15} + \chi^{SU(4)}_{\bf 45} + \chi^{SU(4)}_{\overline{\bf 45}})\Big)  \nonumber \\
 & + &  x^6 \Big(\chi_{\bf 5}[y](1 + \chi^{SU(4)}_{\bf 15}) + \chi_{\bf 3}[y] (2\chi^{SU(4)}_{\bf 20} + 2\chi^{SU(4)}_{\bf 84} 
 + 4\chi^{SU(4)}_{\bf 15} \nonumber \\
 && \quad \mbox{} + \chi^{SU(4)}_{\bf 45} + \chi^{SU(4)}_{\overline{\bf 45}}+2) 
 + \chi^{SU(4)}_{\bf 300} + \chi^{SU(4)}_{\bf 175} + \chi^{SU(4)}_{\bf 45} + \chi^{SU(4)}_{\overline{\bf 45}} 
 + 3\chi^{SU(4)}_{\bf 15}\Big) \nonumber \\
 &+& x^7 \Big(\chi_{\bf 6}[y](1 + \chi^{SU(4)}_{\bf 15})  +  \chi_{\bf 4}[y] (2\chi^{SU(4)}_{\bf 20} 
 + 2\chi^{SU(4)}_{\bf 84} + 6\chi^{SU(4)}_{\bf 15} \nonumber \\
 && \quad \mbox{} +  2\chi^{SU(4)}_{\bf 45} + 2\chi^{SU(4)}_{\overline{\bf 45}} + 2)  +  \chi^{SU(4)}_{\bf 300} + \chi^{SU(4)}_{\bf 256}  
 + \chi^{SU(4)}_{\overline{\bf 256}} + 2\chi^{SU(4)}_{\bf 175} \nonumber \\
&& \quad \mbox{} + 3\chi^{SU(4)}_{\bf 84} + 2\chi^{SU(4)}_{\bf 45} + 2\chi^{SU(4)}_{\overline{\bf 45}} 
+ 2\chi^{SU(4)}_{\bf 20} + 5\chi^{SU(4)}_{\bf 15} + 3\Big) \nonumber \\
 &+& O(x^8) ,
\eea
which shows the enhancement to $SU(4)$.
Note that there are some different $SU(4)$ representations that have the same dimension.
Specifically the ${\bf 20}$ above is the $(0,2,0)$ representation, the ${\bf 84}$ is the $(2,0,2)$, and the ${\bf 300}$ is the $(3,0,3)$ when expressed in terms of the Cartan weights.

\noindent For $(\theta_1,\theta_2)=(\pi,\pi)$ we find:
\bea
I^{SU(2)_\pi\times SU(2)_\pi} & = & 1 + x^2 (2+\chi_{\bf 3}) + x^3 \left( \chi_{\bf 2}[y] (3+\chi_{\bf 3}) 
+ \chi_{\bf 2}(q+\frac{1}{q}) \right) \\ \nonumber 
& + & x^4 \left(\chi_{\bf 3}[y] (3+\chi_{\bf 3}) + \chi_{\bf 5} + \chi_{\bf 3} + 3 
+ \chi_{\bf 2}[y]\chi_{\bf 2}(q+\frac{1}{q}) \right) \\ \nonumber 
& + & x^5 \left(\chi_{\bf 4}[y] (3+\chi_{\bf 3}) + \chi_{\bf 2}[y] (\chi_{\bf 5} + 5\chi_{\bf 3} + 6) 
- b^3 \chi_{\bf 2} - \frac{1}{b^3}\chi_{\bf 2} \right. \nonumber \\
&  & \qquad \mbox{} + \left. (q+\frac{1}{q})(\chi_{\bf 3}[y]\chi_{\bf 2} + \chi_{\bf 4}+\chi_{\bf 2})  \right) \nonumber \\
&+& x^6 \left( \chi_{\bf 5}[y] (3+\chi_{\bf 3})  +  \chi_{\bf 3}[y] (2\chi_{\bf 5} + 9\chi_{\bf 3} + 12) 
-\chi_{\bf 2}[y](b^3\chi_{\bf 2} + \frac{1}{b^3}\chi_{\bf 2}) \right. \nonumber \\
&& \left. \quad \mbox{} + \chi_{\bf 7} +\chi_{\bf 5} + 6\chi_{\bf 3} + 8 
+ (q+\frac{1}{q})\Big[\chi_{\bf 4}[y]\chi_{\bf 2} + 2\chi_{\bf 2}[y](\chi_{\bf 4}+\chi_{\bf 2})  - b^3 - \frac{1}{b^3}\Big] \right. \nonumber \\
&& \left. \quad \mbox{} + (q^2+\frac{1}{q^2})\chi_{\bf 3} \right) \\ \nonumber 
& + & x^7 \left(\chi_{\bf 6}[y] (3+\chi_{\bf 3}) + \chi_{\bf 4}[y] (2\chi_{\bf 5} + 12\chi_{\bf 3} + 16)
- (\chi_{\bf 3}[y]+1)(b^3 \chi_{\bf 2} + \frac{1}{b^3} \chi_{\bf 2}) \right. \\ \nonumber 
& & \left. \quad \mbox{} + \chi_{\bf 2}[y](\chi_{\bf 7} +5\chi_{\bf 5} + 15\chi_{\bf 3} + 19) 
+ (q+\frac{1}{q})\Big[(\chi_{\bf 5}[y]\chi_{\bf 2} + \chi_{\bf 3}[y](3\chi_{\bf 4}+8\chi_{\bf 2})\right.  \\ \nonumber 
& & \left. \quad \mbox{} - \chi_{\bf 2}[y] (b^3 + \frac{1}{b^3}) + \chi_{\bf 6} + 2\chi_{\bf 4}+5\chi_{\bf 2}\Big] 
+ (q^2+\frac{1}{q^2})\chi_{\bf 2}[y](1+\chi_{\bf 3}) \right) \nonumber \\
& + & O(x^8) ,
\eea
where $q = q_1 q_2$ and $b^3 = q_1/q_2$.
This is also the index of $SU(3)$ with two fundamental hypermultiplets. 
>From this point of view $b$ is the baryonic fugacity and $q$ is the instantonic fugacity.

\medskip

\subsection{$Sp(2)\times SU(2)$} 

Here we considered only the case with $\theta_2=0$,
and included only the contributions of the $(0,1)$, $(0,2)$ and $(0,3)$ instantons.
In particular, the result is independent of $\theta_1$.
\bea
I^{Sp(2)_0\times SU(2)_0} & = & 1 + x^2 (1+\chi_{\bf 10}) + x^3 \chi_{\bf 2}[y] (2+\chi_{\bf 10})  \\
&+& x^4 \Big(\chi_{\bf 3}[y] (2+\chi_{\bf 10}) + \chi_{(4,0)}[35] + \chi[14] +  \chi[10] + \chi[5] + 2 \Big)  \nonumber \\
&+& x^5 \Big( \chi_{\bf 4}[y] (2+\chi_{\bf 10}) + \chi_{\bf 2}[y] (\chi_{(4,0)} + \chi_{(2,1)} + \chi_{\bf 14} + 
3\chi_{\bf 10} + 2\chi_{\bf 5} + 1) \Big) \nonumber \\
&+& x^6 \Big( \chi_{\bf 5}[y] (2+\chi_{\bf 10}) + \chi_{\bf 3}[y] (2\chi_{(4,0)} + \chi_{(2,1)}
+   2\chi_{\bf 14} + 4\chi_{\bf 10} + 3\chi_{\bf 5} + 3) \nonumber \\
&& \qquad \mbox{} + \chi_{\bf 84} + \chi_{\bf 81} + \chi_{(4,0)} + 2\chi_{(2,1)} + \chi_{\bf 14} +   5\chi_{\bf 10} + \chi_{\bf 5} + 1 \Big) \nonumber \\
&+& x^7 \Big( \chi_{\bf 6}[y] (2+\chi_{\bf 10}) + \chi_{\bf 4}[y] (2\chi_{(4,0)} + 2\chi_{(2,1)} + 2\chi_{\bf 14}
 +  7\chi_{\bf 10} + 3\chi_{\bf 5} + 4) \nonumber \\
 && \mbox{} + \chi_{\bf 2}[y] (\chi_{\bf 105} + \chi_{\bf 84} + 2\chi_{\bf 81} + 5\chi_{(4,0)} + 6\chi_{(2,1)} 
 + 5\chi_{\bf 14} + 12\chi_{\bf 10} + 5\chi_{\bf 5} + 8) \Big) \nonumber \\
&+& O(x^8) . \nonumber
\eea
The index is written in terms of the characters of the enhanced global $Sp(2) = SO(5)$ symmetry.
This is also the perturbative part for the index of $SU(4)$ with two antisymmetric hypermultiplets.
Note that the two $Sp(2)$ representations with weights $(4,0)$ and $(2,1)$ are both 35-dimensional,
which is why we chose to label them using their weights.

\medskip

\subsection{$Sp(2)\times Sp(2)$}

Here we considered only the case with $(\theta_1,\theta_2)=(\pi,\pi)$,
and included only the contributions of the $(1,0)$ and $(0,1)$ instantons.
\bea
I^{Sp(2)_\pi\times Sp(2)_\pi} & = & 1 + x^2 (2+\chi_{\bf 3}) + x^3 \chi_{\bf 2}[y] (3+\chi_{\bf 3}) \nonumber \\
&+& x^4 \Big(\chi_{\bf 3}[y] (3+\chi_{\bf 3}) + 2\chi_{\bf 5} + 3\chi_{\bf 3} + 5 \Big)  \nonumber \\ 
& + & x^5 \Big(\chi_{\bf 4}[y] (3+\chi_{\bf 3}) + \chi_{\bf 2}[y] (2\chi_{\bf 5} + 8\chi_{\bf 3} + 8 )\Big) \nonumber \\
&+ & x^6 \Big( \chi_{\bf 5}[y] (3+\chi_{\bf 3})  + \chi_{\bf 3}[y] (3\chi_{\bf 5} + 11\chi_{\bf 3} + 15) + 2\chi_{\bf 7} +5\chi_{\bf 5} 
+ 12\chi_{\bf 3} + 10 \Big) 
\nonumber \\
& + & x^7 \Big(\chi_{\bf 6}[y] (3+\chi_{\bf 3}) + 3\chi_{\bf 4}[y] (\chi_{\bf 5} + 5\chi_{\bf 3} + 6) 
+ \chi_{\bf 2}[y](3\chi_{\bf 7} +16\chi_{\bf 5} + 32\chi_{\bf 3} + 27) \nonumber \\
&& \qquad \mbox{} - b^5 \chi_{\bf 4} - \frac{1}{b^5}\chi_{\bf 4} \Big) \nonumber \\
&+& O(x^8) ,
\eea
where $b^5 = q_1/q_2$.
This is also the perturbative part of the index of $SU(5)$ with two antisymmetric hypermultiplets.

\medskip

\subsection{$SU(2)\times SU(2) + 1$}

Only the $\theta$ parameter of the unflavored gauge group is relevant,
and we took that to be $\theta= \pi$. The index is expressed in terms of characters of the enhanced $SU(3)$ global symmetry,
and in terms of the instantonic and baryonic fugacities of the dual 
$SU(3)$ gauge theory with three fundamental hypermultiplets.
\bea
I^{SU(2)\times SU(2)_\pi +1} & = & 1 + x^2 (2 + \chi_{\bf 8}) \nonumber \\
&+& x^3 \Big( \chi_{\bf 2}[y](3 + \chi_{\bf 8}) + b^3 p + \frac{1}{b^3p} 
+ q\sqrt{b^3p}\, \chi_{\bf 3}  +  \frac{1}{q\sqrt{b^3p}}\, \chi_{\bar{\bf 3}} \Big) \nonumber \\
&+& x^4 \bigg[\chi_{\bf 3}[y](3 + \chi_{\bf 8}) + \chi_{\bf 2}[y]\Big(b^3p + \frac{1}{b^3p} + q \sqrt{b^3p}\chi_{\bf 3} + \frac{1}{q\sqrt{b^3p}} \chi_{\bar{\bf 3}}\Big)\nonumber \\
& & \qquad\qquad \mbox{} +  3+2\chi_{\bf 8}+\chi_{\bf 27} \bigg] \nonumber \\
&+& x^5 \bigg[\chi_{\bf 4}[y](3 + \chi_{\bf 8}) + \chi_{\bf 2}[y] (6+6\chi_{\bf 8} + \chi_{\bf 10} + \chi_{\overline{\bf 10}}
+ \chi_{\bf 27}) \nonumber \\
&& \qquad \qquad \mbox{} + (1+\chi_{\bf 3}[y])\Big(b^3p + \frac{1}{b^3p} + q\sqrt{b^3p}\, \chi_{\bf 3} 
+ \frac{1}{q\sqrt{b^3p}} \chi_{\bar{\bf 3}}\Big) \nonumber \\
&  & \qquad\qquad \mbox{} +  q\sqrt{b^3p} \chi_{\bf 15} + \frac{1}{q\sqrt{b^3p}} \chi_{\overline{\bf 15}}  \bigg] \nonumber \\
  &+& O(x^6) .
\eea

\medskip

\subsection{$SU(2)\times SU(2)\times SU(2)$}

Here we take $(\theta_1,\theta_2,\theta_3)=(\pi,0,\pi)$.
We express the result in terms of the enhanced $SU(4)$ global symmetry,
and in terms of the instantonic and baryonic fugacities of the dual $SU(4)$ gauge theory with four fundamentals hypermultiplets.
\bea
I^{SU(2)^3} & = & 1 + x^2 (2+\chi_{\bf 15}) + x^3 \chi_{\bf 2}[y](3+\chi_{\bf 15}) \nonumber \\
&+& x^4 \Big( \chi_{\bf 3}[y](3+\chi_{\bf 15}) + \chi_{\bf 84} + \chi_{\bf 20} + 2\chi_{\bf 15} + 3 + b^4 + \frac{1}{b^4} 
+ (q + \frac{1}{q})\chi_{\bf 6} \Big) \nonumber \\
 &+& x^5 \bigg[ \chi_{\bf 4}[y](3+\chi_{\bf 15}) \nonumber \\
 && \qquad \mbox{} + \chi_{\bf 2}[y] \Big(\chi_{\bf 84} + \chi_{\bf 45} + \chi_{\overline{\bf 45}} 
  +   \chi_{\bf 20} + 6\chi_{\bf 15} + 6 + b^4 + \frac{1}{b^4} + (q + \frac{1}{q})\chi_{\bf 6}\Big) \bigg] \nonumber \\
 &+& O(x^6) .
\eea
There are some different $SU(4)$ representations that have the same dimension.
Specifically the ${\bf 20}$ above is the $(0,2,0)$ representation and the ${\bf 84}$ is the $(2,0,2)$ when expressed in terms of the Cartan weights.


\begin{thebibliography}{40}


\bibitem{Seiberg:1996bd}
  N.~Seiberg,
  ``Five-dimensional SUSY field theories, nontrivial fixed points and string dynamics,''
  Phys.\ Lett.\ B {\bf 388}, 753 (1996)
  [hep-th/9608111].
  
\bibitem{Morrison:1996xf} 
  D.~R.~Morrison and N.~Seiberg,
  Nucl.\ Phys.\ B {\bf 483}, 229 (1997)
  [hep-th/9609070].
  
\bibitem{Douglas:1996xp} 
  M.~R.~Douglas, S.~H.~Katz and C.~Vafa,
  Nucl.\ Phys.\ B {\bf 497}, 155 (1997)
  [hep-th/9609071].
  
\bibitem{Intriligator:1997pq}
  K.~A.~Intriligator, D.~R.~Morrison and N.~Seiberg,
  ``Five-dimensional supersymmetric gauge theories and degenerations of Calabi-Yau spaces,''
  Nucl.\ Phys.\ B {\bf 497}, 56 (1997)
  [hep-th/9702198].
  

  
\bibitem{Aharony:1997ju} 
  O.~Aharony and A.~Hanany,
  Nucl.\ Phys.\ B {\bf 504}, 239 (1997)
  [hep-th/9704170].


\bibitem{KKL}
  H.~-C.~Kim, S.~-S.~Kim and K.~Lee,
  ``5-dim Superconformal Index with Enhanced En Global Symmetry,''
  JHEP {\bf 1210}, 142 (2012)
  [arXiv:1206.6781 [hep-th]].

\bibitem{Bergman:2013ala} 
  O.~Bergman, D.~Rodriguez-Gomez and G.~Zafrir,
  arXiv:1310.2150 [hep-th].

\bibitem{Iqbal:2012xm} 
  A.~Iqbal and C.~Vafa,
  arXiv:1210.3605 [hep-th].
  
\bibitem{Aharony:1997bh} 
  O.~Aharony, A.~Hanany and B.~Kol,
  JHEP {\bf 9801}, 002 (1998)
  [hep-th/9710116].

\bibitem{Bergman:2012kr}
  O.~Bergman and D.~Rodriguez-Gomez,
  ``5d quivers and their AdS(6) duals,''
  JHEP {\bf 1207}, 171 (2012)
  [arXiv:1206.3503 [hep-th]].
  
\bibitem{Bao:2011rc} 
  L.~Bao, E.~Pomoni, M.~Taki and F.~Yagi,
  JHEP {\bf 1204}, 105 (2012)
  [arXiv:1112.5228 [hep-th]].

  
  \bibitem{KMMS}
  J.~Kinney, J.~Maldacena, S.~Minwalla and S.~Raju,
  Commun.\ Math.\ Phys.\ 275:209-254 (2007) 
  [arXiv:0510251 [hep-th]].


\bibitem{Kitao:1998mf} 
  T.~Kitao, K.~Ohta and N.~Ohta,
  Nucl.\ Phys.\ B {\bf 539}, 79 (1999)
  [hep-th/9808111].
  
\bibitem{Bergman:1999na} 
  O.~Bergman, A.~Hanany, A.~Karch and B.~Kol,
  JHEP {\bf 9910}, 036 (1999)
  [hep-th/9908075].
  
  
  \bibitem{BMPTY}
  L.~Bao, V.~Mitev, E.~Pomoni, M.~Taki and F.~Yagi, 
  [arXiv:1310.3841 [hep-th]].
  
  \bibitem{HKT}
  H.~Hayashi, H.~-C.~Kim and T.~Nishinaka, 
  [arXiv:1310.3854 [hep-th]].

\bibitem{Nekrasov:2002qd} 
  N.~A.~Nekrasov,
  Adv.\ Theor.\ Math.\ Phys.\  {\bf 7}, 831 (2004)
  [hep-th/0206161].
 
\bibitem{Nekrasov:2004vw} 
  N.~Nekrasov and S.~Shadchin,
  Commun.\ Math.\ Phys.\  {\bf 252}, 359 (2004)
  [hep-th/0404225].
  

 
  
\bibitem{AGT}
  L.~F.~Alday, D.~Gaiotto, and Y.~Tachikawa,
  Lett. \ Math. \ Phys.\   91:167-197 (2010)
  [arXiv:0906.3219 [hep-th]].

\bibitem{Taki}
  M.~Taki,
  [arXiv:1310.7509 [hep-th]].
  
  
\bibitem{Aharony:2013dha} 
  O.~Aharony, S.~S.~Razamat, N.~Seiberg and B.~Willett,
  JHEP {\bf 1307}, 149 (2013)
  [arXiv:1305.3924 [hep-th]].

  
 
\bibitem{Shad}
  S.~Shadchin, 
  [arXiv:0502180 [hep-th]].


\bibitem{Shad2}
  S.~Shadchin, 
  JHEP {\bf 0603}, 046 (2006)
  [arXiv:0511132 [hep-th]].  

\bibitem{HKS}
  L.~Hollands, C.~A.~Keller, and J.~Song,
  JHEP {\bf 1103}, 053 (2011) 
  [arXiv:1012.4468 [hep-th]].
  

\end{thebibliography}
\end{document}